\documentclass[11pt,a4paper]{article}
\pdfoutput=1
\usepackage{jheppub}
\usepackage{amsfonts,amsbsy,amsmath,amssymb}
\usepackage{graphicx}
\usepackage{booktabs}
\usepackage{caption}
\usepackage{subcaption}
\usepackage{multirow}
\usepackage{algorithm}
\usepackage[noend]{algpseudocode}
\usepackage[utf8]{inputenc}

\newcommand{\HG}{\hat{\Gamma}}
\newcommand{\be}{\begin{equation}}
\newcommand{\ee}{\end{equation}}
\newcommand{\ba}{\begin{array}}
\newcommand{\ea}{\end{array}}
\newcommand{\baa}{\begin{array}}
\newcommand{\eaa}{\end{array}}
\newcommand{\bea}{\begin{eqnarray}}
\newcommand{\eea}{\end{eqnarray}}
\newcommand{\half}{\frac{1}{2}}

\newcommand{\Tr}{\mathrm{Tr}}

\newcommand{\m}{k} 
\newcommand{\kb}{\bar{k}} 

\newcommand{\FSS}{F^2(p,q,-p-q)}

\newcommand{\ttheta}{\tilde \theta}
\newcommand{\hL}{\hat{L}}

\newcommand{\LL}{L}
\newcommand{\kbar}{\bar k}

\newcommand{\tw}{\tilde{W}}
\newcommand{\hw}{\hat{W}}
\newcommand{\RR}{\mathcal{G}}
\newcommand{\Le}{L_{\mathrm{eff}}}
\newcommand{\LLe}{\Lambda_{\mathrm{mom}}}
\newcommand{\Ve}{V_{\mathrm{eff}}}
\newcommand{\pbc}{\mathrm{PBC}}
\newcommand{\tbc}{\mathrm{TBC}}

\newcommand{\hA}{\hat{A}}
\newcommand{\whq}{\widehat{q}}
\newcommand{\sq}{{\ 2}}
\newcommand{\muu}{{\sigma}}
\newcommand{\nuu}{{\delta}}
\newcommand{\V}{V}
\newcommand{\Ls}{\Lambda_\LL}
\newcommand{\comment}[1]{}
\newcommand{\sig}{\mathcal{S}}
\newcommand{\Lv}{{\bf \LL}}
\newcommand{\FI}{{\cal F}}
\newcommand{\WA}{(b,N,\Lv,n_{\mu\nu})}
\newcommand{\WAa}{(b,N,\Lv)}
\newcommand{\WAb}{(N,\Lv,n_{\mu\nu})}
\newcommand{\WAc}{(\kappa, N,\Lv,n_{\mu\nu})}
\newcommand{\gm}{\mathcal{G}}

\title{Perturbative contributions to   Wilson loops in twisted lattice boxes
and reduced models}

\author[a]{Margarita Garc\'{i}a P\'erez,}
\author[a,b]{Antonio Gonz\'alez-Arroyo,}
\author[c,d]{Masanori Okawa}

\affiliation[a]{Instituto de F\'{i}sica Te\'orica UAM-CSIC, Nicol\'as
  Cabrera 13-15, Universidad Aut\'onoma de Madrid, E-28049 Madrid,
  Spain}

\affiliation[b]{Departamento de F\'{i}sica Te\'orica, C-XI Universidad
Aut\'onoma de Madrid, E-28049 Madrid, Spain}
\affiliation[c]{Graduate School of Science, Hiroshima University,
  Higashi-Hiroshima, Hiroshima 739-8526, Japan}
\affiliation[d]{Core of Research for the Energetic Universe, Hiroshima University,
                     Higashi-Hiroshima, Hiroshima 739-8526, Japan}
\emailAdd{margarita.garcia@uam.es}
\emailAdd{antonio.gonzalez-arroyo@uam.es}
\emailAdd{okawa@sci.hiroshima-u.ac.jp}

\abstract{
We compute  the perturbative expression of  Wilson loops up to order 
$g^4$ for SU($N$) lattice gauge theories with Wilson action
on a finite  box with twisted boundary conditions. Our formulas are valid for any dimension and any irreducible twist.  They contain 
as a special case that of the 4-dimensional Twisted Eguchi-Kawai model for 
a symmetric twist with flux $\m$. Our results allow us to analyze the 
finite volume corrections as a function of the flux. In particular, 
one can quantify the approach to volume independence at large $N$ as a
function of flux $\m$. The contribution of  fermion fields in the adjoint representation is also analyzed.} 


\preprint{%
{\flushright
IFT-UAM/CSIC-17-067\\
FTUAM-17-12\\
HUPD-1707\\
}}

\begin{document}


\maketitle

\section{Introduction}
\label{s.intro}

Within lattice gauge theory calculations, finite volume perturbative studies
are interesting for various reasons, one  being that numerical results
are always at finite volume. From the first studies it was  clear that the
periodic boundary conditions (PBC) introduced complications
associated to the existence of infinitely many gauge inequivalent
zero-action configurations: the torons~\cite{GonzalezArroyo:1981vw}. Furthermore, the
toron valley is not a manifold, but rather an orbifold possessing
singular faces and points. Considerable effort was put into setting up a
consistent computational weak coupling expansion~\cite{Luscher:1982ma,Luscher:1983gm,Coste:1985mn,vanBaal:1986ag,Koller:1987fq}.
Other studies  simply ignored the problem by expanding around a single
type of minima~\cite{Heller:1984hx}. The results should approach those
obtained at infinite volume~\cite{DiGiacomo:1981wt,Weisz:1982zw,Weisz:1983bn,Wohlert:1984hk,Alles:1998is} \footnote{Except when explicitly
specified in this paper we will concentrate upon the $d=4$ space-time
dimensional case.}.

`t Hooft realized that PBC  are not the only possible
boundary conditions for SU(N) gauge theories on the torus. He
introduced the concept of twisted boundary conditions (TBC)~\cite{'tHooft:1979uj,'tHooft:1980dx}
that  was soon translated to lattice computations~\cite{Groeneveld:1980tt}.
The new boundary conditions associate to each plane of the torus a certain flux defined modulo $N$, collected into an integer-valued twist tensor $n_{\mu \nu}$.
Already in the first studies it  became clear that TBC  introduced considerable simplification in
perturbative calculations at finite volume~\cite{GonzalezArroyo:1981vw}.

The size of finite volume corrections  was found to be  directly connected to the magnitude of $N$, the number of colours of the theory.
This followed  from the observation made by  Eguchi and Kawai~\cite{Eguchi:1982nm}
when studying the Schwinger-Dyson equations for Wilson loops. Their claim
can be phrased as the statement that, under certain assumptions,
finite volume corrections vanish in the large $N$ limit. This leads to the
large $N$ equivalence of ordinary lattice
gluodynamics with matrix models obtained by collapsing the lattice to a single point: Reduced models. If volume independence holds in the weak coupling region, this should show up in the perturbative expansion of Wilson
loops. This was found to be false~\cite{Bhanot:1982sh} for the original proposal of Ref.~\cite{Eguchi:1982nm} (Eguchi-Kawai model).  The problem arises from the
attraction among the eigenvalues of the Polyakov loops induced by quantum corrections. This invalidates the $Z^4(N)$ center-symmetry assumption of the equivalence proof. This could have been anticipated on the basis of the results of Ref.~\cite{GonzalezArroyo:1981vw}.

Identification of the source of the breakdown
allowed the authors of Ref.~\cite{Bhanot:1982sh} to propose a modification, called the Quenched
Eguchi-Kawai (QEK) model, which could solve the problem. In this proposal the expectation values were computed taking  these eigenvalues as {\em frozen} or quenched. The final results were then averaged over
them. Within the perturbative regime this idea  was analyzed in Ref.~\cite{Gross:1982at} as part of a  general framework called the {\em quenched momentum prescription}. It was shown how the reduced
model reproduced the perturbative expansion of the full theory. Indeed, the aforementioned eigenvalues played the role of effective momentum degrees of freedom. This particular connection between internal and space-time degrees of freedom is quite general as shown by Parisi~\cite{Parisi:1982gp}.

In all the previous works periodic boundary conditions were assumed. However, two of the present authors~\cite{GonzalezArroyo:1982ub} argued that
Eguchi-Kawai proof holds also for TBC, which as mentioned earlier have a very different weak-coupling behaviour. This allowed them to present a reduced model, called  the Twisted Eguchi-Kawai model~\cite{GonzalezArroyo:1982hz} (TEK),
which could achieve the large N volume independence result at all values of the coupling. With a suitable choice of the twist-tensor $n_{\mu \nu}$ one is guaranteed to have zero-action solutions without the zero-modes (torons) which complicate the perturbative expansion in
the absence of twist. One particular simple choice of twist is the so-called {\em symmetric twist} which demands $N=\hL^2$ and has a common flux $|n_{\mu \nu}|= k \hL$. In this case the classical vacua break the
$Z^4(N)$ center symmetry of the  model down to $Z^4(\hL)$.
This remnant symmetry is enough to guarantee the volume independence of loop equations in the large N limit.

The authors of Ref.~\cite{GonzalezArroyo:1982hz} then considered the perturbative expansion of the model by expanding around any of the $N^2$ gauge inequivalent vacua. The Feynman rules were obtained and this iluminated the 
way in which the infinite volume theory is recovered from the matrix model. An important ingredient is the use of a basis of the Lie algebra of the group which has the form of a Fourier expansion. This illustrates a new and more efficient way in which space-time degrees of freedom are obtained from those in the group: the $N^2$ degrees of freedom of the U(N) group show up as the spatial momenta  of an $\hL^4$ lattice (colour momenta). The idea can be used to achieve a volume reduction of theories with scalar or fermionic fields and can be extended to the continuum~\cite{GonzalezArroyo:1983ac}\footnote{This gave rise to Feynman rules later to be recognized as those of field theory in non-commutative spaces.}. A similar treatment was done for $d=2$ and non-gauge theories in Ref.~\cite{Eguchi:1982ta}.

A bonus of this perturbative construction is that it gives a hint of what happens at finite $N$. The propagators are  identical to lattice propagators at finite volume. Thus, finite $N$ corrections appear in part as finite volume corrections. This is not the end of the story because the Feynman rules of the vertices adopt a peculiar form,  including colour-momentum dependent phase factors. 
 In Ref.~\cite{GonzalezArroyo:1982hz}, 
the authors showed how these phase factors cancel out in planar diagrams. It is
these surviving phases that are instrumental  in suppressing
non-planar diagrams and reproducing the perturbative expansion of large $N$ gauge theories. Many years later~\cite{Connes:1997cr} the origin of these peculiar phase factors was clarified as a distinctive feature of field theories in non-commutative space-times (for a review see Ref.~\cite{Douglas:2001ba}). This led some authors~\cite{Ambjorn:1999ts,Ambjorn:2000nb,Ambjorn:2000cs} to propose the use of the TEK model as a regulated version of gauge theories of this type, somehow inverting the path that led to Ref.~\cite{GonzalezArroyo:1982hz}. 

The interest of studying (lattice) gauge theories with TBC beyond the large $N$ reduced model context was soon emphasized  by several authors~\cite{Fabricius:1985jw,Luscher:1985zq,Luscher:1985wf,Coste:1986cb} and the perturbative technique extended to include space-time momenta added to the colour momenta.  Since then,  several perturbative calculations have been performed 
with different  observables  and contexts in mind~\cite{Hansson:1986ia,GonzalezArroyo:1988dz,Daniel:1989kj,Daniel:1990iz,Snippe:1996bk,Snippe:1997ru,Perez:2013dra}. 

Our present work focuses on the perturbative expansion up to order $g^4$ of Wilson loops for an SU(N) lattice gauge theory with  Wilson  action in any dimension and in a box with any irreducible orthogonal twisted boundary conditions. This means that the twist must allow the existence of  discrete zero-action solutions and no zero-modes. This includes  the case of the symmetric twist used in the TEK model. Our formalism is developed for any box size. In this
way we bridge the gap between the infinite volume perturbative results and the $\LL=1$ TEK model. Some preliminary results were presented in the 2016 lattice conference~\cite{GarciaPerez:2016guo}.

There are many interesting issues that our analysis aims at elucidating. These are connected to the interplay between the different parameters entering the game: the box size, the rank of the matrices $N$ and the integer fluxes  defining the twist. One of the aspects has to do with the approach to the large $N$ limit. Volume independence would imply that in this limit the results should not depend on the lattice size. However, it is interesting to ask, as the authors of Ref.~\cite{Kiskis:2003rd,Narayanan:2003fc} did in the periodic boundary conditions case, what is the optimal  balance of spatial and group degrees degrees of freedom that minimizes corrections. This is intimately connected to the important
practical problem of estimating the corrections to volume independence for large but finite $N$. Depending on the results, the usefulness of reduced models as an effective simulation method to compute observables of  the large $N$ theory could be severely limited. From a more conceptual viewpoint one
would like to understand the nature of these corrections. As mentioned earlier, some of these finite $N$ corrections amount to  finite volume corrections with an effective volume which depends on $N$ ($\sqrt{N}$ for the 4 dimensional symmetric twist). However, that is certainly not all. Some effects 
do  depend on the phase factors at the vertices, which
are a function of the  twist tensor. The relevance of monitoring this dependence has  been recognized recently in certain non-perturbative studies of the TEK model. At intermediate values of the coupling, several authors~\cite{ishikawa:2003,Teper:2006sp,Vairinhos:2007qz,Azeyanagi:2007su} reported signals that the center symmetry of the four dimensional TEK  model was broken spontaneously. This is crucial, since in that case the proof of volume independence of Eguchi and Kawai fails. The problem was analyzed in Ref.~\cite{GonzalezArroyo:2010ss}, concluding that to avoid the problem one should scale appropriately the unique flux parameter $\m$ of the model when taking the large $N$ limit. The validity of volume reduction under these premises has been verified  in very precise measurements of Wilson loops~\cite{Gonzalez-Arroyo:2014dua}. Similar constraints are found when analyzing 2+1 dimensional theories defined on a spatial torus with twist~\cite{Perez:2013dra,Perez:2014sqa}. In that case the problem of finding an  optimal flux is related to some recent conjecture in Number Theory~\cite{Chamizo:2016msz}.

Obviously, all these problems do not arise in perturbation theory since centre-symmetry cannot be broken in our finite $N$ and finite volume setting. However, the analytic calculations of perturbation theory can give hints about the origin of the possible transitions occuring when taking the volume or $N$ to infinity. We emphasize that our computation, being of order $\lambda^2$, already includes self-energy and vertex gluonic contributions which contain ultraviolet divergences in the continuum limit. This also relates to problems reported in the perturbative expansion of non-commutative field theories~\cite{Martin:1999aq,Krajewski:1999ja,SheikhJabbari:1999iw,Minwalla:1999px,Matusis:2000jf,Guralnik:2001pv,Guralnik:2002ru} having to do precisely with the self-energy of the gluon. We recall that the twisted theory can be seen as a regulated  version of Yang Mills theory on the non-commutative torus. Indeed, some instabilities also arose when analyzing the model within that context~\cite{Bietenholz:2006cz}.

The lay-out of the paper is as follows. In section~\ref{s.method} we set up the methodology. Our presentation is mostly self-contained and general enough to cover all the twists of the allowed type. At particular points we focus on the specific situation for symmetric twists in 2 and 4 dimensions. To facilitate the reading the Feynman rules necessary to perform the calculation are collected in two appendices.  In section~\ref{s.results} we present our results, first to order $\lambda\equiv g^2 N$, and next to order $\lambda^2$. These results appear as finite sums over a range of momentum values which depends on the twist tensor. In the next section (section~\ref{s.analysis}) we analyze these results. In particular, we split the contributions into sets and study the difference between the computations with twist and those obtained with periodic boundary conditions ignoring the contributions of zero-modes. The focus is on the analysis of the dependence $N$ and the box size specially when any of the two is large. For practical reasons our analysis is concentrated on the case of a symmetric box with a symmetric twist where there is only one size parameter $\LL$ and one flux value $\m$.  This is specially the case in section~\ref{s.numerical} where some of the sums are evaluated numerically and the results analyzed. Comparison of the finite $N$ corrections for the reduced model and other partially reduced options is also addressed. Several formulas that allow the analytic calculation of the leading finite volume corrections are collected in the last two appendices of the paper.

In Section~\ref{s.additional} we try to make our analysis more complete by analyzing a few  extensions. In particular we consider the distinction between U(N) and SU(N) results and the additional contributions to the Wilson loop coming from the  quarks  in the adjoint representation of the group. The latter can be included in a twisted setting in a rather straightforward way, while quarks in the fundamental need the addition of replicas (flavours). The explicit formulas for fermions have been obtained for Wilson fermions at $r=1$ and critical value of the hopping. In that same section we also compare our results with high statistics measurements of the loops with standard Monte Carlo techniques. Our goal is to be able to test extremely tiny effects such as the breakdown of cubic invariance by the twist as well as the non-zero value of the imaginary part. The data match perfectly with the expectations. Furthermore, this analysis also allows for an estimate of the coefficient of order $\lambda^3$ and the determination of the range of couplings for which the truncated perturbative expansion is a good approximation. 

The paper closes with a conclusions section in which we sum up the main results following from our calculation. 

\section{Models and methodology}
\label{s.method}

In this section we will describe the type of models that we will be
considering as well as the tools necessary for the calculation of the
coefficients. 

\subsection{The action with twist}

We will be considering d-dimensional SU(N)  lattice  gauge theory with Wilson action on an hypercubic box of size  $\LL_0\times \cdots \times \LL_{d-1}$. The  $\LL_\mu$ can be taken as the components of a d-dimensional vector $\Lv$. The product of the sizes gives the total lattice volume labelled $\V$.
Different lengths in different  directions break the hypercubic invariance. Hence, for simplicity we will often specify results for a symmetric box $\LL_\mu=\LL$.  
The
action depends on a single coupling $b=\beta/(2 N^2)$.  We will
focus upon the behaviour of the expectation values of 
$R\times T$ rectangular  Wilson loops:
\be
W_{R,T} \WAa=\frac{1}{N}\langle \mathrm{Tr}(U(R,T)) \rangle\, ,
\ee
where $U(R,T)$ is the ordered product of all links around the
perimeter of the $R\times T$ rectangle.
Our goal is to study the behaviour of these 
observables for large values of $b$. In this limit the result is
calculable using perturbative/weak-coupling techniques. As mentioned
in the introduction, this  problem has been addressed earlier by
several authors~\cite{Heller:1984hx,DiGiacomo:1981wt,Weisz:1982zw,Weisz:1983bn,Wohlert:1984hk,Alles:1998is}. The main difference of our work with
others is that  we will consider arbitrary orthogonal irreducible  twisted  boundary conditions on the lattice~\cite{'tHooft:1979uj,review}. In particular this will include symmetric twisted boundary conditions in four dimensions~\cite{GonzalezArroyo:1982hz}.

We do not intend to review the formalism to implement twisted boundary conditions on the lattice~\cite{Groeneveld:1980zx,Groeneveld:1980tt}. We will just remind the readers that after a change of variables on the links one reaches an action of the form 
\be
\label{action}
S = b N \sum_n \sum_{\mu \ne \nu } [N - Z_{\mu \nu }(n) \Tr (U_\mu(n)
U_\nu(n+\hat \mu ) 
U_\mu^\dagger(n+\hat \nu ) U_\nu^\dagger(n)) ]\, ,
\ee
where the link matrices are periodic
$U_\mu(n)=U_\mu(n+\LL_\nu \hat \nu)$, and the plaquette factors $Z_{\mu \nu}(n)=Z^*_{\nu \mu}(n)$ are elements of the center. Not all values of $Z_{\mu \nu}(n)$ amount to twisted boundary conditions. First of all, it is necessary that the product of all $Z$ factors over the faces of every cube (taken with orientation) is equal one. The non-trivial twist follows by multiplying all the $Z$ factors in each $\mu-\nu$ plane, to give an overall center-element $\hat Z_{\mu \nu}$:
\be
\hat Z_{\mu \nu} \equiv \prod_{n_\mu=0}^{\LL_\mu-1} \prod_{n_\nu=0}^{\LL_\nu-1} Z_{\mu \nu}(n)\, .
\ee
Because of the condition on cubes, the $\hat Z_{\mu \nu}$ do not depend on the position of the plane but only on its orientation. Being elements of the center one can write  $\hat Z_{\mu \nu}=\exp\{2\pi i n_{\mu \nu}/N\}$, where $n_{\mu \nu}$ is an antisymmetric tensor of integers defined modulo $N$. It is this twist tensor  that specifies the twist. Redefining the link variables by multiplication with an element of the centre, one can change the value of the individual plaquette  
factors $Z_{\mu \nu}(n)$, but the twist tensor remains unaffected. This allows one to set all the plaquette factors to 1,  
except for a single {\em twisted} plaquette in each $\mu-\nu$ plane.
Conventionally, one can choose that plaquette to be one at the corner ($n_\mu=\LL_\mu-1$; $n_\nu=\LL_\nu-1$).

The change of variables that led to the action Eq.~\eqref{action}, also transforms the Wilson loop expectation variables. These are modified as follows
\be
\label{expectWL}
W_{R,T}\WA=\frac{1}{N} Z(R,T)\langle \Tr(U(R,T)) \rangle
\ee
where $Z(R,T)$ is the product of the $Z_{\mu \nu}(n)$ factors for all
plaquettes which fill up the rectangle. Notice that the result will also depend on the directions defining the plane in which the rectangle is sitting. 

In the next subsections we will derive the perturbative expansion of these quantities up to order $1/b^2$. We will specify the meaning of orthogonal irreducible twists and provide some examples in various space-time dimensions.

\subsection{Classical minima of the action}
As $b\longrightarrow \infty$ the
functional integral is dominated by the configurations that minimize the action. We will restrict ourselves to twist tensors for which the corresponding minimum action vanishes. These are called orthogonal twists~\footnote{The name has its origin in the four-dimensional case.}. The corresponding  zero-action  configuration 
will be named as follows $U_\mu(n)\rightarrow \Gamma_\mu(n)$. The zero-action condition implies  that  the  SU(N) matrices 
$\Gamma_\mu(n)$ satisfy
\be
\label{zeroaction}
\Gamma_\mu(n) \Gamma_\nu(n+\hat{\mu})=Z_{\nu \mu}(n) \Gamma_\nu(n)
\Gamma_\mu(n+\hat{\nu})
\ee
Obviously, the solution  is not unique, since any gauge transformation 
of this one gives also a zero-action solution:
\be
\Gamma_\mu(n) \longrightarrow \Omega(n) \Gamma_\mu(n)  \Omega^\dagger(n+\hat{\mu})
\ee
where $\Omega(n)$ are arbitrary SU(N) matrices periodic on the torus.  New
solutions can also be  obtained by the replacement 
\be
\Gamma_\mu(n) \longrightarrow z_\mu \Gamma_\mu(n) 
\ee
where $z_\mu$ is an element of the center. In some cases these solutions are gauge inequivalent to the previous ones. To study this point one must analyze the remaining gauge invariant observables of this zero-action configurations: the Polyakov loops. 

For every lattice path $\gamma$ with origin in one lattice point,  which we label $0$, and endpoint $n$, one can construct the path-ordered product for this zero-action configuration which we will label $\Gamma(\gamma)$. The closed lattice paths can be classified into subsets according to its winding numbers around each of the torus directions. The condition Eq.~\eqref{zeroaction} implies that the contractible loops (which are associated to vanishing winding) are given by elements of the center. Similarly, those paths having the same number of windings have path-ordered exponentials which are unique up to multiplication by an element of the center. Now we can choose one  representative path having winding 1 in the direction $\mu$ and zero in the remaining torus directions, we will call its associated path-ordered exponential  $\Gamma_\mu$. From   Eq.~\eqref{zeroaction} we deduce that these SU(N) matrices must satisfy
\be
\label{twisteq}
\Gamma_\mu \Gamma_\nu=\hat Z_{\nu \mu} \Gamma_\nu
\Gamma_\mu
\ee
where $\hat Z_{\nu \mu}$ are the elements of $Z_N$ introduced earlier and characterizing  the twist. Notice that the previous equation does not depend on the choice of    representative paths. Indeed, it is the existence of solutions to Eq.~\eqref{twisteq} what guarantees the existence of zero-action solutions and the definition of orthogonal twists. For a more thorough discussion of the conditions on the twist tensor $n_{\mu \nu}$ we refer the reader to Ref.~\cite{review}. Furthermore, we will restrict ourselves to what we call irreducible twists. These are defined by the equivalent of Schur lemma,  stated by saying that the only matrices which commute with all $\Gamma_\mu$ are the multiples of the identity.   If we restrict ourselves to SU(N) matrices, the set of gauge-inequivalent solutions becomes discrete. From a practical viewpoint, the irreducibility  condition eliminates the presence of zero-modes which complicate the perturbative analysis considerably. Hence, the $\Gamma_\mu$   matrices are uniquely defined up to a unitary similarity transformation, which is just a global gauge transformation,  and a multiplication by an element of the centre. The second operation defines the  centre symmetry group. However, not all these transformations produce gauge-inequivalent solutions. This only occurs when the eigenvalues of the $\Gamma_\mu$ matrices, which are gauge invariant quantities, change~\cite{review}.

The irreducibility condition  implies that the algebra generated by multiplication of the  $\Gamma_\mu$ matrices has complex dimension $N^2$. It also implies  that $\Gamma_\mu^{N}$ must be a multiple of the identity for all $\mu$.

\subsection{Derivation of the perturbative expansion}
To proceed  with the  perturbative expansion one has  to expand the link matrices
around the zero-action solutions as follows
\be
\label{expansion}
U_\mu(n)= e^{-i g A_\mu(n)}  \Gamma_\mu(n)
\ee
This is just an expansion around a background field and, as is
well-known (see for example \cite{Dashen:1980vm,GonzalezArroyo:1981ce}), the plaquette becomes
\be
Z^*_{\mu \nu}(n)\Tr(e^{-i g A_\mu(n)}e^{-i g A'_\nu(n+\hat{\mu})} e^{i
g A'_\mu(n+\hat{\nu})} e^{-i g A_\nu(n)}) =Z^*_{\mu \nu}(n) \Tr(e^{-i
g{\cal G}_{\mu \nu}(n) })
\ee
where 
\be
A'_\nu(n+\hat{\mu})=\Gamma_\mu(n)A_\nu(n+\hat{\mu})
\Gamma_\mu^\dagger(n)\equiv A_\nu(n)+ \nabla_\mu^+ A_\nu(n)
\ee
where we have introduced the forward lattice  derivative 
$\nabla_\mu^+$. In our case it is actually  a covariant derivative with 
respect to the background field given by the zero-action solution.
The expression of ${\cal G}_{\mu \nu}(n)$, obtained by applying the
Baker-Campbell-Haussdorf formula, is similar to the one
obtained for periodic boundary conditions with the primes modifying
the translated vector potential. For example, the leading term is 
\be
{\cal G}_{\mu \nu}(n)= \nabla_\mu^+ A_\nu(n) -\nabla_\nu^+ A_\mu(n)
+{\cal O}(g)
\ee
The lattice vector potentials, for each link direction $\rho$,  are $V$ traceless hermitian $N\times
N$ matrices ($\V$ is lattice volume). This is a $\V (N^2-1)$-dimensional real vector space, 
and we can take as its  basis the simultaneous eigenstates of the 
 $\nabla_\mu^+$ operators (notice that they commute). We call these 
 basis vectors $\chi(n;q)$ (which are not necessarily hermitian) and they satisfy 
 \be
 \nabla_\mu^+ \chi(n;q)= (e^{i q_\mu }-1) \chi(n;q)
 \ee
 The form of the eigenvalues comes from the definition of the operator $\nabla_\mu^+$. Spelling out the condition one must have 
\be 
\label{eigenval}
\Gamma_\mu(n)\chi(n+\hat{\mu};q)\Gamma^\dagger_\mu(n)= e^{i q_\mu} \chi(n;q)
 \ee
To solve this equation we choose one reference point on the lattice, which without loss of generality we fix as $n=0$. For any other point we choose a non-winding forward moving path $\gamma(n)$ joining the origin with that point. Then we have 
\be
\label{eigenvaN}
 \chi(n;q)= e^{i q n} \Gamma^\dagger(\gamma(n)) \chi(0;q) \Gamma(\gamma(n)) 
 \ee
 Notice that the solution does not depend on the choice of path $\gamma(n)$, because the corresponding matrices  differ by multiplication by an element of the center. The final requirement is that the eigenvectors satisfy the required periodic boundary conditions. Defining $\HG(q)\equiv \chi(0;q)$, this condition   implies that for any direction $\mu$ we must have
 \be
 \Gamma_\mu \HG(q)\Gamma^\dagger_\mu = e^{i L_\mu q_\mu}  \HG(q) 
 \ee
 This is a well-studied matrix equation. The condition of irreducibility implies that there are $(N^2-1)$ traceless linearly independent solutions. From irreducibility one can conclude that $L_\mu q_\mu$ must be an integer multiple of $2 \pi/N$. Hence, we can write $L_\mu q_\mu = \frac{2 \pi m_\mu}{N}$, where the integers $m_\mu$ are defined modulo $N$. If we remove the condition of vanishing trace there is an additional solution given by a multiple of the identity matrix and having $m_\mu=0$.
 
 Now, coming back to the original eigenvalue equation Eq.~\eqref{eigenval}, and realizing that the $q_\mu$ are defined modulo $2\pi$, we conclude that we have a total of $\V(N^2-1)$ different eigenvalues, each characterized by a different d-dimensional vector $q$. These momentum vectors have the form 
 \be
 \label{momDecomp}
 q_\mu= \frac{2 \pi m_\mu}{N \LL_\mu}+\frac{2 \pi r_\mu}{\LL_\mu} \equiv q_\mu^c  + q_\mu^s
 \ee
 where the $m_\mu$ are the integers introduced earlier, which enter in  the first term which we call colour momentum. The second term has the standard form of momenta  in a periodic lattice and is  thus labelled spatial momentum. It is convenient to include also the $m_\mu=0$ solution because then the set of momenta  has the structure of a finite abelian group, which we will call $\LLe$. It is a subgroup of the group 
 $$ \bigotimes_\mu  \left(2\pi (\mathbb{Z}/(N \LL_\mu \mathbb{Z}) \right) $$
Furthermore, the set of spatial momenta $\Ls$ is a subgroup of $\LLe$ having $\V$ elements. Colour momenta are more rigorously identified with elements of the quotient group $\LLe/\Ls$, having $N^2$ elements. 

Let us now focus on the eigenvector $\chi(n;q)$. The eigenvalue equation only fixes these matrices up to multiplication by a constant. Part of the arbitrarity can be fixed by a normalization condition. 
We will  impose 
\be
\frac{1}{\V} \sum_n \Tr(\chi^\dagger(n;p) \chi(n;q)) = \frac{1}{2}\delta_{p,q}
\ee 
which fixes $\HG(q)$ to be a unitary  matrix divided by $\sqrt{2N}$.
This leaves a phase arbitrarity which can be further reduced by imposing additional conditions on the unitary matrix. For example, one can impose that it belongs to SU(N). 
Alternatively, we can  impose that $(\sqrt{2N}\HG(q))^N=\mathbf{I}$. Any of the two conditions, which  might be incompatible with each other as we will see later, reduces the arbitrarity to multiplication by an element of the center $Z_N$. 
A choice of this element for every $q$ fixes the assignment 
\be
q \longrightarrow \HG(q)
\ee
This  provides a group homomorphism from $\LLe$ to $SU(N)/Z_N$.
For the SU(N) normalization condition, this implies 
\be
\label{Phidef}
 \HG(q) \HG(p) = \frac{e^{i \Phi(q,p)}}{\sqrt{2N}}\ \HG(p+q)
 \ee
 where $\Phi(q,p)$ is an integer multiple of $2 \pi/N$, which depends on the choice of phases. We can restrict $\HG(0)$ by demanding $\Phi(0,q)=\Phi(q,0)=0$. If we adopt the  $(\sqrt{2N}\HG(q))^N=\mathbf{I}$ condition,  Eq.~\eqref{Phidef} also holds, but then $e^{i \Phi(q,p)}$ could be an element of $Z_{2N}$. 
 
Having solved the eigenvalue  and eigenvector problem, we realize that given that our solutions are a collection of linearly independent matrix fields, we can actually decompose our vector potentials as follows
\be
\label{Fourier}
A_\mu(n)=\frac{1}{\sqrt{\V}}\sum_{q \in \LLe\setminus\Ls} \hA_\mu(q) \chi(n;q)\equiv \frac{1}{\sqrt{\V}} \sum_{q}' \hA_\mu(q) \chi(n;q) 
 \ee
 which generalizes the Fourier decomposition. Two comments are necessary at this point. The first affects the condition that the vector potentials are hermitian matrices. This imposes a constraint on the Fourier coefficients $\hA_\mu(q)$ as follows:
 \be
 \label{hermit}
\hA_\mu^*(q)= e^{i \Phi(q,-q)} \hA_\mu(-q)
\ee
The second is the requirement of tracelessness, specific of SU(N). This implies that $\hA(0)=0$ for $q\in\Ls$. Hence, the sum extends over the set difference $\LLe\setminus\Ls$. For simplicity this restriction will be  noted by the prime affecting the summation symbol.

Combining the previous expression we can define 
\bea
\nonumber
D(q,p,k)+iF(q,p,k)&=&4\delta(p+q+k)\, \Tr\left(\HG(q)\HG(p)\HG(-p-q)\right)=\\ 
&=&\delta(p+q+k) \sqrt{\frac{2}{N}}\,  e^{i\Phi(q,p)+i\Phi(q+p,-q-p)} 
\eea
which play the role of the $d$ and $f$ symbols of the SU(N) Lie algebra in our basis. By definition $D$ is completely symmetric and $F$ completely antisymmetric under the exchange of their arguments.

What is the connection between the choice of twist tensor and the value of the lattice of momenta $\LLe$? What is the explicit form of the matrices $\HG(q)$ and of the $F$ and $D$ functions? This can be analysed as follows. As mentioned previously the matrices $\Gamma_\mu$ generate an algebra by multiplication. By irreducibility, this algebra has dimension $N^2$. Hence, the matrices $\HG(q)$ must necessarily have the following form   
\be
\HG(q) = \frac{1}{\sqrt{2N}}\, e^{i \alpha(q)} \, \Gamma_0^{s_0(q)}
\cdots \Gamma_{d-1}^{s_{d-1}(q)}
\ee
where $\alpha(q)$ are integer multiples of $\pi/N$ and $s_\mu(q)$ are integers defined modulo $2N$. Using the commutation relations of the $\Gamma_\mu$ one can find the relation between the integers $s_\mu(q)$ and $q$, given by 
\be
\label{momrel}
q_\mu L_\mu=\frac{2 \pi}{N} \sum_{\nu} n_{\nu \mu} s_\nu(q)
\ (\bmod 2\pi)\ee
The previous equation defines a homomorphism $\mathcal{N}$ from the group
$(\mathbb{Z}/2N\mathbb{Z})^d$ to $\LLe/\Ls$. This cannot be an isomorphism except in two dimensions since the number of elements in $\LLe/\Ls$ is $N^2$, which is smaller that $(2N)^d$. Indeed, using the isomorphism theorem we conclude that 
\be
\LLe/\Ls  \cong (\mathbb{Z}/2N\mathbb{Z})^d/\ker(\mathcal{N})
\ee
This allows the computation of $\LLe$ given the twist tensor. On the other hand the inverse $q\longrightarrow s(q)$ is not uniquely defined and is a matter of convention. This convention dependence is also present in the choice of elements $\alpha(q)$. Indeed, any choice of inverse can always be compensated by appropriately choosing the $\alpha$. The convention dependence extends to the value of $\Phi(q,p)$ and the $F$ and $D$ symbols. A convenient choice is to impose the condition $\Phi(p,-p)=0$. This equation makes the hermiticity condition  Eq.~\eqref{hermit} look just like in the ordinary Fourier decomposition of a real field. It should be noted though, that for even values of $N$ the condition might conflict with the SU(N) normalization condition. In combination with the alternative condition $(\sqrt{2N}\HG(q))^N=\mathbf{I}$, it fixes the value of $\HG(q)$ up to a sign. We stress, nonetheless, that the convention adopted for the definition of $\HG(q)$  affects 
only the corresponding definition of the Fourier coefficients $\hA_\mu(q)$ and has no influence in the results of Wilson loops or other observables. 

Furthermore, it is important to realize that the antisymmetric combination of the phases (sum over repeated indices  implied)
\be
\label{final}
\Phi(q,p)-\Phi(p,q)=\frac{2\pi}{N} n_{\mu \nu} s_\mu(p) s_\nu(q)= -\frac{N}{2\pi} p_\mu L_\mu q_\nu L_\nu \tilde{n}_{\mu \nu}\equiv 2 \theta(q,p)
\ee
is convention independent. The previous equation is an equality among angles, and hence is defined modulo $2 \pi$. The antisymmetric matrix 
$\tilde{n}_{\mu \nu}$ is defined by the relation $n_{\mu \alpha} \tilde{n}_{\alpha \beta} n_{\beta \nu}= n_{\mu \nu} \bmod N$. Its  matrix elements are not necessarily integers, but the inversion formula (see below) should be well-defined. Its existence can be deduced by transforming $n_{\mu \nu}$ to its canonical form (see Ref.~\cite{review}). Although the matrix is not unique, its arbitrarity does not affect Eq.~\eqref{final}. Its non-uniqueness however shows up when using the matrix to define the inverse map $q \longrightarrow s(q)$ as follows
\be
s_\mu(q)=-\frac{N}{2\pi} \tilde{n}_{\mu \nu } q_\nu L_\nu
\ee
The condition that the left-hand side are integers provides the restriction  on the elements $\tilde{n}_{\mu \nu }$, mentioned above.

To summarize, we can say that up to now the presentation has been completely general for the case of irreducible twists\footnote{The last restriction is essential, since otherwise there are zero-modes and the  perturbative expansion becomes very complicated.}. The most important ingredients are the presence of a lattice of momenta $\LLe$ and the convention dependent value of the $D$ and $F$ symbols. In the next subsection we will apply our formalism to the most useful cases in two, three and  four space-time dimensions, and provide explicit formulas for  the different ingredients in terms of the twist tensor. The four-dimensional  case is the most interesting and will be used for most of the numerical analysis that will follow later. 

\subsection{Particular cases of twists in 2 to 4 dimensions} 
\label{twofour}

The two dimensional case is particularly simple since twist tensors are of the form $n_{\mu \nu}=k\epsilon_{\mu \nu}$, where $\epsilon_{0 1}=-\epsilon_{1 0 }= 1$. The condition of irreducibility amounts to constraining  the integer $\m$ to be coprime with $N$. The lattice of momenta $\LLe$ is simply given by all momenta having the form  $q_\mu=\frac{2 \pi m_\mu}{\LL_\mu N}$, where $m_\mu$ are integers modulo $\LL_\mu N$. Notice that this is equivalent to the standard lattice momenta in a box of size $(NL_0)\times (N L_1)$, with an effective volume of $\Ve=N^2\V$. The $\Gamma_\mu$ matrices can be written in terms of 't Hooft matrices $Q$ and $P$ satisfying 
\be 
PQ=zQP
\ee 
where $z=\exp\{2 \pi i /N\}$. The matrices are given by  $Q=\mathrm{diag}(1,z,z^2,\ldots,z^{N-1})$
and $P_{i j}=\delta_{j\ i+1}$. Thus a possible choice of matrices satisfying the algebra would be $\Gamma_0=Q$ and $\Gamma_1=P^k$. Notice, however, that  
for even $N$ the matrices have determinant $-1$. Thus, if we impose that $\Gamma_0$ belongs to SU(N) one should rather take  $\Gamma_0=\pm iQ$
and $\Gamma_1=(\pm i P)^k$, but paying the price that now $\Gamma_0^N=-1$. This is the conflict of normalization conditions that we were mentioning earlier. The same normalization conflict  translates to the choice of $\HG(q)$. We might obviously write 
\be
\HG(q) = e^{i \alpha(q)} \, \Gamma_0^{s_0(q)} \Gamma_1^{s_1(q)}
\ee
where $s_\mu(q)$ should satisfy $L_\mu q_\mu= -(2 \pi k/N) \epsilon_{\mu \nu} s_\nu(q)$. If we choose our $\Gamma_\mu$ matrices such that $\Gamma_\mu^N=\mathbf{I}$, this has a  unique inverse
\be
s_\mu(q)=\frac{\kb N}{2\pi} \epsilon_{\mu \nu} L_\nu q_\nu 
\ee
Here $\kb$ is an integer defined through the relation:
\be
\label{bark2d}
\kb k = 1 \, (\bmod  N )\, ,
\ee
We only need to fix the function $\alpha(q)$. One way to fix it is to demand that the hermiticity condition Eq,~\eqref{hermit} adopts the same form as for ordinary Fourier expansion, namely setting $\Phi(q,-q)=0$.
This leads to 
\be
\alpha(q)+\alpha(-q)= -\frac{\kb N \V}{2 \pi} q_1 q_0=-\frac{2 \pi \kb}{N} m_0 m_1
\ee
where we have written $q_\mu=\frac{2\pi m_\mu}{\LL_\mu N}$. The integers $m_\mu$ are defined modulo $N$. Setting $\alpha(q)=\alpha(-q)$ we obtain 
\be
\alpha(q)= -\frac{\pi \kb}{N} m_0 m_1 + \pi \sig (m_0,m_1)
\ee
where $\sig(m_0,m_1)$ takes the value 0 or 1, giving the two possible values of the square root. This second term is necessary because it compensates for the fact that the first term is not always invariant under the shift $m_\mu \longrightarrow m_\mu +NL_\mu$. We might set it to zero if we restrict $m_\mu$ to lie in a particular range.  If $N$ is odd one can always choose $\kb$ to be even (if $\kb$ was odd, replace it by $\kb+N$) and one can directly set $\sig (m_0,m_1)=0$. In that case one finds 
\be
D(q,p,k)+i F(q,p,k)=\delta(p+q+k) \sqrt{\frac{2}{N}} (\cos\theta(q,p)+i \sin\theta(q,p))
\label{eq.thetapq}
\ee
with 
\be
\theta(q,p)=\frac{\pi \kb}{N}(m'_0m_1-m_0m'_1)=\frac{\kb \Ve}{4 \pi N} (p_0q_1-p_1q_0) 
\ee
For the even $N$ case, the formula is still valid if we set the integers $m_\mu$ to lie in a particular interval.

For the three dimensional case the twist tensor can be written in terms of the completely antisymmetric symbol with three indices as follows: $n_{\mu \nu}= \epsilon_{\mu \nu \rho} r_\rho$, where $\vec{r}$ is a vector of integers modulo $N$. The irreducibility conditions amounts to the fact that the greatest common divisor or $r_\alpha$ and $N$ is 1. The integers $s_\mu(q)$ must satisfy 
\be
m_\mu \equiv \frac{N L_\mu q_\mu }{2 \pi}= \Big(\vec{r}\times \vec{s}(q) \Big)_\mu
\ee
where we have used the standard three-dimensional notation for vector products.
Hence, the space of momenta $\LLe$ can be identified with those that correspond to  
$\vec{m}\cdot \vec{r}=0 \bmod N$. 
The irreducibility condition now guarantees that there exist a vector of integers $\vec{v}$ such that $\vec{r}\cdot\vec{v}=-1 \bmod N$. This allows to define a possible inversion as $\vec{s}(q)=\vec{v}\times \vec{m}$. The rest follows similarly  to the two dimensional case with $\tilde{n}_{\mu \nu}= \epsilon_{\mu \nu \rho} v_\rho$. 

In four dimensions, the orthogonality condition requires a twist satisfying
$\kappa(n_{\mu \nu})\equiv\epsilon_{\mu \nu \rho \sigma} n_{\mu \nu} n_{\rho \sigma}/8 = 0$  (mod $N$).
Irreducibility is granted  provided the greatest common divisor of $N$, $n_{\mu \nu}$, and
$\kappa(n_{\mu \nu})/N$ is equal to 1. The case in which the twist is in one plane or in a three dimensional section proceeds identically to the previous cases with $q^c$ living in a 2 or 3 dimensional subspace as before. Hence, 
we will now focus in the case of the symmetric twists where
$N$ is the square of an integer $N= \hL^2$ and
\be
n_{\mu \nu} = \epsilon_{\mu \nu} \m \hL
\ee
with $\epsilon_{\mu \nu} = 1$ if $\mu < \nu$ and $-1$ if $\mu > \nu$. The twist is irreducible if $\m$ and
$\hL$ are coprime integers. The lattice of momenta $\LLe$ is given by $q_\mu = 2 \pi m_\mu/(L_\mu \hL)$,
with $m_\mu$ integers defined modulo $L_\mu \hL$. This leads, like in the two dimensional case,
to an effective lattice volume $\Ve=N^2\V$. The integers  $s(q)$ are given by
\be
s_\mu (q) =  -\tilde \epsilon_{\mu\nu} \, \bar k \, m_\nu \,  .
\ee
Here $\kb$ is an integer defined through the relation:
\be
\label{bark}
\kb k =  1 \, (\bmod  \hL )\, ,
\ee
and $\tilde \epsilon_{\mu\nu}$ is an antisymmetric tensor satisfying:
\be
\sum_{\rho} \tilde \epsilon_{\mu\rho} \epsilon_{\rho \nu} =
\delta_{\mu\nu} \, .
\ee
Notice that this defines the $\tilde{n}_{\mu \nu}$ matrix to be given by $ \kb \tilde \epsilon_{\mu\nu}/\hL$.  The function $\Phi(p,q)$ becomes:
\be
\Phi(q,p) = \alpha(q) + \alpha(p) - \alpha(q+p) - \frac{2 \pi k}{\hL} \sum_{\mu > \nu} \epsilon_{\mu\nu} s_\mu(q) s_\nu(p)
\ee
As in the two dimensional case, imposing hermiticity by setting  $\Phi (q,-q) = 0$ leads to:
\be
\alpha(q) + \alpha(-q) = \alpha(0) + \frac{2 \pi k}{\hL} \sum_{\mu > \nu} s_\mu(q) s_\nu(q) \, .
\ee
 A particular choice satisfying this condition for a momentum $q_\mu= 2 \pi m_\mu/(L_\mu \hL)$ is:
\be
\alpha(q) = \frac{ \pi k}{\hL} \sum_{\mu > \nu} s_\mu(q) s_\nu(q) + \frac{ \pi \kb }{\hL} (k \kb - 1)
\sum_{\mu > \nu}  \tilde  \epsilon_{\mu\nu} m_\mu m_\nu \, ,
\ee
leading to  $D$ and $F$ functions defined in terms of $\theta(p,q)$ as in 
Eq.~\eqref{eq.thetapq}, with:
\be
\theta(p,q)=\frac{ \theta_{\mu\nu}}{2} \, p_\mu q_\nu 
\ee
where we have introduced the antisymmetric tensor $\theta_{\mu \nu}$ defined as:
\be
\label{theta}
\theta_{\mu \nu} =    \frac{ N L_\mu L_\nu } {4\pi^2} \times  \, \tilde
\epsilon_{\mu \nu} \, \tilde \theta  \, ,
\ee
with the angle $\tilde \theta \equiv 2 \pi \bar k / \hL$.

\subsection{The gauge fixed action at order $\lambda$} 
\label{s_gaugef}

We will be using the standard  covariant gauge fixing term with gauge parameter $\xi$. 
Its contribution to the action is
\be
S_{GF} = \frac{1}{\xi} \sum_n\Tr \{ (\nabla_\mu^- A_\mu(n)) (\nabla_\nu^- A_\nu(n)) \}
\ee
where $\nabla_\mu^-$ is now  minus the adjoint of $\nabla_\mu^+$.
This is a typical background field gauge. To order $g^2$ the ghost action 
corresponding to this gauge fixing is given by:
\bea 
S_{GH} &=&  2 \sum_n \Tr \Big \{ (\nabla_\mu^+ \bar c(n)) \Big ( (\nabla_\mu^+  c(n))
- ig [A_\mu(n), c(n)] - 	\frac{i g}{2} [A_\mu(n), \nabla_\mu^+ c(n) ]\\ 
&-& \frac{g^2}{12} [A_\mu(n), [A_\mu(n), \nabla_\mu^+c(n)]] \Big ) \Big \}\, .  \nonumber
\eea
The ghost fields $c$-$\bar c$ have a similar colour-space Fourier
decomposition as the gauge fields. 

There is also an additional a contribution to the action coming
from the expression of the Haar measure on the group in terms of
integration over the Fourier coefficients $\hat{A}(q)$. To 
order $\lambda$ it is given by
\be
S_{MS} = \frac{\lambda}{24} \sum'_q \hA_\mu (q) \hA_\mu (-q) e^{i \Phi(q,-q)} 
+ {\cal O}(g^4)
\ee
To derive this expression we parameterize a generic group matrix element
as $$U = \exp\{-i g \sum'_{q^c} w(q^c) \HG(q^c) \}\, ,$$  
in terms of the basis of the $SU(N)$ algebra given by $\HG(q^c)$
with $q^c$ the colour momentum taking $N^2-1$ values. The prime
over the sum indicates that zero momentum is excluded.
The volume element of the group in terms of these variables is
\be
dU = (\det \gm)^\half \prod_{q^c} d w(q^c)\, 
\ee
with the metric $\gm$ defined as:
\be
(ds)^2 = \gm(p^c,q^c) dw(p^c) dw^*(q^c)   = 2 \Tr\Big \{ \frac{\partial U} {\partial w(p^c)}  
\frac{\partial U^\dagger}{\partial w^*(q^c)} \Big \} \,  dw(p^c) dw^*(q^c) \, .
\ee
Inserting the expression for the $U$ matrices leads to
\bea
\gm(p^c,q^c) &=&  g^2 \delta (p^c-q^c) - \frac{g^4}{12} \sum_{k^c} w (-k^c-p^c) w^*(-k^c-q^c) \times \\ && 
F(k^c,p^c,-k^c-p^c) F^*(k^c,q^c,-k^c-q^c) + {\cal O}(g^6)\, , \nonumber
\eea
and  from here one computes the $\mathcal{O}(\lambda)$ contribution to the action given by
\be
(\det \gm/\gm_0)^{\half} = \exp\{ \half \Tr \log \gm/\gm_0 \}
= \exp\Big\{-\frac{\lambda}{24} \sum_{q^c\ne 0} w(q^c) w(-q^c) e^{i \Phi(q^c,-q^c)} + {\cal O}(\lambda^2) \Big\}\, . 
\ee
In obtaining this result we have used the hermiticity relation on the coefficients $w(q^c)$ and the equality
\be
\sum_{k^c} |F(k^c,q^c,-k^c-q^c|^2 =N (1-\delta(q^c))
\ee
which is the expression of the quadratic Casimir in the adjoint representation in our basis (see Appendix \ref{appendixD}).

Summarizing, we have obtained the gauge fixed partition function to
order $g^2$ given by
\be
Z = \int D c D\bar c D A_\mu \exp\{-(S+S_{GF}+S_{GH}+S_{MS})\}
\ee
This action can be expanded in powers of $g$ to derive the Feynman 
rules of the theory. For example, in Feynman gauge the 
propagator of the gauge field reads:
\be 
\label{eq:prop}
P_{\mu \nu} (p, q) = 
\delta_{\mu \nu} \,  \delta(q+p) e^{-i \Phi(p,-p)}   \, \frac{1}{ \widehat q^2}  \, ,
\ee
and the ghost propagator is
\be
\label{eq:propgh}
P_{GH} (p, q) =
 \delta(q+p) e^{-i \Phi(p,-p)}  \,\frac{1}{ \widehat q^2} \, ,
\ee
where 
\be 
\label{eq.mom}
\widehat q_\mu = 2 \sin(q_\mu/2)
\ee
Notice that if we adopt the hermiticity condition on the $\HG(q)$, giving $\Phi(q,-q)=0$, the expressions simplify. In what follows we will adopt this convention. Because of the problems associated with this convention and explained in subsection~\ref{twofour}, the momenta are now defined  in a range and not modulo $2\pi$. Correspondingly the momentum conservation delta functions are now strict and not modulo $2 \pi$. In any case these difficulties just affect intermediate expressions and not to the final results.  

\subsection{Expansion of the Wilson loops}

An essential ingredient in the calculation is the expansion of the
Wilson loop in powers of the vector potentials $A_\mu(n)$. This expansion 
for the particular case of the plaquette is also necessary to derive 
the non-quadratic terms in the Wilson action giving the vertices of
the theory. 

We recall the definition of our observable given in Eq.~\eqref{expectWL}.
To process the right-hand side we replace the links by the  expression 
Eq.~\eqref{expansion}. In simplified notation this gives rise to 
\be
Z(R,T) U(R,T) = \prod_{l\in {\cal R}} e^{-i g A'_l}
\ee
where we used the label $l$ to represent a link instead of the
conventional $(n,\mu)$ combination. The product on the right hand 
side is the ordered product of the exponentials around the rectangle
${\cal R}$. Finally $A'_l$ is given by 
\be
A'_l =\Gamma_l A_l \Gamma_l^\dagger 
\ee
where $\Gamma_l$ is the product of the $\Gamma_\mu(n)$ factors from 
one reference point in the square to the origin of the link $l$. The
dependence on the choice of reference point drops out when taking the
trace. Notice that in terms of the $A'$ the $Z(R,T)$ factor has
disappeared from the right-hand side.

One can now use the Baker-Campbell-Haussdorf formula to rewrite this as:
\be
 Z(R,T) U(R,T) =  e^{-i \RR}
\ee
with $\RR$ a hermitian matrix which can be expanded in powers of $g$ as
follows:
\be
\RR = g \RR^{(1)} + g^2 \RR^{(2)} + g^3 \RR^{(3)} + {\cal O}(g^4)
\ee
where:
\bea
\RR^{(1)} &=& \sum_{l\in{\cal R}} A'_l  \\
\RR^{(2)} &=& {-i \over 2}  \sum_{l_1<l_2} [A'_{l_1}, A'_{l_2}]\\
\RR^{(3)} \!&=&\! -{1\over 12} \sum_{l_1,l_2} [A'_{l_1},[ A'_{l_1},A'_{l_2}]] - {1\over 6}
\!\sum_{l_1<l_2<l_3} \!\!\Big( [A_{l_1}, [A'_{l_2},A'_{l_3}]] + [A'_{l_3}, [A'_{l_2},A'_{l_1}]] \Big)
\eea
In  the previous formula the ordering of the links is done along the
perimeter of the rectangle following the plaquette orientation. 

Now we can express the trace  in terms
of $\RR$ as follows:
\bea
{1 \over N} \Tr (Z(R,T) U(R,T)) &=& 1 - {1 \over 2 N} \Tr (\RR^2 ) + {i \over 3!N} \Tr (\RR^3)  + {1 \over 4 ! N}
\Tr (\RR^4)+ \cdots
= \\
& & 1 -{g^2 \over 2 N} \Tr [(\RR^{(1)})^2 ]- {g^3 \over N} \Big ( \Tr [\RR^{(1)} 
\RR^{(2)}] - {i \over 3!} \Tr [(\RR^{(1)})^3] \Big ) 
\nonumber \\
&-& {g^4 \over N} \Big ( \half \Tr [(\RR^{(2)})^2]
 + \Tr [\RR^{(1)} \RR^{(3)}] -\frac{i}{2} \Tr [(\RR^{(1)})^2 \RR^{(2)}] - {1 \over 4 !} \Tr[ (\RR^{(1)})^4]
\Big)
\nonumber
\eea
To perform the calculation we need to substitute the expression for
$\RR$ and use the Fourier decomposition written in the following simplified form
\be
A'_l=  \frac{1}{\sqrt{\V}} \sum'_{q}  A_l(q)  \chi(n;q)
\ee
 where $n$ are the coordinates of the lowest vertex of the rectangle. Notice that if the link $l$ has origin $n'$ and direction $\mu$, the coefficients 
 are given by 
\be
  A_l(q) = e^{i q (n'-n)} \hA_\mu(q)
\ee

We will  also use the following notation 
\be
\bar A (q) = \sum_{l\in {\cal R}} A_l(q)
\ee
After averaging over $n$ and  expressing the traces in terms of  the group constants $F$ and $D$ (using for simplicity the hermiticity condition $\Phi(q,-q)=0$),  we arrive at:
\be
W_{R,T} \WA=  1-
{\lambda \over 2} U^{(2)} - {\lambda \sqrt{\lambda}  \over 3!} (U^{(3)}+i V^{(3)}) - {\lambda^2  \over 4!} (U^{(4)}+ i V^{(4)}) 
\ee
where:
\bea 
\label{eq.u2}
U^{(2)} &=& \frac{1}{2 \Ve} \sum'_{q}  \langle \bar A(q) \bar A(-q)\rangle \\
\label{eq.u3}
U^{(3)} &=&   {3 \sqrt{N}  \over 2\Ve^{3/2}} \sum'_{q_1,q_2} \sum_{l_2<l_3} 
F(q_1,q_2,-q_1-q_2) \langle \bar A(q_1) A_{l_2}(q_2)A_{l_3}(-q_1-q_2)\rangle \\
\label{eq.v3}
V^{(3)} &=&  - { \sqrt{N}  \over 4\Ve^{3/2}} \sum'_{q_1,q_2}  D(q_1,q_2,-q_1-q_2) \langle \bar A(q_1) \bar A(q_2) \bar A(-q_1-q_2) \rangle 
\eea
\bea
\label{eq.u4}
U^{(4)} &=&   {N \over \Ve^2} \sum'_{q_1,q_2,q_3,q_4} \delta (q_1+q_2+q_3+q_4) F(q_1,q_2,-q_1-q_2) F(q_3,q_4,-q_3-q_4)  \\
&\Big (& {3 \over 2} \sum_{l_1<l_2, l_3<l_4} \langle A_{l_1}(q_1) A_{l_ 2}(q_2)A_{l_3}(q_3)A_{l_4}(q_4)\rangle 
+   \sum_{l} \langle \bar A(q_1) A_{l}(q_2)A_{l}(q_3) \bar A(q_4)\rangle \nonumber \\
&+& 2 \sum_{l_2<l_3<l_4} (\langle \bar A(q_1) A_{l_2}(q_2)A_{l_3}(q_3)A_{l_4}(q_4)\rangle 
+ \langle \bar A(q_1) A_{l_2}(q_4)A_{l_3}(q_3)A_{l_4}(q_2)\rangle ) \Big )\nonumber \\
&-&\!\!  { N \over 8 \Ve^2} \sum'_{q_1,q_2,q_3,q_4}
\langle \bar A(q_1) \bar A(q_2)\bar A(q_3)\bar A(q_4)\rangle  \delta (q_1+q_2+q_3+q_4)
\nonumber \\
&& 
 D(q_1,q_2,-q_1-q_2) D(q_3,q_4,-q_3-q_4)  
\nonumber\\
\label{eq.v4}
V^{(4)} &=& \! - {3 N \over 2 \Ve^2} \!\sum'_{q_1,q_2,q_3,q_4}\delta (q_1+q_2+q_3+q_4) D(q_1,q_2,-q_1-q_2) F(q_3,q_4,-q_3-q_4)\\
 &&  \sum_{l_3<l_4} \langle \bar A(q_1) \bar A(q_2) A_{l_3}(q_3)A_{l_4}(q_4)\rangle  \nonumber
\eea
with  $\Ve=\V N^2$ being the effective volume.

The previous formulas express the expectation values of Wilson loops
in terms of the $n$-point Green functions of the vector potentials. 
The latter can be computed as a power series in $g$ using the Feynman 
rules of the theory, given in App.~\ref{appendixA}.

\section{Results of the perturbative expansion of Wilson loops}
\label{s.results}

In the present section we use the machinery developped in the
previous section to compute the perturbative expansion of the expectation 
values of rectangular Wilson loops.  In particular, we  
consider the coefficients of the expansion up to order $\lambda^2=1/b^2$
as follows:
\be
W_{R,T} \WA= 1- \lambda \hat{W}^{(R\times T)}_1\WAb - \lambda^2
\hat{W}^{(R\times T)}_2\WAb + \ldots
\ee
Alternatively, we might consider the expansion of the logarithm instead
\be
\log\left( W_{R,T} \WA \right)=-\lambda
\tilde{W}^{(R\times T)}_1\WAb - \lambda^2
\tilde{W}^{(R\times T)}_2\WAb + \ldots
\ee
The two sets of coefficients are related as follows
\bea
\nonumber
\tilde{W}^{(R\times T)}_1 \WAb &=&\hat{W}^{(R\times T)}_1 \WAb \\
\tilde{W}^{(R\times T)}_2 \WAb&=&\hat{W}^{(R\times T)}_2 \WAb +\frac{1}{2}
(\hat{W}^{(R\times T)}_1 \WAb)^2
\eea

To obtain these coefficients we start by the expressions given in the
previous section and expand the $U^{(n)}$ and $V^{(n)}$ terms in powers of $g$:
\be
U^{(n)}= \sum_a g^a U^{(n)}_{a}, \ \ V^{(n)}= \sum_a g^a V^{(n)}_{a},
\ee
Using this terminology, similar to that followed in Ref.~\cite{Weisz:1982zw,Weisz:1983bn,Wohlert:1984hk},
we arrive at the following expression for the 
coefficients of the logarithm of the Wilson loop at ${\cal O} (g^4)$:
\bea
\tilde{W}^{(R\times T)}_1 \WAb&=& {1 \over 2 }  U_0^{(2)} \\
\label{secondord}
\tilde{W}^{(R\times T)}_2\WAb &=& \frac{1}{4!}\left( 
3 (U_0^{(2)})^2 + 4 U_1^{(3)} + 4 i V_1^{(3)} + U_0^{(4)} + i V_0^{(4)}  + 12
U_2^{(2)}\right)
\eea
Notice that coefficient of order $\lambda^2$ 
requires the calculation at one-loop of the two point function
$U_2^{(2)}$. 

In the following subsections we will spell out the calculation of 
these coefficients.

\subsection{ The Wilson loop  at ${\cal O} (\lambda)$}

Combining the previous results we obtain the expression of the 
first coefficient as follows:
\be
\tilde{W}^{(R\times T)}_1 \WAb =
{1 \over 2} U_0^{(2)} = { 1 \over 4 \Ve } \sum'_{q}  
\langle  \bar{A}(q)  \bar{A}(-q)\rangle_0  
\ee
To compute this expression we must write down explicitly the expression 
of $\bar{A}(q)=\sum_{l \in {\cal R}} A_l(q)$ in terms of the Fourier
coefficients $\hA_{\rho}(q)$. To simplify notation we will specify that 
the rectangle is sitting in the $\mu-\nu$ plane, with $R$ and $T$
being the length of the edges in the $\mu$ and $\nu$ directions
respectively. We can then separate $\bar{A}(q)$ as a sum of the 
contributions of its four edges. Noting these contributions as 
$A^{(i)}$ with $i=1,3$ ($\mu$ direction) and  $i=2,4$ ($\nu$ direction), we get:
\bea
A^{(1)}(q) &=& \sum_{n=0}^{R-1} e^{i n  q_\mu }  \hat{A}_\mu (q) =
e^{i (R-1) q_\mu/2} Q_\mu(q) \hat{A}_\mu (q) \\
A^{(3)} (q) &=& - \sum_{n=0}^{R-1} e^{i n  q_\mu } e^{i T q_\nu}
\hat{A}_\mu (q)= - e^{i (R-1) q_\mu/2} Q_\mu(q)  e^{i T q_\nu} \hat{A}_\mu (q) \\
A^{(2)} (q) &=& \sum_{n=0}^{T-1}e^{i n  q_\nu } e^{i R q_\mu}
\hat{A}_\nu (q) = e^{i (T-1) q_\nu/2} Q_\nu(q) e^{i R q_\mu} \hat{A}_\nu (q)\\
A^{(4)} (q) &=& -\sum_{n=0}^{T-1} e^{i n   q_\nu } \hat{A}_\nu (q) = -
e^{i (T-1) q_\nu/2} Q_\nu(q)  \hat{A}_\nu (q)
\eea
where we have introduced the symbols $Q_\mu(q)$ and $Q_\nu(q)$ given
implicitly in terms of finite geometric sums. Performing these sums
explicitly we have   
\be
Q_\mu(q_\mu\ne0)= \frac{S_\mu(q)}{\widehat q_\mu}
\ee
with 
\be
\label{eq.smu}
S_\mu (q) = 2 \sin \Big( {R q_\mu \over 2} \Big)
\ee
and $\widehat q_\mu$ the lattice momentum introduced in Eq.~\eqref{eq.mom}. This
expression is singular for $q_\mu=0$ in which case the result if
$Q_\mu(0)=R$. Replacing $\mu$ by $\nu$ and $R$ by $T$, we get the
remaining symbols. 

With this notation we finally get
\be
\bar A(q) = i e^{i {R q_\mu+T q_\nu \over 2}} (-Q_\mu(q)
S_\nu(q) e^{-i \frac{q_\mu}{2}} \hA_\mu (q)+ Q_\nu(q) 
S_\mu(q) e^{-i \frac{q_\nu}{2}}\hA_\nu (q))
\ee
which can be rewritten in a more symmetric fashion as 
\be
\bar A(q) =  e^{i {R q_\mu+T q_\nu \over 2} } Q_\mu(q) Q_\nu(q) F^{(0)}_{\mu \nu} (q) 
\ee
with:
\be
 F^{(0)}_{\mu \nu} (q) =  i \widehat q_\mu e^{-i \frac{q_\nu}{2}}\hA_\nu (q) -  i  \widehat q_\nu e^{-i \frac{q_\mu}{2}}\hA_\mu (q)
\ee

Using the expression for the propagator, gives the final result:
\be
\tilde{W}^{(R\times T )}_1 \WAb=    
{1 \over  4 \Ve } \sum'_q \tilde S^2_{\mu \nu} (q)  \, 
\, {\widehat q_\mu^2 + \widehat q_\nu^2 \over \widehat q^2}\, ,
\ee
where
\be
\tilde S_{\mu \nu} (q) = Q_\mu(q) Q_\nu(q)
\ee
The result agrees with the tree level result for the standard Wilson action on an infinite 
lattice derived in \cite{Weisz:1982zw} if one replaces appropriately the momentum sums by integrals.

For the particular case of the plaquette ($R=T=1$) the result
simplifies and we get 
\be
{1 \over 4 \Ve}\,  \sum'_q {\widehat q_\mu^2 + \widehat q_\nu^2
\over \widehat q^2}\, ={(N^2-1) \over 2d N^2 }
\ee
The last equality is true for the average of the plaquette over all $\mu-\nu$ planes in   $d$  space-time dimensions.  It coincides with the plaquette in each plane  if  there is symmetry among all directions. Otherwise the plaquette expectation value at this order depends on the plane.  

\subsection{ The Wilson loop  at ${\cal O} (\lambda^2)$}

To compute the coefficient of the logarithm of the Wilson loop
expectation value  to the next order $\tilde{W}^{(R\times T)}_2\WAb$, we need to evaluate
$U^{(n)}$, $n=1,\cdots, 4$, and  $V^{(n)}$, $n=3, 4$ in Eqs.~\eqref{eq.u2}-\eqref{eq.v4}
and substitute them in expression \eqref{secondord}. The computation of
the different terms can be done using the Feynman rules given in
App.~\ref{appendixA}. 

In the following paragraphs we list the expression of the $U^{(n)}_a$ 
and $V^{(n)}_a$ terms entering in Eq.~\eqref{secondord}. As we did at leading order,  we use the
label $\mu$ to indicate the direction of the loop having length $R$ and 
$\nu$ that having length $T$. We also use  the simplifying symbols
given below: 
\bea
C_\mu(k) &=& \cos(R k_\mu/2)\\
C_\nu(k) &=& \cos(T k_\nu/2) 
\eea

We arrive at:
\bea
U^{(2)}_2 &=& { 1 \over 2 \Ve} \sum'_q \tilde S_{\mu \nu}^2 (q)
(\widehat q_\mu \delta_{\nu \tau} - \widehat q_\nu \delta_{\mu \tau})
(\widehat q_\mu \delta_{\nu \sigma} - \widehat q_\nu \delta_{\mu \sigma})
{\Pi_{\tau \sigma}(q) \over (\widehat q^\sq)^2}
\label{eq.u22}
\\
U^{(3)}_1 &=& -{3 N \over 2 \Ve^2} \sum_{q_1,q_2} \delta (q_1+q_2+q_3 ) \, 
{F^2(q_1,q_2,q_3) \over \whq_1^\sq \whq_2^\sq \whq_3^\sq }\\
&\Big \{&  Q_\nu(q_1) \cos\Big ({q_{1 \mu} \over 2}\Big ) \widehat{(q_2-q_3)}_\nu  
 \Big [Q_\mu(q_2) Q_\mu(q_3)  C_\mu(q_1)
S_\nu(q_2-q_3)  \,    \nonumber \\
&+&  {  \whq_{1 \mu}  \whq_{1 \nu} \over 2 \whq_{2\mu} \whq_{3\mu}} \, \tilde S_{\mu \nu} (q_1)  
(\widehat{q_3-q_2})_\mu  
\Big( Q_\mu (q_3-q_2)-  Q_\mu (q_3+q_2)\Big)
 \Big ]  \nonumber \\
&+& (\mu \leftrightarrow \nu) (R\leftrightarrow T) \Big\} \nonumber \\
V^{(3)}_1  &=&  { 3 N \over 4 \Ve^2} \sum_{q_1,q_2} \delta (q_1+q_2+q_3 ) D(q_1,q_2,q_3) F(q_1,q_2,q_3) {1 \over
\widehat q_1^2 \whq_2^2 \whq_3^2 }\\
&& S_{\mu \nu } (q_1)  S_{\mu \nu } (q_2)  S_{\mu \nu } (q_3) 
\Big ( \whq_{1\mu}  \whq_{2\mu}  \whq_{3\nu} \cos ( q_{3\nu} /2) (\widehat{q_2 -q_1})_\mu - (\mu \leftrightarrow \nu) (R\leftrightarrow T) \Big ) \nonumber\\
U^{(4)}_0 &+& 3 \Big(U^{(2)}_0\Big)^2 =  -{N \over  \Ve^2} \sum_{q_1, q_2} 
{F^2(q_1,q_2, -q_1-q_2) \over \whq_1^\sq \whq_2^\sq } \\ 
&\Big\{& 
-  \half S_{\mu \nu}^2(q_1) (\whq_{1\mu}^\sq + \whq_{1\nu}^\sq)
\nonumber \\
&+& 6 C_\mu(q_2) S_\nu^2(q_1) Q_\mu(q_1) (Q_\mu(q_1+q_2) - Q_\mu(q_1-q_2)) {\cos(q_{1\mu}/2) \over \whq_{1\mu} \whq_{2\mu}}
\nonumber \\
&-& 3 {(\widehat{q_2+q_1})_\mu^2+(\widehat{q_2-q_1})_\mu^2  \over 8 \whq_{1\mu}^\sq \whq_{2\mu}^\sq}
S_\nu^2(q_1+q_2)
\Big( Q_\mu (q_2-q_1)-  Q_\mu (q_2+q_1)\Big)^2 \nonumber \\
&+& {3 \over 4} Q_\mu^2(q_1) Q_\mu^2 (q_2) S_\nu^2(q_1+q_2)
- 3 C_\mu(2q_2) C_\nu(2q_1) Q_\mu^2(q_1) Q_\nu^2(q_2) \nonumber
\\
&+& (\mu \leftrightarrow \nu) (R\leftrightarrow T) \Big\}
 \nonumber \\
V^{(4)}_0 &=& -{ 3 N \over  \Ve^2 }  \sum_{q_1, q_2} D(q_1,q_2,-q_1-q_2) F(q_1,q_2,-q_1-q_2) {1 \over \whq_1^2 \whq_2^2}  \\
&&  S_{\mu \nu } (q_1)  S_{\mu \nu } (q_2)
\Big \{  \whq_{1\nu}  \whq_{2\mu}  Q_\mu(q_1) Q_\nu(q_2)  C_\mu(q_2) C_\nu (q_1) \nonumber \\
&-& \whq_{1\nu}  \whq_{2\nu} S_\nu (q_1+q_2)  {(\widehat{q_2-q_1})_\mu  \over 2 \whq_{1\mu} \whq_{2\mu}}
\Big( Q_\mu (q_2-q_1)-  Q_\mu (q_2+q_1)\Big)
 \nonumber
\\
&-& (\mu \leftrightarrow \nu) (R\leftrightarrow T) \Big\}
 \nonumber
\eea

The expression for the vacuum polarization $ \Pi_{\alpha \beta} $
 can be found in App.~\ref{appendixA}.

The corresponding expressions for the plaquette simplify considerably:
\bea
U^{(2)}_2 &=& { 1 \over 2 \Ve } \sum'_q {1 \over (\whq^{\ 2} )^2} \,\Big(\widehat
q_\mu^2 \Pi_{\nu\nu} - \whq_{\mu} \whq_\nu\Pi_{\mu\nu} + (\mu \leftrightarrow \nu)\Big) 
\\
U^{(3)}_1 &=& - {3 N \over  2\Ve^2 } \sum_{q_1,q_2,q_3}  \delta (q_1+q_2+q_3 ) \,
{F^2(q_1,q_2,q_3) \over \whq_1^\sq  \whq_2^\sq  \whq_3^\sq } \\
&\Big(&\cos^2\Big({q_{1\mu}\over 2}\Big) \widehat {(q_2-q_3)}_\nu^2  + (\mu \leftrightarrow \nu)\Big) \nonumber
\\
V^{(3)}_1  &=&  { 3 N \over 4 \Ve^2} \sum_{q_1,q_2} \delta (q_1+q_2+q_3 ) D(q_1,q_2,q_3) F(q_1,q_2,q_3) {1 \over
\whq_1^2 \whq_2^2 \whq_3^2 }  \\
&\Big (& \whq_{1\mu}  \whq_{2\mu}  \whq_{3\nu} \cos ( q_{3\nu} /2) (\widehat{q_2 -q_1})_\mu - (\mu \leftrightarrow \nu) \Big ) \nonumber \\
U^{(4)}_0&+& 3 \Big(U^{(2)}_0\Big)^2 =  -{N\over  4 \Ve^2} \sum_{q_1, q_2} {F^2(q_1,q_2,-q_1-q_2)
\over \widehat q_1^\sq  \widehat q_2^\sq }\\
 &\Big (&
 3   \widehat {(q_1+q_2)}_\nu^2 - 4 \whq_{1\mu}^\sq  - 12 \cos(q_{2\mu}) \cos(q_{1\nu})+ (\mu \leftrightarrow \nu)
\Big) \nonumber\\
V^{(4)}_0 &=& -{ 3 N \over  \Ve^2 }  \sum_{q_1, q_2} D(q_1,q_2,-q_1-q_2) F(q_1,q_2,-q_1-q_2) {1 \over \whq_1^2 \whq_2^2}  \\
&& 
 \Big(\sin(q_{1\nu}) \sin(q_{2\mu})  -
(\mu \leftrightarrow \nu)\Big) \nonumber
\eea

Notice that all the previous expressions are valid for arbitrary space-time dimension and for an arbitrary irreducible twist. All the dependence on the twist is contained in the $F$ and $D$ factors and in the ranges of the momentum sums. Notice also that in some of these  sums we have dropped the prime affecting the summation symbol. As explained below, this is because the  $F$ factor vanishes for the excluded momenta in the primed summation. 

\section{Analysis of the results}
\label{s.analysis}

In the previous section we have presented  the result of the calculation expressed as single and double sums over discrete momenta. To help 
in understanding the implications it is interesting to analyze the $N$
and $L$ dependence and to understand the connection with the case of
periodic boundary conditions. For the case of the real part,  we can naturally separate out two types
of contributions to the coefficients: those proportional to the
structure constant square $F^2$ and those that are not. They will be
called non-abelian and abelian respectively.  The imaginary parts are always proportional to $F$ (indeed they are also proportional to the anticommutator $D$) so they can be classified also as non-abelian. 

We recall that the momentum sums range over a finite lattice labelled $\LLe\setminus \Ls$, where $\LLe$ is a finite abelian group and $\Ls$ the subgroup of spatial momenta. The zero momentum $q=0$ (neutral element)  is contained in both sets. It is convenient to exclude it from both  and use a prime symbol to label the 
resulting set: $\LLe' =\LLe -\{0\}$. Notice that this restriction does not affect the set difference: $\LLe \setminus \Lambda_{\LL} = \LLe' \setminus \Lambda'_{\LL}$.  The removal of the zero-momentum from the sum eliminates the apparent ill-definition of the expressions. 
An interesting observation for the analysis that follows is that
$F$ vanishes when any of its arguments belongs to  $\Lambda_{\LL}$. Thus, in all contributions of non-abelian type, we can drop the prime in the sum and extend the sums to  $\LLe'$.

Now we are in position to discuss the relation between  our results and
those obtained for periodic boundary conditions, more precisely 
with the finite volume periodic results  obtained by Heller and Karsch~\cite{Heller:1984hx} by 
neglecting the contribution of zero-modes and explicitly excluding 
zero momentum in the sums. According to our previous
considerations, the integrands of the different contributions to the real part of the Wilson loops for
periodic and twisted boundary conditions are identical.  The 
main difference  is that for the periodic case, the 
momentum sums are now  over $\Lambda'_{\LL}$, and the colour sums are 
performed independently. Hence, it is possible to transform our
formulas into those of Ref~\cite{Heller:1984hx}  by the following substitutions:
\bea
 {N\over \Ve} \sum_q F^2  \ & \longrightarrow  & \hspace{1.3cm} {1 \over V} \sum_{q
\in \Lambda'_L} \hspace{1.3cm}  \longrightarrow \  \int dq \\
{N\over \Ve^2} \sum_{q_1} \sum_{q_2} F^2  & \longrightarrow &  {1 \over
V^2} {N^2-1 \over N^2}  \sum_{q_1 \in \Lambda'_L} \sum_{q_2 \in
\Lambda'_L}  \longrightarrow  {N^2-1 \over N^2} \int dq_1 \int dq_2
\label{typeNA}
\eea
for the non-abelian terms proportional to $F^2$, and 
\bea
{1\over \Ve} \sum_q' & \longrightarrow \ & \hspace{.8cm} \ {N^2-1 \over N^2 V}
\sum_{q \in \Lambda'_L} \hspace{.8cm}  \longrightarrow  \ {N^2-1 \over N^2}   \int dq
\label{typeMes}
\\
{1\over \Ve^2} \sum_{q_1}' \sum_{q_2}' & \longrightarrow&  \Big({N^2-1
\over N^2 V}\Big)^2  \sum_{q_1 \in \Lambda'_L} \sum_{q_2 \in
\Lambda'_L} \ \longrightarrow \  \Big({N^2-1 \over N^2}\Big)^2 \int dq_1 \int dq_2
\label{typeW2}
\eea
for the abelian ones. In the previous formulas we have added a third 
column  corresponding to the infinite volume limit. It
reproduces the results of Weisz, Wetzel and Wohlert~\cite{Wohlert:1984hk} for the four dimensional case.

For simplicity let us now focus specifically in the case of a symmetric  box of length $\LL_\mu=\LL$ and upon symmetric twists for which all directions  appear on symmetrically on $\LLe$ (in more detail for the d=2 and 4 cases). 
At leading order, the coefficient corresponding to periodic boundary
conditions (PBC) is given by 
\be
\tw^\pbc_1(N, \LL)= \Big(1-\frac{1}{N^2}\Big) F_1(\LL) 
\ee
which spells out its dependence on $N$ and $L$. The function 
$F_1(\LL)$ is given by a single sum over momenta in the set
$\Lambda_{\LL}'$. It vanishes for the one-point box $F_1(1)=0$.
The case of two dimensions for $R,T \le \LL$ is particularly simple since the sums can be evaluated exactly and they give
\be
F_1(\LL) = \frac{RT}{4}\Big(1-\frac{RT}{\V}\Big)\, .
\ee
For general dimension $d$ and at large $\LL$,  $F_1(\LL)$ behaves as (see
Appendix~\ref{appendixC}) 
\be
F_1(\LL) = F_1(\infty)- \frac{R^2 T^2}{2d \V} +{\cal O}(1/\LL^{d+2})
\ee
For the particular case of the plaquette, higher order corrections vanish and the d-dimensional result is given by 
$F_1(\LL) = (1-1/\LL^d) /2d$.

Using the formulas given earlier  the result for twisted
boundary conditions and symmetric twist can be expressed in terms of the same function as follows
\be
\tw^\tbc_1(N, \LL, k\ne 0)= F_1(\Le)-  \frac{1}{N^2} F_1(\LL)
\ee
where $\Le=L \hL$ is the effective size parameter, with  $\hL=\sqrt{N}$ in 4 dimensions and equal to $N$ in two. It is interesting
to observe that the result for the TEK model ($\LL=1$) is just 
$F_1(\hL)$, the large $N$ result on a box of size $\hL^d$. In particular, for large $\LL$
the $d=4$ twisted result approaches infinite volume limit value with corrections that go like  $(N-1)/(N^3 \LL^6)$.

A similar analysis can be done at the next order. For that purpose we
have to separate the different contributions into those that we called
abelian and non-abelian. Within the former  there are two different 
$N$ dependencies corresponding to the measure term Eq.~\eqref{eq.mes} and the abelian contribution from the tadpole
Eq.~\eqref{eq.abel} respectively (corresponding to $\Pi^{\rm mes}$ and  $\Pi^{W_2}$ in App. \ref{appendixB}). 
Finally, the  (PBC) result (zero-mode contribution excluded) can be expressed as follows
\be
\tw^\pbc_2(N, \LL)=\Big(1-\frac{1}{N^2}\Big) F_2(\LL)+ \Big(1-\frac{1}{N^2}\Big)^2 F_W(\LL)
\ee
in terms of two functions of $\LL$. The first function $F_2(\LL)$ can
be split as follows
\be
F_2(\LL)= F_{NA}(\LL) + F_{\mathrm{mes}}(\LL) 
\ee
where  $F_{NA}(\LL)$ includes the non-abelian part and
$F_{\mathrm{mes}}(\LL)$ the contribution from the measure. 
The other function $F_W(L)$ comes from another abelian contribution 
to the vacuum polarization and, using the symmetry of all directions, can be expressed in terms of 
$F_1(L)$ as follows:
\be
F_W(L)= \frac{1}{2d}\Big(1-\frac{1}{V}\Big)F_1(L)
\ee
Our functions can be connected to those given by  Heller and
Karsch~\cite{Heller:1984hx} for 4 dimensions by the following relations:
\bea
\bar{W}_2(\LL)&=&F_1(\LL)\\
Y(\LL)&=&2 F_W(\LL)-F_1^2(\LL)\\
X(\LL)&=& F_2(\LL)+ \frac{1}{6} Y(\LL)
\eea

Now we will analyse the expression for the symmetric twist case. The different contributions can be split as follows:  
\bea
\label{MAIN}
\tw^\tbc_2(N,\LL, k)&=& \Big(F_2(\Le)-\frac{1}{N^2}F_2(\LL)\Big) + 
\frac{1}{2d}\Big(1-\frac{1}{N^2}\Big)\Big(F_1(\Le)-\frac{F_1(L)}{N^2}\Big)\\ &+&
F_{2T}(N,\LL,k) + i G_{2T}(N,\LL,k) \nonumber
\eea
where the first two terms contain the same two functions that enter the
periodic formula and the last two are specific of the twisted case. Let us see how this decomposition comes about. The term involving $F_1$ comes from the abelian tadpole $W_2$, which has the general structure Eq.~\eqref{typeW2}. The measure contribution conforms to type  Eq.~\eqref{typeMes}, from which the corresponding PBC and TBC contributions can be easily read out. 
The remaining terms contributing to the real part involve double sums of type Eq.~\eqref{typeNA}:
\bea
&&{N\over \Ve^2} \sum_{p} \sum_{q} F^2(p,q,-p-q)\ {\cal A}(p,q) = \label{eq.fsq}\\ 
&&{1\over \Ve} \sum_{p\in\Lambda'_{\Le}} \sum_{q\in\Lambda'_{\Le}} \ {\cal A}(p,q)-{1\over \Ve} \sum_{p\in \Lambda'_{\Le}} \sum_{q \in \Lambda'_{\Le}} \cos(\theta_{\mu \nu} p_\mu q_\nu)
\ {\cal A}(p,q)  \equiv \nonumber \\
\nonumber
& & F_{NA}(\Le) + {\cal F}_{NP}(N,\LL,k)\quad \quad \quad \quad \nonumber
\eea
where ${\cal A}(p,q)$ is a smooth function of its arguments, whose explicit form can be read out from our formulas. The important part of the previous chain of equations is the substitution 
$NF^2=1-\cos$, which leads to its decomposition into two functions. The first $F_{NA}$ was already present in the periodic case.   The new function ${\cal F}_{NP}(N,\LL,k)$  is the only one in which the $N$, $\LL$, and $\m$ dependence are mixed up. The analysis of all diagrams that was carried out in Ref.~\cite{GonzalezArroyo:1982hz} implies that this term  contains  the contribution of non-planar diagrams, and  this explains the name given to it. Notice that for volume reduction to hold in perturbation theory ${\cal F}_{NP}(N,\LL,k)$ should go to zero in the large $N$ limit. In appendix~\ref{appendixD} we analyze the behaviour of this type of sums as a function of $N$ and $L$.  In particular we show that when  $L$ goes to infinity ${\cal F}_{NP}(N,\LL,k)$ tends towards  $-\frac{1}{N^2}  F_{NA}(\LL)$.  Thus, we define 
\be
F_{2T}(N,\LL,k)= \frac{1}{N^2}  F_{NA}(\LL) + {\cal F}_{NP}(N,\LL,k)
\label{eq.f2t}
\ee
which then goes to zero when $\LL$ goes to infinity. The added piece is substracted out and combined with the $-1/N^2F_{\mathrm{mes}}$ term to produce our final expression  Eq.~\eqref{MAIN}. 
The usefulness of making this arrangement goes beyond this simplicity. Indeed, the new function  $F_{2T}(N,\LL,k)$ contains all the twist dependence and goes to zero in the infinite volume limit and in the large $N$ limit.  

The imaginary part of the coefficient has been collected into the function $G_{2T}(N,\LL,k)$ which has no periodic counterpart. Its presence is due to a violation of CP symmetry induced by the twist vector. Obviously, it  vanishes for $k=0$ as well as in the infinite volume limit. Volume independence implies that it should also vanish in the large N limit.  

Our goal is to use the decomposition Eq.~\eqref{MAIN},
to analyse the $N$ and $\LL$ dependence of the coefficients. 
One particular case is that of the TEK model ($\LL=1$).
The formula simplifies since $F_i(1)$ vanishes. The resulting
expression is quite appealing 
\be
\tw^\tbc_2(N,1, k)=\tw^\pbc_2(N=\infty,\hL)+{\cal F}_{NP}(N,1,k) +
i G_{2T}(N,1,k)
\ee
It means that the TEK coefficient  is equal to the periodic one at 
large $N$ computed at an effective lattice size of $\hL^d$ plus an 
additional complex contribution coming from non-planar diagrams. The two terms correspond nicely with the two main effects of the Feynman rules  of the TEK model: a propagator equivalent to that of  $\hL^d$ lattice and a modified vertex which affects the non-planar diagrams only.
In the general case, as mentioned earlier, if the  non-planar term ${\cal F}_{NP}(N,1,k)+i G_{2T}(N,1,k)$ goes to zero in the large $N$
limit, one recovers the volume reduction result: The large $N$ limit of the twisted theory coincides with the infinite volume large $N$ result. On the contrary, it is very clear that reduction does not work  for periodic boundary conditions. In  the large $N$ limit the coefficient is given by
\be
\tw^\pbc_2(\infty,\LL)= F_2(\LL) +  F_W(\LL)
\ee
which is still size dependent. We emphasize nevertheless that what we call PBC is not the correct weak coupling expression for periodic boundary conditions. Our calculation does not take into account zero-modes. This is known to leading order, but not to the $1/b^2$ order that we are calculating. 

On the other extreme we can consider the behaviour of the perturbative coefficients for large values of $\LL$.  The infinite volume coefficients coincide for periodic and twisted
boundary conditions provided $F_{2T}+iG_{2T}$ vanishes at large $\LL$  (see appendix~\ref{appendixD}).  This result was to be expected,  since at infinite volume boundary conditions should not matter.

A complete  analytic study of  the approach to large $\LL$  
is hard to do  due to the complicated structure of the coefficients 
${\cal A}(p,q)$. Functions involving single sums can be easily analysed though along the guidelines given in appendix~\ref{appendixC}. Apart from the behaviour of $F_1(\LL)$ given earlier, we also 
study the  $\LL$-dependence of  $F_{\mathrm{mes}}(\LL)$ for large
$\LL$. Using the formulas developped in  appendix~\ref{appendixC} we show that in four dimensions the leading correction 
is given by:
\be
\bar{\gamma} \frac{R^2 T^2}{L^2} 
\label{eq.gammabar}
\ee
where 
\be
\bar{\gamma}= -\frac{1}{192 \pi} \left(-1+\int_1^\infty dz \,
(\vartheta^4(0;i z)-1)\right)=0.0014631352661\, .
\ee
In two dimensions  $F_{\mathrm{mes}}(\LL)$ diverges logarithmically with $\LL$ as:
\be
-\frac{R^2 T^2}{96 \pi} \, \log L
\label{eq.mes2d}
\ee
The  complicated structure of $F_{NA}(\LL)$ involving double sums has prevented us from obtaining its expansion in inverse powers of $\LL$. Numerically it seems that, to a high precision, the leading $1/\LL^{2}$ correction is equal and opposite to that  of $F_{\mathrm{mes}}(\LL)$. This is presumably associated to the vanishing of the vacuum polarization at zero momentum  which occurs through a similar cancellation. This is another reason for expressing the results in terms of $F_2$  rather than separating out its two components.  In summary, in the four dimensional case the function $F_2$  can be fitted at large $\LL$ to  a functional form 
\be
\label{f2formula}
F_2(\LL)=F_2(\infty)- \frac{R^2
T^2\, (\gamma_2+\gamma_2'\, \log(\LL))}{\LL^4}+ \ldots
\ee
A similar fit for the case of the plaquette was also advocated  by other
authors earlier~\cite{Bali:2014fea}. 
The explicit $R^2 T^2$ dependence is not 
intended to be exact, i.e. $\gamma_2$ and $\gamma_2'$ can also depend
on $R$ and $T$. However, our best fit values for square loops to be presented later on in table~\ref{table2}, give values which are of similar size.

Notice  that in the  perturbative coefficients for twisted boundary conditions 
the functions  $F_i(\LL)$ appear in the combination $F_i(\Le)-F_i(\LL)/N^2$. This 
cancels the $1/\V$ subleading terms but not the ones containing a 
logarithm. 

Computing the large $\LL$ behaviour of the functions
$F_{2T}(N,L,k)$ and $G_{2T}(N,L,k)$ is even more difficult, beyond the fact that they should
vanish both in the large $N$ limit and in the large $\LL$ limit. The situation is analyzed in appendix~\ref{appendixD}.  A numerical study will be presented in the next section for the four-dimensional case with symmetric twist.

\section{Numerical evaluation of the coefficients in four dimensions}
\label{s.numerical}

In parallel with the analysis performed in the previous section,
it is interesting to study the numerical values of the perturbative 
coefficients for some values of the parameters ($N$, $\LL$ and twist $\m$).
To obtain the numbers one has to perform 
the momentum sums. These are finite sums (for finite $\LL$ and $N$) 
that are encoded in the functions given in the previous section.  
The leading order coefficient depends on the function $F_1(\LL)$, 
expressible as single four-momentum sum. To the next order we need the 
functions $F_2$, $F_{2T}$ and $G_{2T}$. The first one is the sum of two terms: 
$F_{\mathrm{mes}}(\LL)$ which is  also given by a single four-momentum sum,
and  $F_{NA}(\LL)$  given  by a double four-momentum sum instead.  This
involves $\LL^8$ sums, limiting  the maximum value of $\LL$ that can be achieved. Some numerical values were obtained in Ref.~\cite{Heller:1984hx}. The functions $F_{2T}$ and $G_{2T}$
which are specific of the twisted case, share the same difficulty plus 
the additional one of depending on several variables: $N$, $\LL$ and $\m$.
\begin{table}
\renewcommand{\arraystretch}{1.3}
\setlength{\tabcolsep}{3pt}
\begin{tabular}{||l||c|c|c|c|c||}\hline \hline
LOOP &  $F_1(\infty)$ & $F_2(\infty)$ & 
$\tw_2(\infty,\infty)$ & $\hat{W}_2(\infty,\infty)$ & K(R,R) \\ \hline \hline
$1 \times 1$ & 0.125 & -0.0027055703(3) &   0.0129194297(3)  &
0.0051069297(3)
&0.12013262(2)\\ \hline
$2 \times 2$ & 0.34232788379 & -0.00101077(1) &   0.04178022(1)  &
-0.01681397(1)
&0.6361389(5)\\ \hline
$3 \times 3$ & 0.57629826424  & 0.00295130(2) &   0.07498858(2)  &
-0.09107126(2)
&1.294258(1)\\ \hline
$4 \times 4$ & 0.81537096352 & 0.0076217(1) & 0.1095431(1)  &
-0.2228718(1) 
&1.996582(5)\\ \hline
\end{tabular}
\caption{ Values at infinite volume of the functions $F_1$ and $F_2$
defined in the text. The next three columns are combinations of these 
numbers giving the second order coefficients $\tw_2$ and $\hat{W}_2$ at
large $N$, as well as the parameter $K$ of Ref.~\cite{Alles:1998is}
  }
     \label{tableI}
     \end{table}
     
\begin{figure}
\includegraphics[width=.9\linewidth]{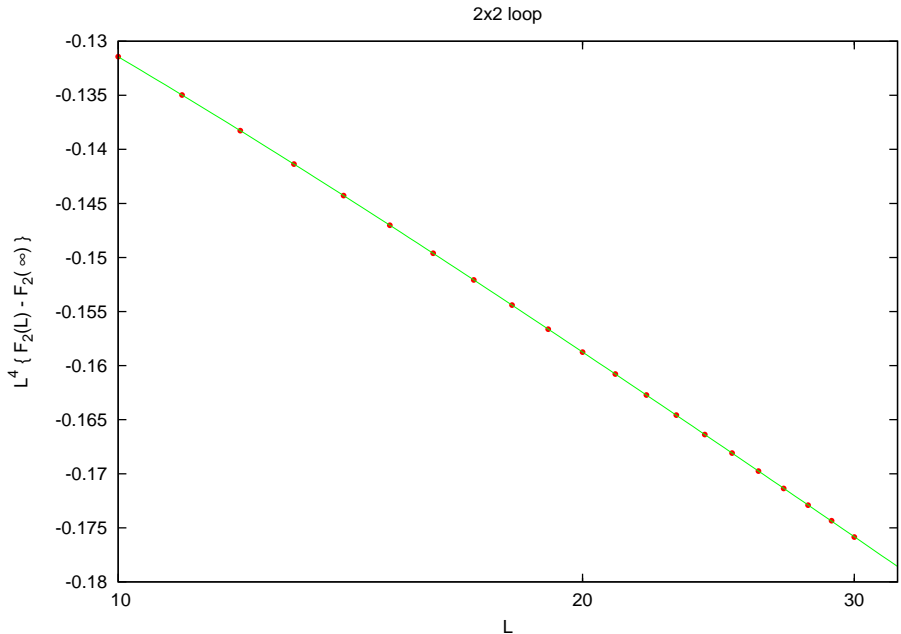}
\caption{The function  $\LL^4(F_2(\LL)-F_2(\infty))$ is plotted as a function of $\LL$ for
a $2\times 2$ Wilson loop. The solid line is the result of our fit Eq.\eqref{f2formula}.}
\label{figF2}
\end{figure}

Let us start by presenting our results for $F_i(\LL)$. As mentioned
earlier $F_1(\LL)$ can be computed for large values of $\LL$ ($\sim
100$). Since the leading coefficients in inverse powers of $1/\LL^2$
are known analytically, one can extrapolate the results to
infinite volume with very high precision. The values of $F_1(\infty)$
are given in table~\ref{tableI} for square Wilson loops up to $4
\times 4$. In the case of $F_2(\LL)$ we have been able to compute it
up to $\LL=34$. Going slightly beyond this point is feasible but
unnecessary. As mentioned earlier the behaviour for large $\LL$ is well  fitted by Eq.~\eqref{f2formula}. In the case of larger loops, stable fits require the inclusion 
of $1/\LL^6$ and $1/\LL^8$ terms. The  infinite volume value is given in 
table~\ref{tableI}. Errors are obtained from the variation of the
parameters with the fitting range. As a example of the  quality of the fit 
we display $\LL^4(F_2(\LL) - F_2(\infty))$   for a $2\times 2$ loop in Fig.~\ref{figF2}.  
Notice that while the infinite volume coefficients 
$F_i(\infty)$ grow moderately in size with $R$, the leading correction coefficient goes
rather like $R^4$. This is apparent from the similar magnitude of 
$\gamma_2$ and $\gamma_2'$ for all $R$ as seen in table~\ref{table2}.

Using $F_1(\infty)$ and $F_2(\infty)$ we can compute 
the infinite volume perturbative coefficients at any $N$. The
values at $N=\infty$ of the second coefficient in the expansion of the
Wilson loop expectation value and its logarithm ($\hat{W}_2$ and $\tw_2$
respectively) are also given in table~\ref{tableI}. We also add the
coefficient $K(R,R)$ used in Ref.~\cite{Wohlert:1984hk}. Our calculations are
consistent with the  precise  results of Ref.~\cite{Alles:1998is} for the plaquette and improve
by many significant digits the published results for larger loops ~\cite{Wohlert:1984hk,Heller:1984hx,Bali:2002wf}.

\begin{table}
\renewcommand{\arraystretch}{1.3}
\setlength{\tabcolsep}{17pt}
\begin{tabular}{||l||c|c|c|c||}\hline \hline
LOOP &$1 \times 1$ &$2 \times 2$ &$3 \times 3$ &$4 \times 4$\\
\hline \hline
$\gamma_2$ &-0.00325(10) & 0.0023(7)& 0.0036(8)&0.005(3)\\
\hline
$\gamma_2'$ & 0.0026(1)&0.0025(2)   &0.00275(20)& 0.0025(8)\\ 
\hline 
\end{tabular}
\caption{ Values of the parameters $\gamma_2$ and $\gamma_2'$ entering in the 
large $L$ expansion of the function $F_2$ given by Eq.~\eqref{f2formula}.
  }
     \label{table2}
     \end{table}

Now let us proceed to study the new functions $F_{2T}(N,\LL,k)$ and $G_{2T}(N,\LL,k)$, which appear for the twisted case. The difficulty in computing these functions numerically for large values of $\Le$ is the large number of sums involved. Within reasonable computer resources we could reach values of $\Le \sim 35$. Furthermore, these functions depend on three integer arguments making its study more demanding.
The functions also depend on the plane in which the Wilson loop is sitting. This breakdown of rotational invariance, similarly to the case  of CP,  is induced by the introduction of the twist vector. 
In any case, the symmetry is not completely broken and the residual symmetry in the case of the symmetric twist implies that all planes group into two different sets: $S_1\equiv \{01, 12,23,30\}$ and $S_2=\{02,13\}$.   
\begin{figure}
\includegraphics[width=.9\linewidth]{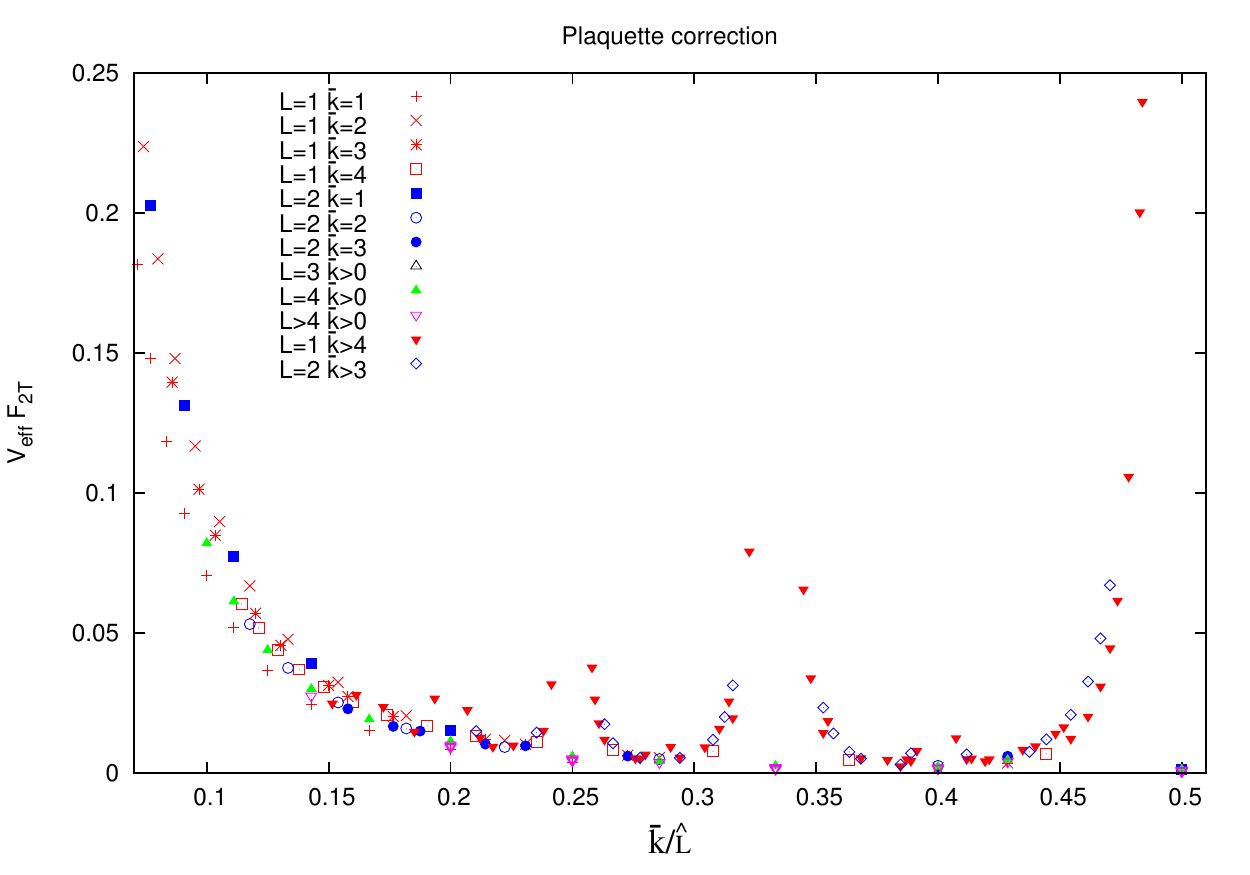}
\caption{The function $\Ve F_{2T}(\LL)$ for the plaquette is plotted as a function of
$\kbar/\hL$. 
}
\label{figf2tA}
\end{figure}

In
a recent work~\cite{Perez:2013dra}, the present authors advocated that
physical results for SU(N) gauge theories on twisted boxes 
depend on these variables only through the
combinations $\Le=\LL\hL$ and $\tilde \theta= 2 \pi \kbar /\hL$.  This
applies rather well to the 2+1 dimensional case both in perturbation
theory and non-perturbatively~\cite{Perez:2013dra,Perez:2014sqa,Perez:2014jra}
and to the non-perturbative calculation of the twisted gradient flow running coupling in $SU(\infty)$\cite{Perez:2014isa}.

\begin{figure}
 \centering    
 \begin{minipage}{.5\textwidth}
   \centering
 \includegraphics[width=0.9\linewidth]{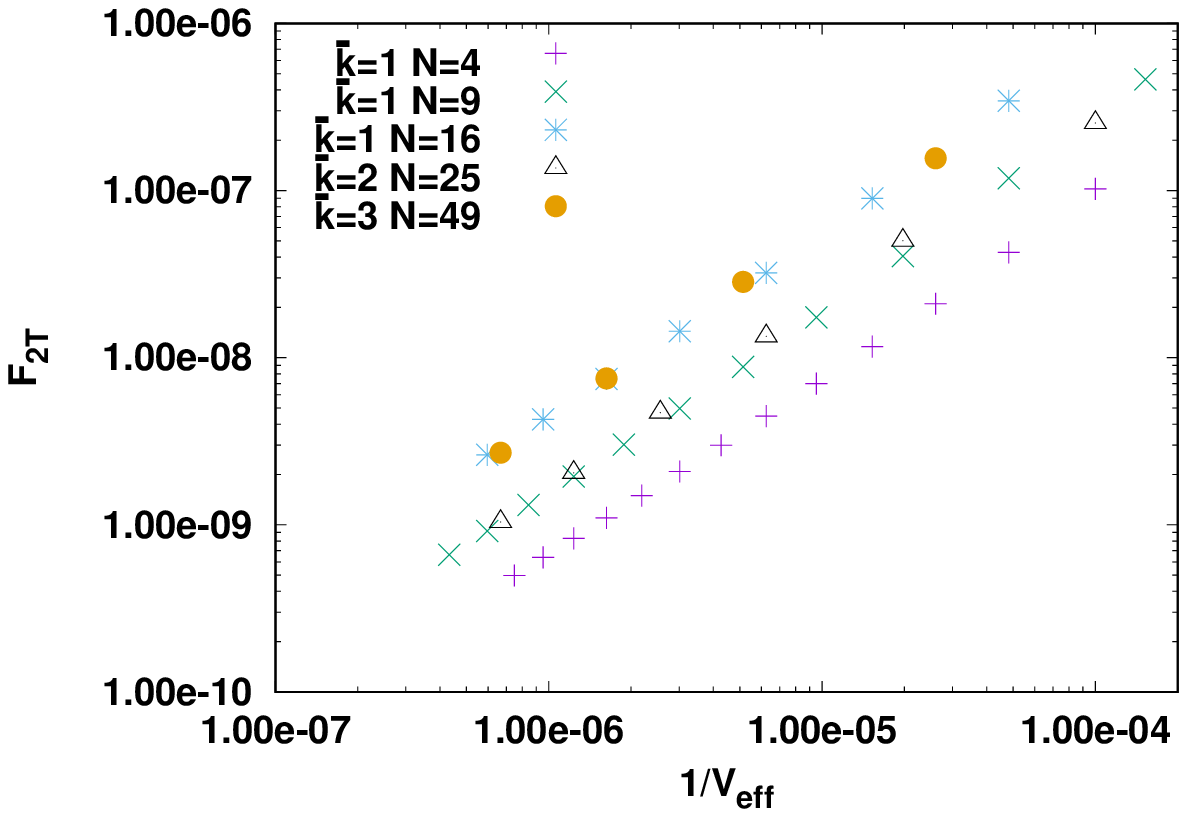}
      \captionof{figure}{$\LL$-dependence of $F_{2T}$  	   for various\\ $N$ and $\m$ values }			  
			  \label{LDEPF2T}
			  \end{minipage}%
			  \begin{minipage}{.5\textwidth}
			  \centering
\includegraphics[width=.9\linewidth]{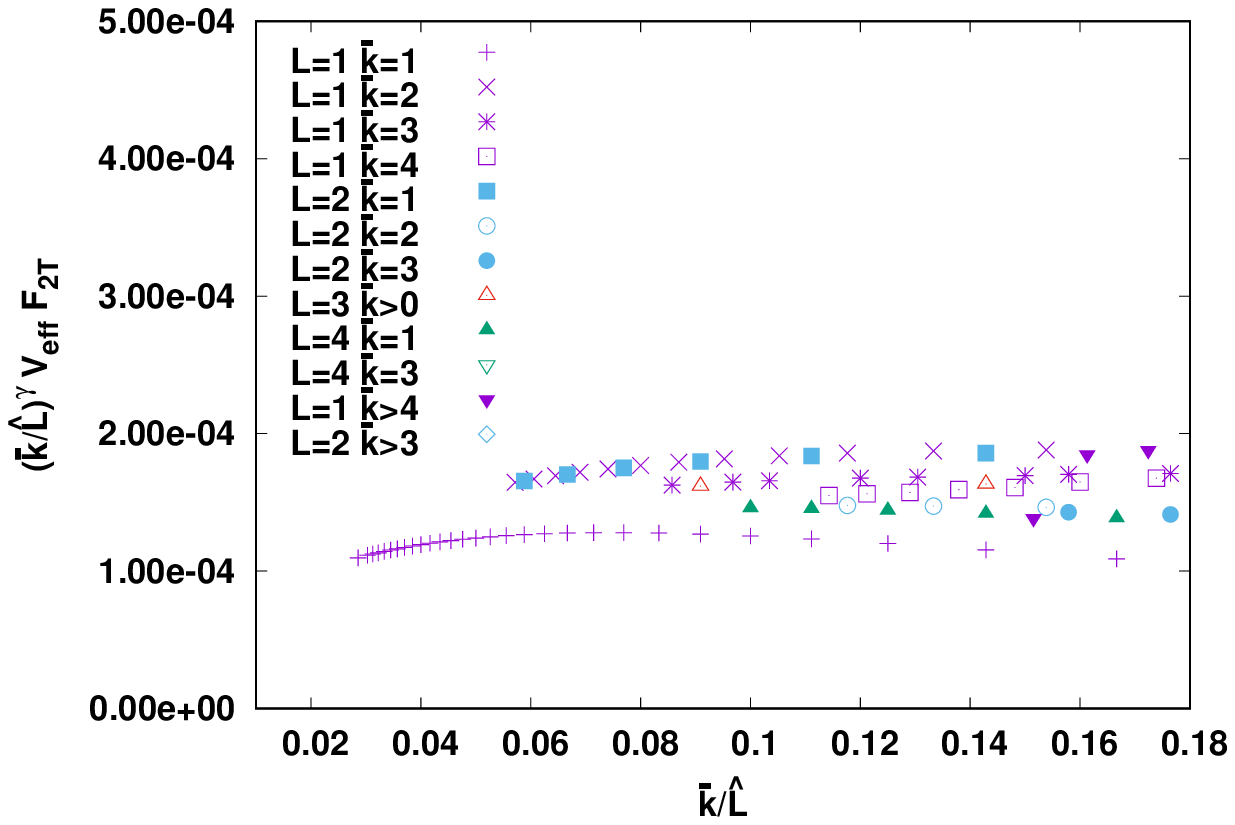}
              \captionof{figure}{Behaviour of $F_{2T}$ for small values of $\kbar/\hL$}
 \label{F2Tsmallthet}
			\end{minipage}
         \end{figure}

The previous observation suggests that we display the functions multiplied by the   effective volume $\Ve= \LL^4 N^2$ versus $\kbar/\hL$. All functions have a similar behaviour so that we will 
focus on  $F_{2T}$ for the plaquette for a plane in $S_1$. This is 
given in Fig.~\ref{figf2tA}.  Different symbols describe the
different values of the independent arguments $\kbar$ and $\LL$. The
plot contains a lot of information that we will now spell out. First
of all, the data does not show any growth with rising $\Le$ at fixed
values of $\kbar/\hL$. This is very important since it validates the
two main expectations of our previous discussion: that the function 
$F_{2T}$ goes to zero when either $\LL$ or $N$ go to infinity.
Furthermore, it tells us that when the limit is taken at fixed
$\tilde{\theta}$ the approach to zero goes roughly as $1/\Ve$. We
cannot exclude logarithmic or other mild dependencies, but this would
hardly change the conclusion. The result can be easily confirmed by 
studying the $\LL$ dependence of the values at fixed $\kbar$ and $N$. 
Our data at $N=4,9,16,25,49$ cover a sufficiently large number  of $\LL$
values to get a good fit to a $1/\LL^4$ dependence (see
Fig.~\ref{LDEPF2T}). 

Concerning the $N$ dependence the test is complicated by the fact that 
when we change $\hL$ we are also changing $\kbar/\hL$. However, as we
slightly change the value of $\kbar/\hL$ the value changes only by factors
of 2 or so. It is unclear at this stage whether as $N$ gets larger  one 
approaches a smooth oscillatory function or not. In any case,  these
changes are small compared to the large changes in values of $\Ve$.
Indeed, the value of $F_{2T}$ itself at neighbouring points sometimes changes 
by three orders of magnitude. As an example, let us discuss the results
for the range $\kbar/\hL\in [0.27,0.3]$.  We have 13 different values of $\kbar$
, $\LL$ and $\hL$ which give data in this region. The values of 
$F_{2T}$ themselves change considerably within this set. The result 
for $\LL=1$, $\hL=7$, $\kbar=2$ is $2.28\ 10^{-6}$, which multiplied 
by the effective volume gives $0.0055$. On the other extreme we have 
values as low as $7.42\ 10^{-9}$, $4.05\ 10^{-9}$, $6.71\ 10^{-9}$ 
for $(\hL, \LL,  \kbar)$ values of $(29,1,8)$, $(17,2,5)$ and $(7,4,2)$, 
which multiplied by the effective volume give $0.0053$, $0.0054$ and
$0.0041$ respectively. Similar results (within a factor of 2) are obtained
for the remaining data points. We believe this is enough to put our
main conclusion on robust grounds.

A different perspective is obtained if instead of fixing $\ttheta$ we
fix $\kbar$. For example if we fix $\kbar=1$ as we increase the value 
of $\hL=\sqrt{N}$ we are moving toward lower values of $\ttheta$ and
the coefficient begins to rise. This phenomenon can be seen in
Fig.~\ref{figf2tA} and continues for the data points not  shown in the plot  at smaller $\kbar/\hL$. A similar but hierarchically less pronounced increase is
observed for data points approaching $\kbar/\hL=\frac{1}{2},
\frac{1}{3}, \frac{1}{4}, \frac{1}{5}, \frac{2}{5}$. The increase for 
small values of $\kbar/\hL$ flattens out if we multiply the coefficient 
by $(\kbar/\hL)^\gamma$ with $\gamma$ in the range 2.5 to 3 (see
Fig.~\ref{F2Tsmallthet}). Since the 
value of $F_{2T}$ has been multiplied by $\hL^4$, the conclusion is
that even if we fix $\kbar=1$ the function $F_{2T}$ goes to zero in the 
large $N$ limit although at a slower rate $1/\hL^{4-\gamma}$. This
question recalls  the problems observed in the non-perturbative
simulations of the TEK model at $\kbar=1$. Simulations at intermediate
values of the coupling show a breakdown of center symmetry, which 
disappears when taking the large N limit at fixed~$\ttheta$~\cite{GonzalezArroyo:2010ss}.
At fixed order in  perturbation theory the breaking does not take place,
but the size of the corrections also points towards the benefits of 
keeping  $\ttheta$ within a reasonable range. A similar analysis can
be carried with respect to the potential divergences of 
$\Ve F_{2T}$ approaching the main harmonics $p/q$ of an analogous musical
scale (small $q$). Again the rise flattens when multiplying by $(\kbar/\hL-p/q)^\gamma$
with the same $\gamma$ as before. Once more, one concludes  from this 
that $F_{2T}$ does not diverge when taking a sequence $\kbar/\hL$ of values
converging to $p/q$. Curiously, if one takes $\kbar=p$ and $\hL=q$
with small $q$,  the  results have a  similar size to the rest. For example for $N=4$
and $\kbar=1$ or $N=9$ and $\kbar=1$, which correspond to
$\kbar/\hL=1/2,1/3$, the values one gets for various $\LL$ are not particularly large.

 \begin{figure}
 \centering    
 \begin{minipage}{.5\textwidth}
   \centering
 \includegraphics[width=\linewidth]{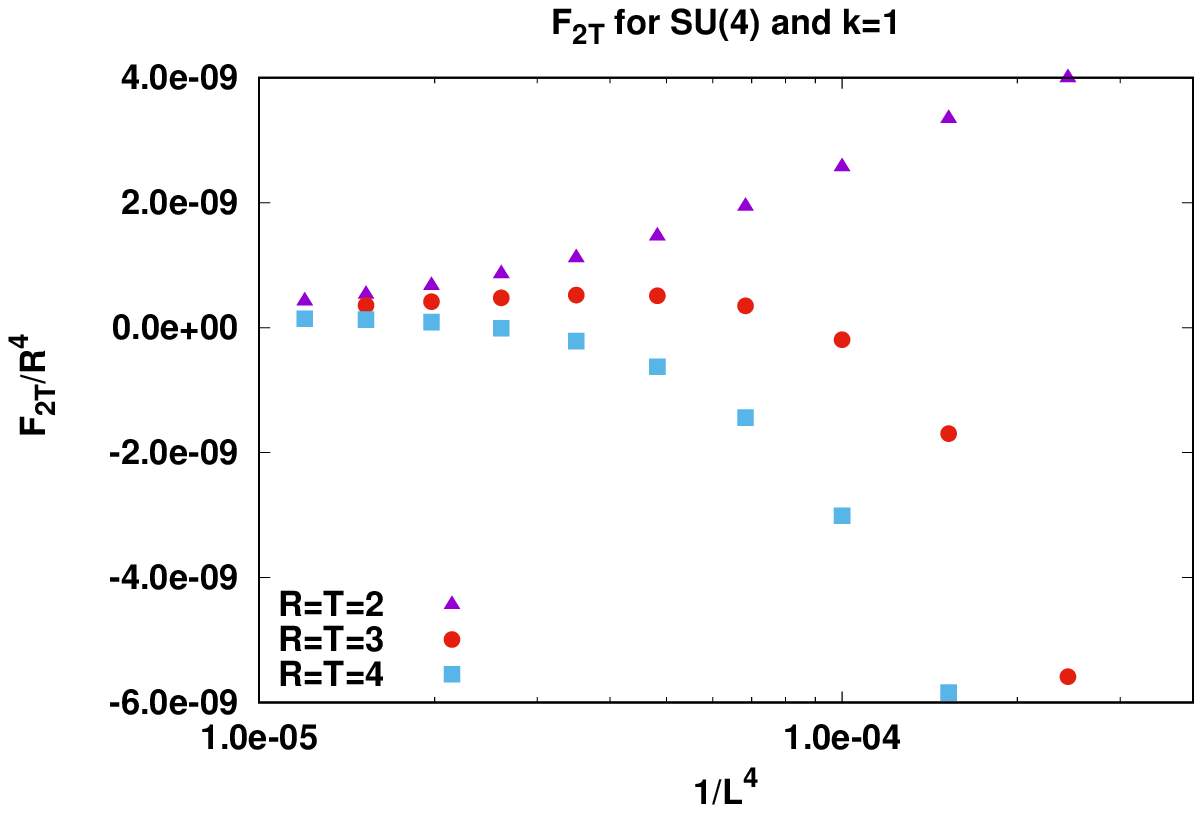}
      \captionof{figure}{$F_{2T}$ divided  by $R^4$ for an  $R\times R$ \\loop for SU(4) and $k=1$ as a function	\\
      of volume.}	     	      	  
			  \label{F2TIR}
			  \end{minipage}%
			  \begin{minipage}{.5\textwidth}
			  \centering
\includegraphics[width=.9\linewidth]{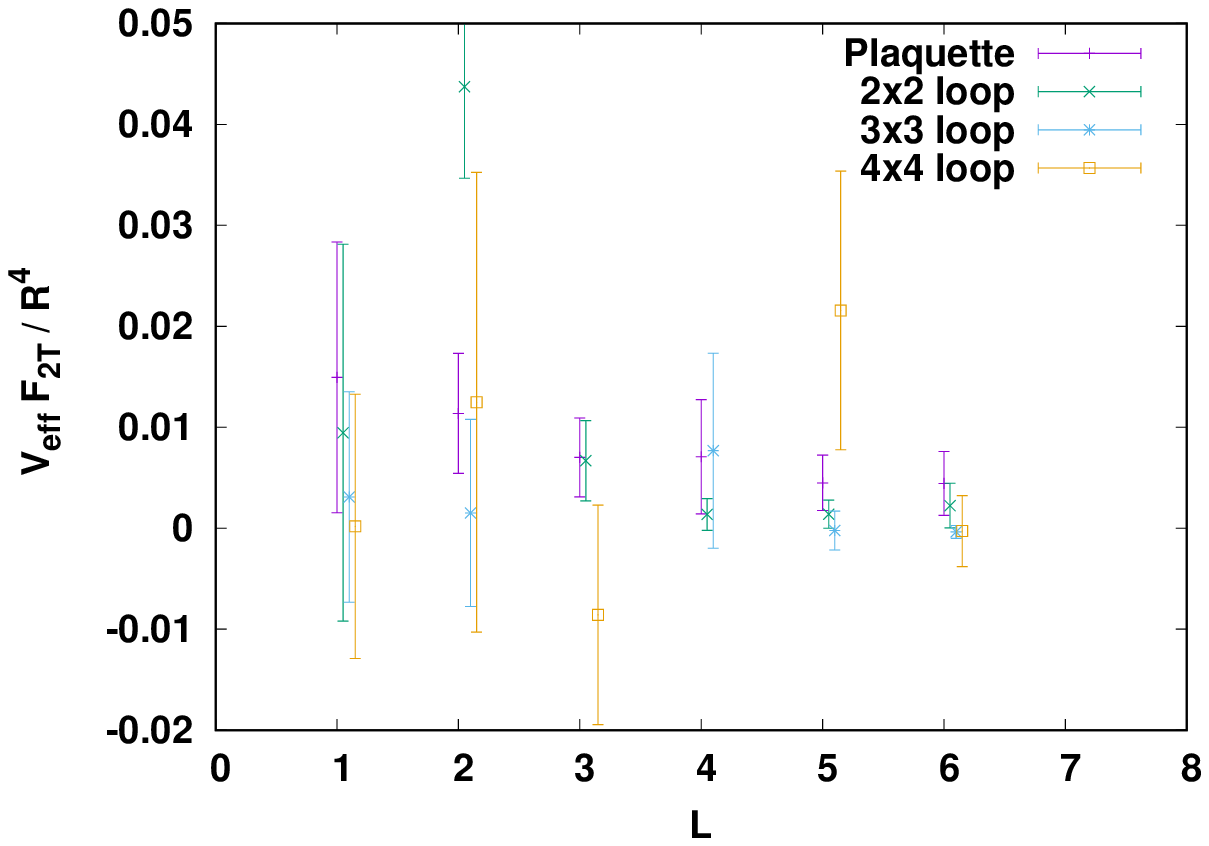}
              \captionof{figure}{$F_{2T}$ for all square loops multiplied by the effective volume and divided by the square of the area of the loop. The result is an average over various groups and sizes (see text)}
	    \label{F2Taver}	    
			\end{minipage}
         \end{figure}
         
Now we proceed to analyze the situation for larger  $R\times R$ loops. 
The results are consistent with $F_{2T}$  decreasing  with the
effective volume, but the $\kbar/\hL$ scaling of  $\Ve F_{2T}$ is much
less clear. One possible explanation is that as we increase $R$ the
asymptotic regime is achieved at larger values of $L$. As an example we 
we display in Fig.~\ref{F2TIR}  the case of SU(4). It is clear that
for  $R=4$ only for the largest sizes one can observe a linear
approach to zero. Another aspect  is also clearly illustrated by this 
figure: the growth of $F_{2T}$ with $R$. Re-scaling the data by $1/R^4$ 
we can put all data in the same plot. To see if this phenomenon
extends to all values of $\LL$, $\hL$ and $\kbar$ we studied  $\Ve F_{2T}/R^4$.
At fixed value of $L$ we averaged this quantity over all values of 
$\hL$ and $\kbar$ such that $\kbar/\hL \in [0.15, 0.45]$. The filter
eliminates the growth effects reported earlier for the plaquette. 
The final average is presented in Fig.~\ref{F2Taver} as a function 
of $L$. The results for different $R$ are slightly displaced for 
visualization purposes. The error bar is the dispersion of the set 
of averaged values. The main conclusion is that  all the values 
of $R$ and $L$ give results which are roughly of the  same size of order
$0.01$.  This is non-trivial given that the average values of $N^2 F_{2T}$
have been multiplied by $L^4/R^4$ ranging from 1296 to 0.0039.  
This leaves no doubt that the function $F_{2T}$ for larger loops also
goes to zero when either $L$ or $N$ go to infinity.  

\begin{figure}
\includegraphics[width=.9\linewidth]{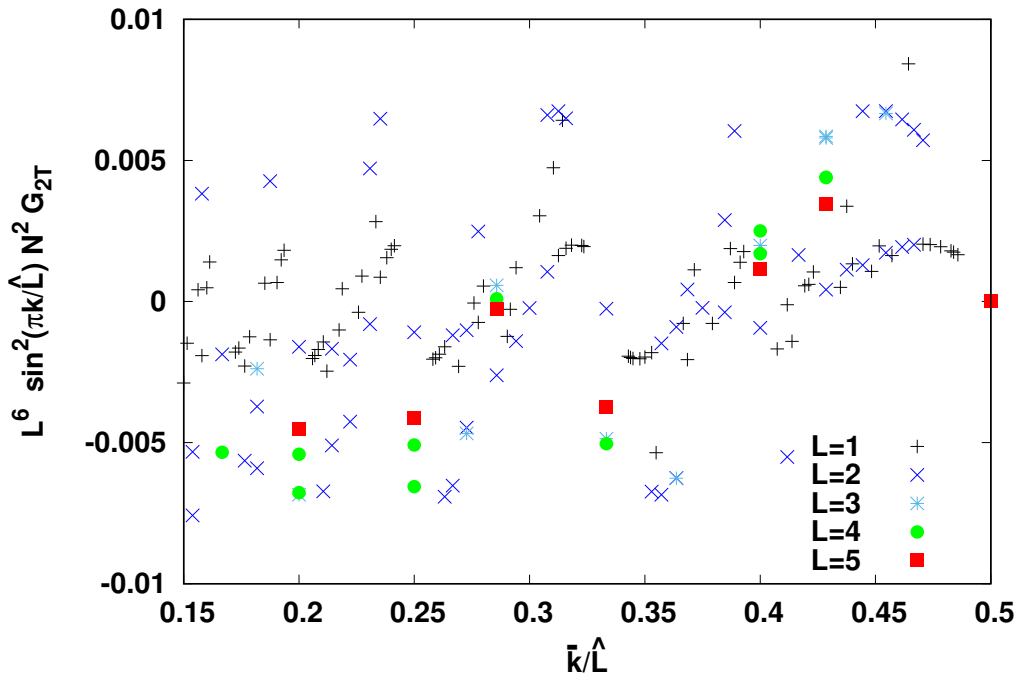}
\caption{The function $L^6 N^2 \sin^2(\pi \m/\hL) G_{2T}(\LL)$ for the plaquette is plotted as a function of
$\kbar/\hL$. 
}
\label{figg2tA}
\end{figure}
	 
Concerning the behaviour of the loops in the planes belonging to the $S_2$ set (02 and 13), the results are qualitatively the same as those reported previously, but the corresponding $F_{2T}$ function is typically a factor two or three smaller than the one for the planes in the $S_1$ set.      

Finally we should comment about the imaginary part of the Wilson loop coefficient described by the function $G_{2T}$. The conclusion is that this function also vanishes for either large volumes or large $N$. In the first case the values  drop at a faster rate compatible with $L^{-6}$. Another difference, is that while the real part $F_{2T}$ is typically positive, the imaginary part alternates in sign for the different values of $L$, $\hL$ and $\m$. The sign flips seem to coincide with the points where $\kbar/\hL$ approaches a  rational fraction $p/q$ with small denominator, where as we saw earlier $F_{2T}$ had peaks. These points coincide with those corresponding to small values of $\m$. A good way to display our results is that given in Fig.~\ref{figg2tA}. We multiplied the $G_{2T}$ function by the combination of factors given below:
\begin{equation}
L^6 N^2 \sin^2(\pi k/\hL) G_{2T}
\end{equation}
The result for all values of $L$, $\hL$ and $\kbar$ lies within a band  stretching from -0.01 to 0.01. The aforementioned $L$ dependence can be deduced from this plot. Notice that if we take the large $N$ limit keeping   $\m/\hL$ fixed, $G_{2T}$ goes to zero as $1/N^2$. However, of one keeps $\m$ fixed and takes $\hL=\sqrt{N}$ to infinity, the function is going to zero at a slower rate $\frac{1}{N \m^2}$. This matches our conclusion~\cite{GonzalezArroyo:2010ss} driven from non-perturbative considerations that it is better to take the large $N$ limit keeping $\m/\hL$ and $\kbar/\hL$  fixed and sizable.
		      
Our final discussion affects the   sum of all contributions and 
the comparison of the  numerical value of the second order coefficients 
$\tw(\LL,N,k)$ with that for infinite volume and number of colours. 
For the PBC case the  leading corrections are associated to finite 
$N$ or finite volume $V$, which  go as $1/N^2$ and $1/V$ (modulo logarithmic
corrections).  The coefficients are  $-(F_2(\infty)+F_1(\infty)/4)$ and 
$-F_1(\infty)/8- R^4(\frac{1}{64}+(\gamma_2+\gamma'_2\log(\LL)))$ respectively.  For the
plaquette the first coefficient is $-0.028$ and the second one is of
similar size for typical values of $\LL$.  As $R$ grows, the relative 
importance of the finite volume  correction grows since it contains a
term  that  goes as $R^4$, instead of $R^{1.35}$ which is roughly the 
dependence of the  $1/N^2$ coefficient. 

In the twisted case the finite volume and $N$ corrections are blended. 
Keeping only the leading terms in $1/N^2$ one gets 
\be
C_V(\Le) \frac{R^4}{\Ve} + C_N(L) \frac{1}{N^2} +F_{2T} + i G_{2T}
\ee
where $C_V(\Le)=-\frac{1}{64}-\gamma_2-\gamma'_2\log(\Le))$ and
$C_N=-F_2(\LL)-F_1(\LL)/8-F_1(\infty)/8$.  Notice that for large $\LL$
the first term goes to zero while the second term converges to the
finite $1/N^2$ correction of the periodic case. In the opposite extreme 
for the TEK model ($\LL=1$) the first and second terms combine to give
the finite volume correction of the periodic case. Let us now consider 
the general case. We may  ask ourselves what configuration gives the
smallest corrections at fixed number of degrees of freedom $\Ve \equiv VN^2$.
It is clear that the first term  does not depend on how  we
split these degrees of freedom onto spatial and colour ones. 
According to the analysis presented  in the previous paragraphs,
$F_{2T}$  has a similar structure to the first term 
$\propto \xi/(N^2 L^4)$ with a coefficient $\xi$ which 
varies slightly with the $L$-$\hL$ splitting and the value of $\kbar$.
On the contrary  the second term gets smaller for larger $N$.  
In conclusion,  the smallest corrections are 
obtained with the fully reduced TEK model, although the benefits
decrease as $R$ grows. To give a  quantitative idea of the implications 
we see that for the plaquette expectation value the correction is 
$4\ 10^{-8}$ for the TEK model and $\hL=29$, $7\ 10^{-7}$ for $L=2$-$\hL=14$, 
$10^{-5}$ for $L=4$-$\hL=7$ and $10^{-4}$ for $L=7$-$\hL=4$. In the
case of the $4\times 4$ loop all corrections for the previous cases, except 
the last one, are of order $10^{-5}$.    

\section{Additional considerations}
\label{s.additional}

\subsection{Comparison with numerical simulations}

\begin{figure}
\centering
\begin{subfigure}[b]{0.5\textwidth}
\includegraphics[width=\textwidth]{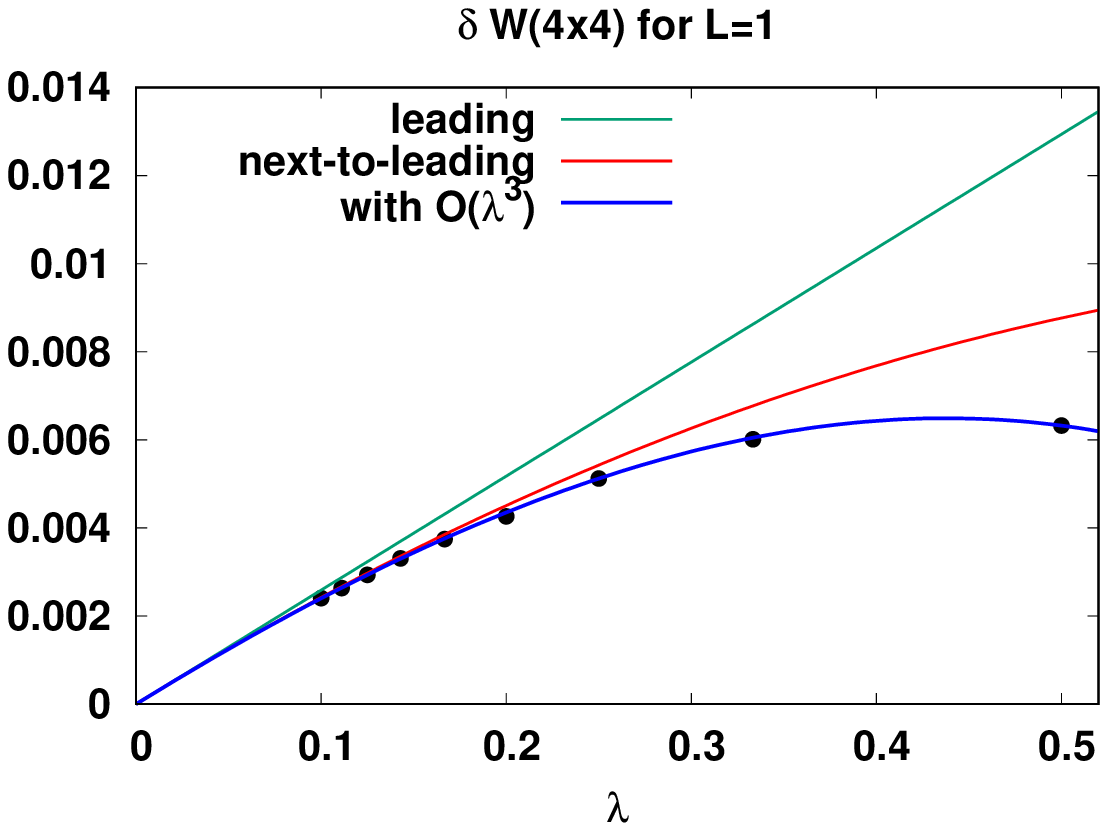}
\caption{$\LL=1$ (TEK model) $k=2$ and N=49.  }
\label{fig:8a}
\end{subfigure}%
\begin{subfigure}[b]{0.5\textwidth}
\includegraphics[width=\textwidth]{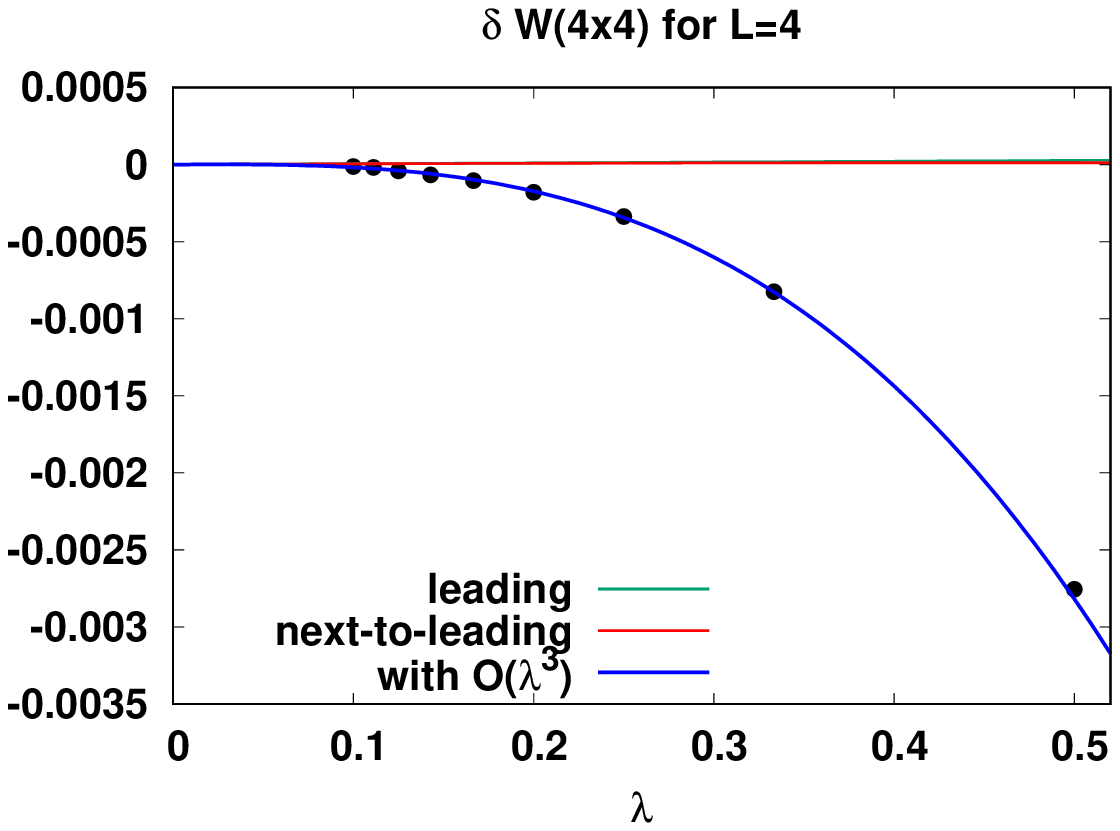}
\caption{$\LL=4$ $k=2$ and $N=49$. }
\label{fig:8b}
        \end{subfigure}
\caption{We display the Monte Carlo measured value of the real part of the $4\times 4$ loop for the twisted $N=49$ model on an $\LL^4$ box minus the prediction of perturbation theory at infinite volume and infinite N up to order $\lambda^2$ (Eq.~\eqref{PertDiff}). The difference is compared with the predictions of this paper at leading and next to leading order and a fit including a $\lambda^3$ correction. The left and right subfigures correspond to $\LL=1$ (TEK  model) and $\LL=4$ respectively.}
\label{fig:8T}
\end{figure}

Apart from the perturbative calculation we also measured the
expectation value of square  Wilson loops using Monte Carlo
simulations.  The purpose is to determine the region of values of
$\lambda=1/b$, for which this truncated perturbative expansion is a good
approximation.  Our methodology is based upon the auxiliary  field method~\cite{Fabricius:1984wp} followed by overrelaxation~\cite{Perez:2015ssa}. 
The numerical values of the perturbative coefficients for the
twisted case are very close to those of infinite $N$ and volume. To
notice a significant effect one has to consider small values of $N$,
$L$ and large values of the loop size $R$.
We first studied   $N=49$ with $k=2$ and measured the spatial average of the Wilson loops. To display our results instead of 
plotting the expectation value
of the Wilson loop directly, we substract its perturbative contribution
for infinite $N$ and volume as follows:
\be
\delta W(R\times R)\equiv W_{R,R}(b,N,L,k)-1+\lambda \hat{W}_1^{R\times R}(\infty,\infty) +
\lambda^2 \hat{W}_2^{R\times R}(\infty,\infty)  
\label{PertDiff}
\ee
Thus, this quantity measures both the difference between the
coefficients at finite and infinite $N$, $L$, as well as the effects 
of higher terms in the perturbative expansion. In Fig.~\ref{fig:8T}
we display the result for $N=49$  together with the analytic corrections to order 
$\lambda$, $\lambda^2$ and $\lambda^3$. The first two come from our
calculation in this paper. The latter is the result of a fit leaving
the coefficient $\hat{W}_3$ free.  The result for the TEK model $L=1$ (Fig.~\ref{fig:8a})
for the $4\times 4$ loop shows that the data follow our perturbative 
calculation up to $\lambda\sim 0.15$. For higher values of $\lambda$ a 
non-zero value of $\hat{W}^{(4\times4)}_3\sim 0.0195(4)$ is needed to match the 
measured value. On the other hand for $L=4$ (Fig.~\ref{fig:8b}) one sees that the numerical 
results are unable to distinguish the first two coefficients from those 
of infinite $N$ and $L$. The numerical value of $\hat{W}^{(4\times4)}_3\sim 0.0223(2)$
is close  to the one of the TEK model. 

The same analysis can be done for the 
smaller loops but
the difference between the finite and infinite $N$-$L$ is smaller. The
corresponding fitted values of the third order coefficients for $L=4$
are  $\hat{W}^{(1\times1)}_3=0.00087(2)$,
$\hat{W}^{(2\times2)}_3=0.00011(8)$ and
$\hat{W}^{(3\times3)}_3=0.00347(11)$. The errors do not include  systematics from neglecting higher orders. Unfortunately, these
coefficients are not known at infinite value of $N$ and $L$ except for
the plaquette~\cite{Alles:1993dn,Alles:1998is} giving  $\hat{W}^{(1\times1)}_3=0.000794223$.

We attempted a more detailed analysis in order to  verify the breakdown of CP and cubic invariance induced by the twist. These effects can be seen in  our calculated coefficients displayed in Table~\ref{table3a} for the aforementioned $N=49$ $k=2$ case and for $N=16$ $k=1$. Even for this low values of $N$ the effects are so tiny that one needs a very high statistics study to be able to observe this  breaking explicitly. For that purpose, we generated   500000 configurations of the TEK model in each case for 5 values of $b$ (2,4,6,8 and 10). The effect is of course more pronounced the smaller the value of $N$ and the bigger the value of $R=T$. 

\begin{table}
\renewcommand{\arraystretch}{1.3}
\setlength{\tabcolsep}{13pt}
\begin{tabular}{|| l |c|c|c|c||} \hline \hline
Plane & $k$ & $N$  & $S_1$ & $S_2$  \\
\hline \hline
Re $\hw_2^{11}$ &  1&16&0.505598546516147E-02& 0.504242209710592E-02\\
\hline
Re $\hw_2^{22}$ &1&16&  -0.146731255248501E-01&-0.147369863827783E-01\\
\hline
Re $\hw_2^{33}$ &1&16&-0.459930796958422E-01& -0.459201765925436E-01\\
\hline
Re $\hw_2^{44}$ &1&16&0.158384331597222\, \, & 0.157896050347222\, \, \\
\hline
$|{\rm Im} \hw_2^{11}|$ & 1&16&0.851447349773243E-05& 0.160804339096750E-04\\
\hline
$|{\rm Im}\hw_2^{22}|$ &1&16&0.208333333333333E-03& 0.448495370370370E-04 \\
\hline
$|{\rm Im} \hw_2^{33}|$ &1&16&0.135865088222789E-02& 0.105671370110544E-02\\
\hline
$|{\rm Im}\hw_2^{44}|$ &1&16&0&0  \\
\hline
\hline
Re $\hw_2^{11}$ &  2&49&0.510102929561289E-02&0.510026134690107E-02\\
\hline
Re $\hw_2^{22}$ &2&49& -0.166780624449552E-01 &-0.166807555892582E-01\\
\hline
Re $\hw_2^{33}$ &2&49&-0.882280314255889E-01& -0.882350911570933E-01\\
\hline
Re $\hw_2^{44}$ &2&49&-0.206182506618171\, \, & -0.206240116435204\, \, \\
\hline
$|{\rm Im} \hw_2^{11}|$ & 2&49&0.287740792864228E-06&0.101618022893695E-05 \\
\hline
$|{\rm Im}\hw_2^{22}|$ &2&49&0.142439450854039E-04& 0.199080420953164E-05 \\
\hline
$|{\rm Im} \hw_2^{33}|$ &2&49&0.182495006628446E-04&0.649747149140617E-04 \\
\hline
$|{\rm Im} \hw_2^{44}|$ &2&49&0.383145588068173E-03&0.447564739887152E-03  \\
\hline
\end{tabular}
\caption{ Second order coefficients of the $R\times R$ Wilson loops with $R=1 \cdots 4$,  for the TEK model and: $N=16$ ($k=1$) and $N=49$ ($k=2$). The coefficients have been computed with quadruple precision.}
\label{table3a}
\end{table}

We fitted the results of our Monte Carlo to a polynomial of third degree in $\lambda=1/b$, but fixing the first two coefficients to the analytic result. This was done for the real and imaginary parts of the Wilson loops in each plane separately. The  two free parameters of the fit measure the quadratic and cubic coefficients of the polynomial in $\lambda$. For the $N=49$ case, the results for the quadratic piece coefficient agrees with the results of table~\ref{table3a}. Unfortunately, the errors are of the same size as the breaking of the cubic symmetry so that this aspect could not be tested with the only exception of the imaginary part of the $4\times 4$ loop. The value of this coefficient obtained for the S1 planes was 0.000385(16), and for the S2 planes 0.000488(36). This shows clearly both the CP and cubic invariance violation with statistical significance in agreement with table~\ref{table3a}. In the case of the real part, although  unable to show a clean plane dependence, the results were perfectly in agreement with the same table. The fitted coefficients for the S1 planes were 0.005092(12),-0.01664(6), -0.08829(8) and -0.20619(12) for $R=1$,$2$,$3$ and $4$ respectively. 

In order to see the violation of cubic invariance more neatly we also studied the $N=16$ $k=1$ case. Here the imaginary part (which vanishes for $R=4$) shows clearly the breaking for $R=1$, $2$ and $3$. For example for the $3\times 3$ loop , the fitted coefficient is 0.00137(2) for the S1 planes  and  0.00099(3) for the S2 planes. In the case of the real part there is a signal of breaking for the $4\times 4$ loop, giving 0.15830(3) and 0.15778(6) for S1 and S2 respectively. 

\subsection{Addition of fermions in the adjoint}
A very simple extension of our work is that of including fermions. There is a difficulty in including fermions in the fundamental representation since the twisted boundary conditions are singular for them. There are two ways to circunvent this problem. One is to include flavour to compensate for the boundary conditions. The other one is to allow the fermions to live in a larger lattice where they are insensitive to the boundary conditions. On the other hand there is no problem in adding fermions in the adjoint representation. There are many reasons for considering this theory interesting. One is certainly supersymmetry, but another one is the proposal done by several authors of restoring volume independence for the periodic boundary conditions case~\cite{Kovtun:2007py}. 

Another incentive for considering fermions is the simplicity of adding them. At the order that we are working the contribution to Wilson loop expectation values comes through a fermion loop term in the vacuum polarization, which is rather simple to add. However, the addition also induces a proliferation of options: fermion masses, number of flavours, type of lattice Dirac operator, etc. The comparison and analysis  is very interesting, no doubt, but it opens up a non-trivial addition to this, already long and complex work. Hence, we opted for a mild inclusion in which we simply stick to Wilson fermions with a fixed value of the hopping parameter. The contribution of fermions to the Wilson loop amounts to the addition of a new term to  the second order coefficient $\tw^\tbc_2\WAb$, which we label $N_f H_2 \WAc$, with $N_f$ the number of adjoint flavours. Given that there is no contribution to first order there is an apparent conflict with the claim that this addition restores volume independence. However, we recall that the calculation in the case of periodic boundary conditions is not complete. We have expanded around the non-trivial holonomy ground-state and ignored the contribution of zero-modes. The addition of fermions is expected  to affect the degeneracy of classical vacua which is responsible of the zero-modes. 

We will now present our result for $H_2\WAc$ and discuss its structure. 
For simplicity we will focus on the case  of a symmetric box and symmetric twist in  4 dimensions. The Fourier decomposition of the adjoint fermion fields is similar to that of the gauge fields and the Feynman rules, presented in App.~\ref{appendixA}, are easily derived. 
They lead to two extra terms in the vacuum polarization, given 
in App.~\ref{appendixB2}. Our expressions can be mapped to the standard ones for
fundamental Wilson fermions in infinite volume~\cite{Kawai:1980ja} by performing the substitution given in Eq.~\eqref{typeNA} and taking into account the change in the trace normalization of the fermion representation. One of the self-energy terms is a lattice tadpole, given by $\Pi^{f_1} (q)$ in Eq.~\eqref{eq.vacf1}. The other, $\Pi^{f_2} (q)$ in Eq.~\eqref{eq.vacf2},  is the lattice analog of the fermionic contribution to the gluon self-energy. 
These two terms contribute at second order in $\lambda$ to $U_2^{(2)}$ through Eq.~\eqref{eq.u22}. They are proportional to $F^2$ and hence of purely non-abelian nature. Following the same strategy as in Eqs.~\eqref{eq.fsq} -~\eqref{eq.f2t}, they can be decomposed in two functions in terms of which the fermionic contribution to $\tw_2^\tbc$ reads
\be 
H_2^\tbc(\kappa,N,\LL,k) = \Big(F_{2}^f(\kappa,\Le)-\frac{1}{N^2}F_{2}^f(\kappa,\LL)\Big) +  F_{2T}^f(\kappa,N,\LL, k)\quad ,
\ee
with
\be 
F_{2T}^f(\kappa,N,\LL, k)= \frac{1}{N^2}F_{2}^f(\kappa,\LL) + {\cal F}_{NP}^f(\kappa,N,\LL,k)\quad.
\ee
As for the pure gluonic case, $F_{2T}^f$ should go to zero both in the large $N$ and in the infinite volume limit. 

We will briefly discuss below the results of the numerical evaluation of $F_2^f$ and $F_{2T}^f$ for $r=1$ massless Wilson fermions. Note that we can directly work with massless adjoint fermions due to the absence of zero-modes in the twisted box. Let us start by analyzing the behaviour of $F_2^f$ at large volume. As mentioned above, this function comes from the contribution of two fermion self-energy terms. Both of them have a leading $1/\LL^2$ correction that arises from a constant, volume-independent, term in the vacuum polarization. The structure of the correction is hence identical, modulo an overall coefficient, to the one coming from the measure. The leading $1/\LL^2$ correction to $F_2^f$ takes thus the form:
\be
-12 \, (C_{f_1}+ C_{f_2}) \, \Big (\bar{\gamma} \, \frac{R^2 T^2}{\LL^2} \Big )
\ee
with the same constant $\bar{\gamma}$ appearing in Eq.~\eqref{eq.gammabar}. An easy way to determine the coefficients $C_{f_i}$ is to compute the vacuum polarization at vanishing external momentum. This is a single momentum sum whose volume expansion can be obtained following the strategy described in Appendix~\ref{appendixC}. The constant, volume-independent, term is given by the infinite volume expression. In the particular case of massless $r=0$ Wilson fermions, it is easy to see that it vanishes~\cite{Kawai:1980ja}. The same holds for other values of $r$, implying that $C_{f_1}+C_{f_2}=0$. Although not required for computing the expectation value of the Wilson loop, one can easily determine the tadpole coefficient analytically in the massless case from the infinite volume formula:
\bea
C_{f_1}(r) &=& -1 + \frac{1}{d} \sum_\mu \, \Pi^{f_1}_{\mu,\mu} (q=0)\Big |_{r,\LL=\infty} - \frac{1}{d} \sum_\mu \, \Pi^{f_1}_{\mu,\mu}(q=0)\Big |_{r=0,\LL=\infty}\\
&=& -1 + \int D\alpha\, \, \frac{r\, d\, M(\alpha) }{M^2(\alpha) + \sum_\mu \sin^2(\alpha_\mu)}\nonumber
\eea
with $M(\alpha) = rd - r \sum_\mu \cos(\alpha_\mu)$.
The integral can be estimated numerically. In four dimensions for $r=1$ we obtain $C_{f_1}(r=1)=-0.0612733799(1)$.

\begin{table}
\renewcommand{\arraystretch}{1.3}
\setlength{\tabcolsep}{8pt}
\begin{tabular}{||l||c|c|c|c||}\hline \hline
LOOP & $1 \times 1$  & $2 \times 2$ & $3 \times 3$ & $4 \times 4$\\ \hline\hline
  $F_2^f(\infty)$ & -0.0013858405(1)   & -0.004721988(1) &-0.00877155(1) & -0.01312182(1)\\ \hline 
\end{tabular}
\caption{ Values at infinite volume of the function $F_2^f(\kappa_c, \infty)$
defined in the text for massless $r=1$ Wilson fermions.}
     \label{table3}
     \end{table}

\begin{table}
\renewcommand{\arraystretch}{1.3}
\setlength{\tabcolsep}{17pt}
\begin{tabular}{||l||c|c|c|c||}\hline \hline
LOOP & $1 \times 1$  & $2 \times 2$ & $3 \times 3$ & $4 \times 4$\\ \hline
\hline
  $\gamma_2^f$ & 0.014(1) & 0.0018(1) & -0.0009(2) & -0.0015(3)\\ \hline 
  $\gamma_2^{f \prime }$ & -0.0020(1) & -0.0020(1) & -0.0018(1) & -0.0019(1) \\ \hline
\end{tabular}
\caption{ Values of the parameters $\gamma_2^f$ and $\gamma_2^{f \prime }$ entering the 
large $\LL$ expansion of the function $F_2^f(\kappa_c, \LL)$ given by Eq.~\eqref{f2fformula} for massless $r=1$ Wilson fermions.
  }
     \label{table4}
     \end{table}
With the cancellation of the leading $1/\LL^2$ correction, the large $\LL$ expansion of $F_2^f$ in four dimensions is given by:
\be
\label{f2fformula}
F_2^f(\kappa_c,\LL)=F_2^f(\kappa_c,\infty)- \, \frac{R^2
T^2\, (\gamma_2^f+\, \gamma_2^{f \prime }\, \log(\LL))}{\LL^4}+ \ldots
\ee
The infinite volume values for loops up to $4\times 4$ are presented in table \ref{table3}. The results for the $1\times 1$ and $2\times2$ loops are consistent with the less precise results by Bali and Boyle \cite{Bali:2002wf}. Our best fit values for $\gamma_2^f$ and $\gamma_2^{f \prime }$ are given in table \ref{table4}. They are similar in magnitude to the gluonic ones,  previously presented in table \ref{table2}. Notice that the coefficients of the leading fermionic and gluonic logarithmic corrections are in both cases almost independent of the loop size and opposite in sign, with the fermions counteracting as expected the gluonic contribution.

\begin{figure}
\includegraphics[width=.95\linewidth]{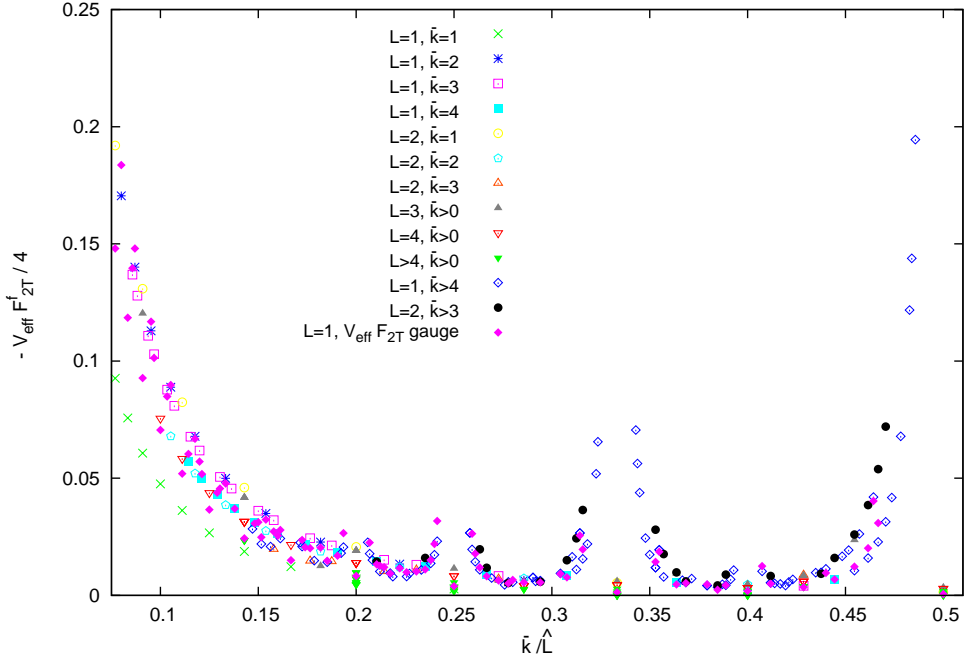}
\caption{The function $-\Ve F_{2T}^f(\LL)/4$ for the plaquette in a $S_1$ plane is plotted as a function of $\kbar/\hL$ for massless $r=1$ Wilson fermions. For comparison we also display the pure gauge results for the TEK, $L=1$ model. }
\label{figf2tfA}
\end{figure}

The remaining function $F_{2T}^f$ is very similar in structure to its gluonic counterpart but with opposite sign.
It tends to zero in the same way when either $N$ or $L$ go to infinity. 
As an illustration, we plot in Fig. \ref{figf2tfA} the quantity $- \Ve  F_{2T}^f/4$ as a function of $\kbar/\hL$, for the plaquette in a $S_1$ plane. The plot corresponds to massless $r=1$ Wilson fermions. The factor 1/4 has been chosen to obtain a result comparable to the gluonic contribution. This is illustrated by displaying in the plot the pure gauge results for the $L=1$ TEK model from Fig.~\ref{figf2tA}. At a given value of $\kbar/\hL$, the two functions have the same magnitude. As a last remark, we point out that the function $F_{2T}^f$ for other square loops scales as $R^4$ like in the pure gauge case. 

Although it would be interesting to explore the dependence on the fermion mass and extend this analysis to other kind of lattice fermions, this is a lengthy project that is beyond the scope of this paper and will be addressed elsewhere.

\subsection{U(N) versus SU(N)}
It is interesting to compare the perturbative expansion of these two groups.  In the large $N$ limit the two groups differ only by $1/N^2$
corrections. In principle the U(N) group is neater as exemplified by the `t Hooft double-line notation. Our calculation was done for the SU(N) group, so that it would be interesting to know which of the $1/N^2$ corrections are attributable to the restriction to this group. At leading order the result is rather simple: all $1/N^2$ dependence disappears when studying U(N) instead of SU(N). Thus, 
the leading order coefficient is $F_1(L)$ for periodic boundary conditions and $F_1(\Le)$ for twisted ones. This is consistent with 't Hooft topological expansion which holds for U(N). All corrections in powers of $1/N^2$ are associated to non-planar diagrams, which are absent at leading order. 

If we proceed to next-to-leading order and focusing on the coefficient of the logarithm of the Wilson loop, one sees that the additional U(1) gluon present in U(N) only contributes to tadpole-like terms. Revising our calculation we can easily identify 
in what places we omitted the possible contribution of that gluon. Indeed, this was implemented by the restriction in the sum over momentum denoted by a prime. This only appeared in the terms that we called abelian: the measure contribution and the tadpole. Only the latter is affected by the addition of the U(1) mode.  In summary, the second order coefficients of the logarithm of the Wilson loop for the U(N) theory are  given by
\be
\tw^\pbc_2(N,\LL)= \Big(1-\frac{1}{N^2}\Big) F_2(\LL)+ \frac{1}{2d}\Big(1-\frac{1}{V}\Big) F_1(\LL)
\ee
and
\be
\tw^\tbc_2(N,\LL, k)= \Big(F_2(\Le)-\frac{1}{N^2}F_2(\LL)\Big) + 
\frac{1}{2d}F_1(\Le) +
F_{2T}(N,\LL,k) + i G_{2T}(N,\LL,k)
\ee
This matches nicely with the identification of non-planar diagrams in the U(N) theory with the only exception of the measure insertion. 

\section{Conclusions}
In this paper we have studied  the  perturbative
expansion of Wilson loops up to order $\lambda^2=g^4 N^2$ for lattice Yang-Mills fields (with Wilson action)  in a finite box with  irreducible twisted boundary conditions.~\footnote{A web tool  will allow  other scientists to obtain the value of the coefficients for their particular setting.} 
Contrary to the case of periodic boundary conditions, this perturbative expansion at finite volume is perfectly well-defined. This is due to the absence of zero-modes.  Our general presentation is valid for any irreducible twist and any dimension. The final formulas are given in terms of finite momentum sums. The effect of the different twists  sits in the range over which these momentum sums run and the particular form of the momentum-dependent structure constants. 

We have then studied with special focus the case of symmetric twist in a symmetric box. In particular we have analyzed the difference between the results with twist and those obtained in a simplified version of periodic boundary conditions in which the effect of zero-modes is neglected. These results depend on some common functions, $F_1(\LL)$ and $F_2(\LL)$,  of the lattice size. Their large $\LL$ dependence and infinite volume value have been determined for the four-dimensional case with an increased precision  with respect to previous determinations. For the twisted boundary conditions case, the coefficient of the perturbative expansion to order $\lambda^2$ for each type of loop also depends on a complex  function $F_{2T}(\LL,N,\m)+i G_{2T}(\LL,N,\m)$. This function contains the  contribution of non-planar diagrams and vanishes when either $\LL$ or $N$ go to infinity.  The latter fact being a manifestation of the phenomenon of {\em volume independence}, while the former signals the independence of boundary conditions for large volumes $\V$.  These functions contain all the dependence of the result on the common flux $\m$ of the symmetric twist and, hence, are the only terms where CP and cubic symmetry violations show up. For the four-dimensional case we have evaluated the functions for a large number of values of the arguments ($\LL$, $N$ and $\m$) in order to determine their value and the rate of decrease to zero with either $N$ or $\LL$. The best way to describe our findings is by plotting the values as a function of $\kbar/\hL$, where  $\kbar$ is the congruent inverse of $\m$ modulo $\hL\equiv \sqrt{N}$. It turns out that for generic values of this ratio,  the function decreases as one over the effective volume $\Ve=N^2 \V$. The coefficient multiplying $1/\Ve$ grows  when $\kbar/\hL$ tends to rationals with small denominators, effectively reducing the power of $N$ at which the functions vanish. These results are  consistent with the requirement, established in Ref.~\cite{GonzalezArroyo:2010ss}  on the basis of non-perturbative arguments, that both this ratio as well as $k/\hL$ should be kept large  enough when taking the large $N$ limit. It is interesting to realize that, although centre symmetry cannot be broken at finite $N$ and $\LL$, our analytic calculations reinforce the interest of  approaching  the large $N$ limit following our criteria.

All our results apply as well for the one-site Twisted Eguchi-Kawai reduced model ($\LL=1$). Indeed, the best determination of the infinite $N$ and infinite volume plaquette expectation value for a fixed finite number of degrees of freedom is obtained by using this fully reduced model. For expectation values of large loops, this advantage with respect to partial reducion ($\LL>1$) diminishes.

 We should emphasize that many of the calculations of this paper have been performed independently and using different programs by a subset  of the authors. This minimizes the possibility of errors. Furthermore, we have also compared our results with Monte Carlo simulations at large values of $b$.
A very high statistics study is necessary to verify some of the specific predictions of our calculation, such as the pattern of cubic symmetry breaking and the non-vanishing imaginary parts for each fixed plane of the loop. We found perfect agreement. Furthermore, these numerical results allow us to find out  the range of values of $b$ for which $\mathcal{O}(\lambda^2)$ provide a good approximation.  An estimate of the $\mathcal{O}(\lambda^3)$ coefficients has also been obtained.

A bunch of additions have been included to make this paper as complete as possible. In particular, we have analysed the difference between the U(N) and SU(N) cases, and more importantly we have also computed  the effect of  including  fermions in the adjoint representation. These fermions are fully compatible with twisted boundary conditions and have been subject of great interest because of their role in supersymmetry, volume independence~\cite{Kovtun:2007py,Catterall:2010gx,Azeyanagi:2010ne,Hietanen:2010fx,Hietanen:2009ex,Lohmayer:2013spa,Basar:2013sza,Bringoltz:2009kb,Koren:2013aya,Bringoltz:2011by,Gonzalez-Arroyo:2013bta}, infrared fixed points~\cite{Sannino:2004qp,Luty:2004ye,Bursa:2009we,DelDebbio:2010hx,Perez:2015yna}, etc. In relation with our perturbative results, their contribution  only enters via the self-energy of the gluons and is proportional to the number of flavours $N_f$. For the four-dimensional symmetric twist case, we studied its effect on $F_{2T}$ for massless Wilson fermions. The result is similar qualitatively and quantitatively to the purely gluonic results.    Given their interest a more detailed analysis using different versions of lattice fermions and different masses is well justified but falls away from the main scope of this paper.

\section*{Acknowledgments}

We would like to thank Gunnar Bali, José L. F. Barbón, Fernando Chamizo, Herbert Neuberger, Carmelo P. Martín and Alberto Ramos for useful comments, discussions and suggestions.
M.G.P. and A.G-A acknowledge financial support from the MINECO/FEDER grant
FPA2015-68541-P  and the MINECO Centro de Excelencia Severo Ochoa Programs SEV-2012-0249 and SEV-2016-0597.
M. O. is supported by the Japanese MEXT grant No 17K05417 and the MEXT program for promoting the enhancement of research universities. We acknowledge the use of the Hydra cluster at IFT and  Hitachi SR16000 supercomputer
at High Energy Accelerator Research Organization(KEK)
supported by the Large Scale Simulation Program No.16/17-02.

\appendix
\section{The Feynman rules at order $\lambda$}
\label{appendixA}

Here we collect the Feynman rules of the Wilson action on a twisted box to order $\lambda$. They can be found in explicit form in  papers of other authors, such as  Snippe in Ref.~\cite{Snippe:1997ru}. As explained in the text, 
they are valid for any  twisted box with arbitrary irreducible twist if one takes  into account the corresponding modifications in the set of lattice momenta $\LLe$ and the convention dependent group constants $F$ and $D$. We include in addition the Feynman rules for Wilson fermions, at the same order, in the adjoint representation.
For simplicity we adopt the hermiticity condition  $\Phi(p, -p) =0$. 
As discussed in Sec.~\ref{s_gaugef}, under this condition the $\delta$ functions imposing momentum conservation at the vertices should be understood in strict sense and not modulo $2\pi$. 

The different vertex terms, up to order $\lambda$, arising from the Fourier expansion of minus the gauge fixed action $-(S+S_{GF}+S_{GH}+S_{MS})$ described in Sec.~\ref{s_gaugef} are given by:

\begin{itemize}
\item
Vertex coming from the measure term
\be
-{\lambda \over 24 } \sum'_q \hA_\mu (q) \hA_\mu(-q)\, ,
\ee
\item
Ghost-gluon vertices
\bea
 &-&i\,  \sqrt {\frac{\lambda N}{\Ve} } F(q_1,q_2,-q_1-q_2) \, \widehat q_{1\mu} \cos (q_{2\mu}/2) \\ &&\delta(q_1+q_2+q_3) \, \bar c (q_1)\, c (q_2) \, \hA_\mu (q_3) e^{-i \frac{q_{3\mu}}{2}} \, ,\nonumber
\eea
and 
\bea
&-&{\lambda N \over 12\Ve }   F(q_1,q_3,-q_1-q_3)  F(q_2,q_4,-q_2-q_4)\, 
 \widehat q_{1\mu} \widehat q_{2\mu} \\
&& \delta(q_1+q_2+q_3+q_4)\, \bar c (q_1) \, c (q_2) \, \hA_\mu (q_3) \, \hA_\mu (q_4) e^{-i \frac{q_{3\mu}+q_{4\mu}}{2}}  \nonumber \, .
\eea

\item
3-gluon vertex
\bea
&&\frac{i}{3!}\sqrt {\frac{\lambda N}{\Ve} } F(q_1,q_2,q_3) V_{\mu_1\mu_2\mu_3} (q_1,q_2,q_3) \\
&&\delta(q_1+q_2+q_3) \hA_{\mu_1} (q_1)\hA_{\mu_2}(q_2) \hA_{\mu_3}(q_3)
e^{-\frac{i}{2} (q_{1\mu_1}+q_{2\mu_2}+q_{3\mu_3})}\, ,
\nonumber
\eea
where
\be
V_{\mu_1\mu_2\mu_3} (q_1,q_2,q_3) = \Big(\cos({q_{3 \mu_1}\over 2}) (\widehat{q_1-q_2})_{\mu_3} \delta_{\mu_1 \mu_2}
+ {\rm 2 \ cyclic\ permutations} \Big) \, .
\ee
\item
4-gluon vertex
\bea
&-&{\lambda N \over 4!\Ve }\delta(q_1+q_2+q_3+q_4) \hA_{\mu_1}(q_1)\hA_{\mu_2}(q_2) \hA_{\mu_3}(q_3) \hA_{\mu_4}(q_4) \\ && e^{-\frac{i}{2} (q_{1\mu_1}+q_{2\mu_2}+q_{3\mu_3}+q_{4\mu_4} )}
\Big\{ F(q_1,q_2,-q_1-q_2) F(q_3,q_4,-q_3-q_4) \nonumber \\
&\Big(&V_{\mu_1 \mu_2 \mu_3\mu_4}(q_1,q_2,q_3,q_4) -
V_{\mu_2 \mu_1 \mu_3\mu_4}(q_2,q_1,q_3,q_4)\Big)\nonumber \\
&-& {1\over 24 } W_{\mu_1 \mu_2 \mu_3\mu_4}(q_1,q_2,q_3,q_4) \nonumber \\
& \Big (&D(q_1,q_2,-q_1-q_2) D(q_3,q_4,-q_3-q_4)+ (q_2\leftrightarrow q_3) +
(q_2\leftrightarrow q_4) \Big) \Big \} \, ,\nonumber
\eea
where
\bea
V_{\mu_1 \mu_2 \mu_3 \mu_4}(q_1,q_2,q_3,q_4) &=& f_{\mu_1} \delta_{\mu_1 \mu_2 \mu_3 \mu_4} + (g_{\mu_1 \mu_4} \delta_{\mu_1 \mu_2 \mu_3 } + {\rm 3 \ cyclic\ perms})\\
&+& h_{\mu_1 \mu_2} \delta_{\mu_1 \mu_3 } \delta_{\mu_2 \mu_4 }
+(h_{\mu_1 \mu_3}^{'} \delta_{\mu_1 \mu_2 } \delta_{\mu_3 \mu_4 } + {\rm 1 \ cyclic\ perm}) \quad , \nonumber
\eea
with:
\bea
f_{\mu_1} &=& {1 \over 6} \Big ( \widehat{(q_1+q_3)}^2
- \half \widehat{(q_1+q_2)}^2 -\half \widehat{(q_1+q_4)}^2 + \sum_\rho \widehat q_{1\rho} \widehat q_{2\rho}
\widehat q_{3\rho}\widehat q_{4\rho} \Big)\quad , \\
g_{\mu \nu} &=& {1 \over 6} \Big(\cos\Big({q_{3\nu}\over 2}\Big) \widehat{(q_1-q_2)}_\nu
-\cos\Big({ q_{1\nu}\over 2}\Big) \widehat{(q_2-q_3)}_\nu
- \widehat q_{1\nu } \widehat q_{2\nu }\widehat q_{3\nu }\Big)\widehat q_{4\mu } \ \ \quad ,\\
h_{\mu \nu} &=& 2 \cos\Big({q_{2\mu}-q_{4\mu}\over 2}\Big) \cos\Big({ q_{1\nu}-q_{3\nu}\over 2}\Big) \quad ,\\
h_{\mu \nu}^{'} &=& -\cos\Big({q_{3\mu}-q_{4\mu}\over 2} \Big) \cos\Big({q_{1\nu}-q_{2\nu}\over 2} \Big)
+{1 \over 4} \widehat q_{3\mu } \widehat q_{4\mu }\widehat q_{1\nu}\widehat q_{2\nu} \quad .
\eea
and
\bea
&&W_{\mu_1 \mu_2 \mu_3\mu_4}(q_1,q_2,q_3,q_4) = 2\,  \delta_{\mu_1 \mu_2 \mu_3\mu_4}  \sum_\rho \widehat q_{1\rho}
\widehat q_{2\rho} \widehat q_{3\rho} \widehat q_{4\rho} \\
&+& \Big( - 2\, \delta_{\mu_1 \mu_2 \mu_3}  \widehat q_{1\mu_4} \widehat q_{2\mu_4}
\widehat q_{3\mu_4} \widehat q_{4\mu_1} + {\rm 3 \ cyclic\ permutations} \Big) \nonumber \\
&+& \Big (2\, \delta_{\mu_1 \mu_2}  \delta_{\mu_3 \mu_4}  \widehat q_{3\mu_1} \widehat q_{4\mu_1}
\widehat q_{1\mu_3} \widehat q_{2\mu_3} + (q_2\leftrightarrow q_3) +
(q_2\leftrightarrow q_4) \Big)
\nonumber
\eea
\end{itemize}

With adjoint Wilson fermions and up to order $\lambda$, one should include two additional vertices, given by:
\be
 \sqrt {\frac{\lambda N}{\Ve} } F(q_1,q_2, q_3)\, \delta(q_1+q_2+q_3) \, \bar \psi^j(q_1)V_\mu^{ffg} (q_1,q_2,q_3) \hA_\mu(q_2) \psi^j(q_3) e^{-i \frac{q_{2\mu}}{2}} \, ,
\ee
and 
\bea
&&{\lambda N \over 2\Ve } \, 
F(q_1,q_2, -q_1-q_2) \, F(q_3,q_4, -q_3-q_4) \, \delta(q_1+q_2+q_3+q_4)\\
&&\bar \psi^j(q_1)\, V_{\mu\nu}^{ffgg} (q_1,q_2,q_3,q_4)\,  \psi^j(q_3) \,  \hA_\mu(q_2) \hA_\nu(q_4) e^{-i \frac{q_{2\mu}+q_{4\nu}}{2}}\, ,
\nonumber
\eea
where
\be
V_\mu^{ffg} (q_1,q_2,q_3) =    
\Big ( i r \sin{(q_3-q_1)_\mu \over 2} - \gamma_\mu \cos {(q_3-q_1)_\mu \over 2} \Big )\, ,
\ee
and 
\be
V_{\mu \nu}^{ffgg} (q_1,q_2,q_3,q_4) =  - \delta_{\mu \nu}  
 \Big ( r \cos {(q_3-q_1)_\mu \over 2} - i \gamma_\mu \sin {(q_3-q_1)_\mu \over 2} \Big ) \, .
\ee
The adjoint Wilson fermion propagator reads:
\be
S_f(q) = {M(q) - i \sum_\mu \gamma_\mu \sin(q_\mu) \over M^2(q) + \sum_\mu \sin^2(q_\mu) }\, ,
\ee
with
\be
M(q) = {1 \over 2 \kappa} - r \sum_\mu \cos q_\mu \equiv 
m + {r \over 2} \, \widehat q^2 \, .
\ee
where $m= 1/(2\kappa) - r d$.

\section{ The vacuum polarization at ${\cal O}(\lambda^2)$}
\label{appendixB}

For completeness we give the expression for the vacuum polarization for the Wilson action up to order $\lambda^2$ as  derived by Snippe in Ref. \cite{Snippe:1997ru}, generalized to a twisted box with an arbitrary irreducible twist.

\bea
\Pi_{\muu \nuu}^{\rm mes } (q)&=&  - {1 \over 12 }  \delta_{\muu \nuu}\quad \label{eq.mes},\\
\Pi_{\muu \nuu}^{\rm gh1} (q) &=&- {N \over 6 \Ve}    \delta_{\muu \nuu} \sum_{p} \FSS\, \,{\widehat p_\muu^2 \over \widehat p^2} \quad ,\\
\Pi_{\muu \nuu}^{\rm gh2} (q)&=&
 {N \over 4 \Ve}   \sum_{p} \FSS \, \,
{\widehat q_\muu \widehat q_\nuu - \widehat{(2p +q)}_\muu \widehat{(2p + q)}_\nuu \over \widehat p^2 \widehat 
{(p+q)}^2}\quad ,\\
\Pi_{\muu \nuu}^{ \rm V3} (q)&=&  {N \over 2 \Ve}   
\sum_{p} \FSS\,\,  {V_{\muu \lambda \rho}(q,p,-p-q) V_{\nuu \lambda \rho}(q,p,-p-q) \over \widehat p^2 \widehat {(p+q)}^2} \quad ,\\
\Pi_{\muu \nuu}^{ \rm V4}(q) &=& {N \over 3 \Ve}  \sum_{p} \FSS
\times \nonumber \\
&&{1 \over \widehat p^2}   \Big(V_{\lambda \lambda \muu \nuu}(p, -p, q,-q) - V_{\lambda \muu \lambda \nuu}(p, q,-p,-q)\Big) \quad ,\\
\Pi_{\muu \nuu}^{ \rm W_1}(q) &=& -{N \over 12 \Ve } 
\sum_{p} 
\FSS    \times \nonumber \\
&&{1 \over \widehat p^\sq}  \, \widehat q_\rho \, (\widehat q_\rho \delta_{\muu \nuu} -\widehat q_\muu \delta_{\rho \nuu}) \,
(\widehat p_{\muu}^\sq  + \widehat p_{\rho}^\sq ) \quad , \\
\Pi_{\muu \nuu}^{ \rm W_2}(q) &=& {1 \over 4 \Ve } 
\sum_{p}' {1 \over \widehat p^\sq}  \, \widehat q_\rho \, (\widehat q_\rho \delta_{\muu \nuu} -\widehat q_\muu \delta_{\rho \nuu}) \,
(\widehat p_{\muu}^\sq  + \widehat p_{\rho}^\sq ) \quad ,
\label{eq.abel}
\eea
with $V_{\mu_1 \mu_2 \mu_3}(p_1,p_2,p_3)$ and $V_{\mu_1 \mu_2 \mu_3 \mu_4}(p_1, p_2,p_3,p_4)$ given by the 3 and 4-gluon vertices 
in App.~\ref{appendixA}.

\subsection{ The contribution of adjoint Wilson fermions}
\label{appendixB2}

The fermionic contribution to the vacuum polarization is proportional to the number of fermion flavours $N_f$. For $N_f=1$ adjoint Wilson fermions and up to order $\lambda^2$, it includes two terms given by:

\bea
\Pi_{\muu \nuu}^{f_1} (q) &=&   { d N \, \over \Ve }  \,  \delta_{\muu \nuu} \sum_{p} \FSS\, 
{ M(p) \, r \,\cos p_\muu - \sin^2 p_\muu \over M^2 (p) + \sum_\rho \sin^2 p_\rho}\, ,\label{eq.vacf1}\\
\Pi_{\muu \nuu}^{f_2} (q) &=&   {d N \, \over \Ve} \,    \sum_{p} { \FSS \, N_{\muu \nuu}(q,p) \over (M^2 (p) + \sum_\rho \sin^2 p_\rho)(M^2 (p+q) + \sum_\rho \sin^2 (p+q)_\rho)} \label{eq.vacf2}\, ,
\eea
where 
\bea
N_{\muu \nuu}(q,p) &=&
 \delta_{\muu \nuu} \cos^2 {(2p+q)_\muu \over 2}  \Big ( M(p) M(p+q) + \sum_\rho \sin p_\rho \sin (p+q)_\rho \Big ) \\
&-&\frac{r^2}{4}  \widehat{(2 p +q)}_\muu  \widehat{(2 p +q)}_\nuu \Big (  M(p) M(p+q) - \sum_\rho \sin p_\rho \sin (p+q)_\rho \Big ) \nonumber\\
&-& \Big \{ \frac{r}{2} \widehat{(2 p +q)}_\muu \cos {(2p+q)_\nuu \over 2} \Big ( M(p) \sin(p+q)_\nuu + M(p+q) \sin(p_\nuu) \Big ) \nonumber\\
&+&  \cos {(2p+q)_\muu\over 2} \cos {(2p+q)_\nuu\over 2} \sin (p_\muu) \sin(p+q)_\nuu + \muu \leftrightarrow \nuu \Big\}\quad .\nonumber
\eea

Under the substitution Eq.~\eqref{typeNA} and the change in trace normalization of the fermion representation, our formulas reproduce the infinite volume results by Kawai, Nakayama and Seo in Ref.~\cite{Kawai:1980ja}.

\section{Evaluating finite volume corrections}
\label{appendixC}

Here we consider the evaluation of quantities of the type
\be
\mathcal{I}(\Lv)=\frac{1}{\V} \sum_{p\in \Ls'} \FI (p) 
\ee
where the momenta $p$ are d-dimensional vectors where each component has the form $p_\mu=\frac{2 \pi m_\mu}{L_\mu}$ with integer $m_\mu$ ranging from $0$ to $L_\mu-1$. The sum extends over all values of $m_\mu$ except when all of them vanish simultaneously. Finally $\V$ stands for the volume, which is equal to the product of all $L_\mu$. The goal would be to obtain the large volume behaviour of  $\mathcal{I}(L)$, and in particular the corrections to the infinite volume limit.

Traditionally in dealing with sums of our type one makes use of the Euler-MacLaurin formula. Here however our  integrands are periodic (with period $2\pi$) in each variable.  
To treat this type of integrals we use the following expression
$$ \mathcal{I}(\Lv)= \int D\alpha\,  \FI (\alpha) \frac{1}{\V}\sum_{p\in \Ls'} \delta(\alpha-p)= \int
D\alpha\,  \FI (\alpha) \frac{1}{\V} \sum_{p\in \Ls'} \sum_{l\in \mathbb{Z}^d} e^{i l(\alpha -p)} $$
where $D\alpha$ is the product of $d\alpha_\mu/2\pi$ over all
directions. The next step is to sum over $p$, giving
$$ 
\frac{1}{\V} \sum_{p\in \Ls'}\, e^{-i lp} = \prod_\mu \left(\sum_{n_\mu \in \mathbb{Z}} \delta(l_\nu-n_\mu L_\mu)\right) - \frac{1}{\V}
$$
Plugging the expression onto the formula we get 
\be
\mathcal{I}(\Lv)=\int D\alpha\,  \FI (\alpha)  \left(\prod_\mu \sum_{n_\mu\in \mathbb{Z} }   e^{i L_\mu n_\mu \alpha_\mu } -\frac{1}{\V} \sum_{l \in 
\mathbb{Z}^d} e^{i l\alpha}\right)
\ee
The last term is equal to $\FI (0)/\V$, unless $\FI (0)$ diverges. Notice
that the term in the first sum corresponding to $n_\mu=0$ for all $\mu $ gives 
the infinite volume limit of the expression, which we assume to be
convergent. This is the leading contribution in the Euler-MacLaurin formula. Thus, we get an exact expression for the finite volume corrections to our integral as follows
\be 
\delta \mathcal{I}(\Lv)= \mathcal{I}(\Lv)-\mathcal{I}(\infty)= \int D\alpha\,  \FI (\alpha) \sum'_n
e^{i \Lv n\alpha} -\frac{\FI (0)}{V} 
\ee
where the prime means the sum over all d-dimensional vectors $n \in \mathbb{Z}^d$, excluding $n=0$. The argument of the exponential is a simplified form meaning $\sum_\mu n_\mu L_\mu \alpha_\mu$.

\subsection{The finite volume propagator}
\label{appendixCC}
Now we can apply our formalism to the study of the expectation values of Wilson loops. Our main integral under consideration comes from using as integrand the following expression 
\be 
\FI (\alpha)=\frac{e^{i\alpha l}}{(\tilde{D}(\alpha))^\beta}
\ee 
where $\tilde{D}(\alpha)=4 \sum_{\mu}  \sin^2(\alpha_\mu/2) +m^2 =2d
-2\sum_{\mu} \cos(\alpha_\mu)+ m^2 $. For $\beta=1$ the corresponding sum $\mathcal{I}$ is nothing but the propagator of a scalar particle of mass $m$ on a finite lattice. The mass is necessary to have a well defined value of $\FI (0)$. In all the main expressions that we will use later it would be possible to take the limit $m\longrightarrow 0$.
Working with $\beta$ different from 1 allows to evaluate other expressions (like the measure contribution to the vacuum polarization given by $\beta=2$)  and  can also act as a regulator.

Now we can use Schwinger trick to recast the integrand as an integral as follows
\be 
\FI (\alpha)= \int_0^\infty dx\,  \frac{x^{\beta-1}}{\Gamma(\beta)}\ 
e^{i\alpha l-x\tilde{D}(\alpha)}=  \frac{1}{\Gamma(\beta)}\int_0^\infty dx\,x^{\beta-1} e^{-x m^2} \prod_\mu \left( e^{i \alpha_\mu l_\mu -2x (1-\cos(\alpha_\mu))} \right) 
\ee
The factorized exponentials can then be treated as the  integrand and we can apply our formalism to it. The infinite volume limit result is then given by
\be 
P(l,\beta)\equiv \int_0^\infty dx\,  \frac{x^{\beta-1}}{\Gamma(\beta)}\ \exp\{-x (2d +m^2)\}\ \prod_{\mu} I_{l_\mu}(2x) 
\ee
where $I_l$ is the modified Bessel function. For $\beta=1$ this is just the lattice propagator at infinite volume. The finite volume propagator has then the form of the sum of the propagators to all replica points $l_\mu+n_\mu L_\mu$. The lattice propagator at large distances is well-approximated by the continuum one. This follows from the asymptotic expansion of the Bessel functions at large values of $x$. The leading term is as follows:
\be
e^{-2x} I_{l_\mu}(2x) = \frac{1}{\sqrt{4 \pi x}} e^{-l_\mu^2/(4 x) }+ \ldots
\ee
Thus, the leading finite volume correction to  $P(l,\beta)$ is given by  
\be
\label{finiteprop}
\delta P( l,\beta,\Lv)= -\frac{1}{m^{2 \beta}\V} +\int_0^\infty dx\, \frac{x^{\beta-1}}{\Gamma(\beta)}\ e^{-x m^2}\ e^{-l^2/(4x)}
\left(\prod_\mu G(x,z_\mu, L_\mu) - \frac{1}{(4 \pi x)^{d/2}} \right) +\ldots
\ee
where $z_\mu=l_\mu/L_\mu$ and the function $G(x,z,L)$ is given by
\bea 
\nonumber
G(x,z,L)=\frac{e^{z^2 L^2/(4x)}}{\sqrt{4 \pi x}} \sum_{n \in \mathbb{Z}} e^{-(n+z)^2L^2/(4x)} =\\ \frac{1}{\sqrt{4 \pi x}}  \vartheta(i\frac{z L^2}{4 \pi x};i \frac{L^2}{4 \pi x})= \frac{e^{z^2L^2/(4 x)}}{L}\vartheta(z;i\frac{4 \pi x}{L^2}) 
\eea
 where $\vartheta(z;i\tau)$ is the Jacobi theta function, whose duality relation has been used for the last equality. The first term on the right-hand side of Eq.~\eqref{finiteprop} is just the $\FI(0)/V$ subtraction. When the sizes go to infinity uniformly as follows $L_\mu=\lambda_\mu \bar{L}$, it is clear that the integrand of Eq.~\eqref{finiteprop} is strongly suppressed whenever $x/\bar{L}^2 \ll 1$. 
 Thus, we can change variables to $y=x/\bar{L}^2$ and restrict the integral to go from $\epsilon$ to infinity. This gives
\be
-\frac{1}{m^{2 \beta}\V} + \frac{1}{\bar{L}^{d-2\beta}} \int_\epsilon^\infty dy\, \frac{y^{\beta-1-d/2}}{\Gamma(\beta)(4 \pi )^{d/2}}\ e^{-y \bar{L}^2 m^2- \sum_\mu \frac{\lambda_\mu^2 z_\mu^2}{4y}}\ 
 \left(\prod_\mu \vartheta(i z\frac{\lambda_\mu^2}{4 \pi y};i \frac{\lambda_\mu^2}{4 \pi y})) - 1 \right) +\ldots
\ee
The convergence of the integral at small $y$ is guaranteed by the behaviour of the quantity inside parenthesis irrespective of the other factors. At large $y$ the mass term guarantees convergence. Taking the mass to zero produces a divergence of the integral, coming from the large $x$ behaviour  of $\vartheta(z;i\frac{4 \pi x}{L^2}) \longrightarrow 1$. Its value is subtracted by the leading first term giving a convergent result. Hence one can combine these  terms and give a 
result after take the limit of vanishing mass and $\epsilon$ in this combination. 
\be 
\label{propformula}
\delta P( l,\beta,\Lv)= \frac{\bar{L}^{2\beta}}{\V} \int_0^\infty dy\, \frac{y^{\beta-1}}{\Gamma(\beta)}\  
 \left(\prod_\mu \vartheta(z_\mu;i \frac{4 \pi y}{\lambda_\mu^2})) - 1 -\frac{\V e^{-l^2/(4y\bar{L}^2)} e^{-m^2 \bar{L}^2 y} }{(4\pi y)^{d/2}\bar{L}^d}\right) 
\ee
We have still preserved the mass dependence in the last term. It is necessary to ensure its convergence at large $y$ whenever $d\le 2 \beta$. This is the same range for which the massless infinite volume limit does not exist. Outside this range one can savely set $m=0$ in Eq.~\eqref{propformula}.

\subsection{Leading order of Wilson loop}
Now we can apply the formalism to the expectation value of Wilson
loops at lowest order. The integrand is given by 
$$
\FI(\alpha)= \frac{\sin^2(\alpha_0T/2) \sin^2 (\alpha_1
R/2)}{\sin^2(\alpha_0/2)\,  (\tilde{D}(\alpha))^\beta} + (\stackrel{R\leftrightarrow T}{1\leftrightarrow 0}) 
$$
with $\beta=1$.
It is possible to write the expression as follows 
$$
F(\alpha)=  \frac{|e^{i \alpha_1 R}-1|^2 |\sum_{k=0}^{T-1} e^{i
k \alpha_0}|^2}{ 4 (\tilde{D}(\alpha))^\beta} + (\stackrel{R\leftrightarrow T}{1\leftrightarrow 0})
$$
which in the presence of a mass term vanishes at $\alpha=0$. Notice that we take the loop in the $0-1$ plane with size $T$ and $R$ respectively.

We can relate the calculation to the previous one of the $P(l;\beta,\Lv)$
We introduce the 
displacement operator $\delta_1$ which adds $1$ to $l_1$. We call 
$\delta_1^{-1}$ the inverse operator which displaces by $-1$. In an
analogous fashion $\delta_0$ displaces $l_0$. With this notation 
the finite volume correction to the leading contribution to the wilson
loop is given by 
$$
\frac{1}{4}\left(  (-\delta_1^R-\delta_1^{-R}+2 ) \sum_{k=-T+1}^{T-1} (T-|k|)
\delta_0^k\,\right)\,  P(l;\beta,\Lv)  + (\stackrel{R\leftrightarrow T}{1\leftrightarrow 0})
$$
In order to obtain the leading result it is interesting to consider the limit in which  $z_1=R/L_1$ is treated as a small quantity.  The operator can then be expanded as 
$$ (-\delta_1^R-\delta_1^{-R}+2 )= -z_1^2 \partial_1^2
-\frac{1}{12} z_1^4 \partial_1^4 +\ldots $$
where $\partial_1$ is the derivative with respect to $z_1$ (treated as
a continuum variable). If we now apply the same procedure to the operator along the time direction, we get 
$$
\sum_{k=-T+1}^{T-1} (T-|k|) \delta_0^k = T^2 + \frac{1}{L^2} \partial_0^2 \sum_{k=1}^{T-1} k^2 (T-|k|)
+\ldots= T^2\left( 1+ \frac{T^2-1}{L^2}\partial_0^2 +\ldots\right) 
$$
We are now in position to compute the leading correction to the Wilson loop. 
All we have to is to apply the operator 
$$-\frac{R^2T^2}{4 L_1^2} \partial_1^2 -\frac{R^2T^2}{4 L_0^2} \partial_0^2 $$
to our previous expression $P(l,\beta;\Lv)$ and then set $z=0$. We can make use of the result
$$ \partial_1^2 \vartheta(z_1;i\frac{4 \pi y}{\lambda_1^2})\big|_{z_1=0}= \lambda_1^2 \frac{\partial}{\partial y}    \vartheta(0;i\frac{4 \pi y}{\lambda_1^2})\equiv \lambda_1^2 \vartheta'(0;i\frac{4 \pi y}{\lambda_1^2}) $$
Thus, the correction to the Wilson loop ($\beta=1$) is 
\be
\delta W(\beta)= 
-\frac{R^2 T^2 \bar{L}^{2\beta}}{4 L_1^2 \V} \int_0^\infty dy\, \frac{y^{\beta-1}}{\Gamma(\beta)}\  
 \left(\lambda_1^2 \vartheta'(0;i\frac{4 \pi y}{\lambda_1^2})\prod_{\mu\ne 1} \vartheta(0;i \frac{4 \pi y}{\lambda_\mu^2}))  + \frac{\V \lambda_1^2 }{2 y(4\pi y)^{d/2}\bar{L}^d}\right) +0
\leftrightarrow 1
\ee
in which the mass has been set to zero, assuming that $\beta-1 < d/2$.
In the particular case in which all lengths are equal $L_\mu=\bar{L}$, there is considerable simplification since the expression inside the parenthesis becomes a total derivative
\be
\frac{1}{d}\frac{\partial}{\partial y}\left( \vartheta^d(0;i 4 \pi y)-1-\frac{1}{(4\pi y)^{d/2}}\right)
\ee
For the Wilson loop  ($\beta=1$) the result is just given by the value of the function at the limits ( -1 at $y=0$) giving 
\be
\delta W(1)=-\frac{T^2 R^2}{2 d \V}
\ee

For the measure term ($\beta=2$) the formula for the symmetric case can be obtained by  integration by parts  
and gives 
\be 
F_{\mathrm{mes}}(L)-F_{\mathrm{mes}}(\infty)=-\frac{\delta W(2)}{12}=-\frac{T^2 R^2}{24 d L^{d-2}} \int_0^\infty dy \ \left( \vartheta^d(0;i 4 \pi y)-1-\frac{1}{(4\pi y)^{d/2}}\right)
\ee 
The two dimensional case is the only physical case for which the previous expression is not valid. This is so because there is no massless  infinite volume limit. To deal with this case one has to go back to the expression including a non-zero mass.  The divergence comes from the last term which when integrated from $y=1$ gives the incomplete gamma function $\Gamma(0,m^2 L^2)=-\log(m^2)-\log(L^2)-\gamma+ \ldots$. The mass singularity cancels with that coming from the infinite volume limit quantity and shows that 
\be 
F_{\mathrm{mes}}(L)=-\frac{R^2 T^2}{96\pi}\log(L)+\mathrm{constant}+ \ldots 
\ee
which is used in the text. 

\section{Non-abelian contributions}
\label{appendixD}

In this section we develop the methodology to study $L$ and $N$
dependence of expressions of the form
\be
\mathcal{I}'(\mathbf{L})=\frac{1}{N^3 V^2}\sum_{p_c,q_c\in \LLe/\Ls}\sum_{p_s\in\Ls}\sum_{q_s\in\Ls} 
{\cal A}(p,q) F^2(p_c,q_c,-p_c-q_c)
\ee
where the momenta $p=p_s+p_c$, with $p_s$ and $p_c$ are the spatial and color momenta respectively. The integrand is an unspecified function ${\cal A}(p,q)$, which is assumed to be periodic of period $2\pi$ in each of the arguments and regular everywhere.  Finally $F(p_c,q_c,-p_c-q_c)$ is the characteristic  structure constant in the colour momentum basis.  We recall that with  the hermiticity convention that we are using one has
\be 
F^2(p_c,q_c,-p_c-q_c)= \frac{1}{N} (1-\cos(2\theta(p_c,q_c)))=\frac{1}{N} (1-\cos(\Phi(p_c,q_c)-\Phi(q_c,p_c)))
\ee
It is convenient to consider the two terms separately as we did in the analysis of the results performed in section~\ref{s.analysis}. The first part contributed to all functions labelled $NA$. They are called like that because they arise from terms involving the $F^2$, which are structure constants of the non-abelian group. In this case there is no colour factor and the corresponding sum will be given by 
\be
\mathcal{I}_{NA}(\mathbf{L})=\frac{1}{\Ve^2}\sum_{p,q\in \LLe} {\cal A}(p,q) 
\ee

The part containing the cosine was labelled NP, standing for non-planar, since this twist dependent factors only occurs in the non-planar part of the diagram. The resulting expression is 
\be
\mathcal{I}_{NP}(\mathbf{L})=-\frac{1}{\Ve^2}\sum_{p_c,q_c\in \LLe/\Ls}\sum_{p_s\in\Ls}\sum_{q_s\in\Ls} 
{\cal A}(p,q) \cos(2 \theta(p_c,q_c))
\ee

To start with, let us consider the infinite volume limit of the non-planar part. We can use the Euler MacLaurin result, or the analysis performed in the previous appendix to conclude that it is given by 
\be
-\frac{1}{N^4} \int D\alpha\, \int D\beta\, \sum_{p_c,q_c\in \LLe/\Ls}  {\cal A}(\alpha+p_c,\beta+q_c)  \cos(2 \theta(p_c,q_c))
\ee
where the symbol $D\alpha$ stands for the product of $d\alpha_\mu/(2\pi)$ over all directions, and the integrals extend from $0$ to $2 \pi$. Now using the translational invariance of the integration measure and the periodicity property one can rewrite 
the result in factorized form
\be 
\mathcal{I}_{NP}(\infty)=-\int D\alpha\, \int D\beta\,  {\cal A}(\alpha,\beta) \left( \frac{1}{N^4} \sum_{p_c,q_c\in \LLe/\Ls} \cos(2 \theta(p_c,q_c)) \right)
\ee
Let us now evaluate the colour factor. The cosine can be written as the sum of two phases. The first  can be related to the $\HG$ matrices as follows  
\be 
\label{PHASE}
2 \Tr(\HG(p_c)\HG(q_c) \HG^\dagger(p_c)
\HG^\dagger(q_c))= \frac{1}{2N}   e^{2 i \theta(p_c,q_c)} 
\ee 
Now one can make use of the completeness relation 
\be
\label{completeness}
\sum_{p_c\ne 0} (\HG(p_c))_{ij} (\HG^\dagger(p_c))_{kl} = \frac{1}{2}
\delta_{jk} \delta_{il} -\frac{1}{2N} \delta_{ij} \delta_{kl}
\ee
which can be easily proven by tracing the expression with the generators $\Gamma_\mu$. Applying  this result to Eq.~\eqref{PHASE} we conclude 
$$
\sum_{p_c\ne 0}  e^{2 i \theta(p_c,q_c)} = -1 +N^2 \delta(q_c)
$$ 
Adding the contribution of $p_c=0$ kills the $-1$ on the right hand side. Now summing over $q_c$ we conclude 
\be
\label{sumcos}
\frac{1}{N^4}\sum_{p_c}\sum_{q_c} \cos(2\theta(p_c,q_c))= \frac{1}{N^2}
\ee
which as expected is suppressed with respect to the planar part.
Using the previous result one can deduce the following result 
\be
\sum_{p_c\ne 0} F^2(p_c,q_c,-p_c-q_c)= N (1-\delta(q_c))
\ee
which is just the well-known value of the quadratic Casimir in the adjoint representation. Using Eq.~\eqref{sumcos} one can obtain  the infinite volume limit of the non-planar part, given by 
\be
\mathcal{I}_{NP}(\infty)=-\frac{1}{N^2}\int D\alpha\, \int D\beta\,  {\cal A}(\alpha,\beta)
\ee
which is just the infinite volume limit of the original expression but omitting the colour degrees of freedom. 

One can go beyond this result and try to evaluate the finite volume corrections to this non-planar part. We can use the formalism introduced in the previous appendix to replace all sums over space momenta by integrals.
Finally,  we get 
\be
\mathcal{I}_{NP}(\mathbf{L})= -\int D\alpha\, \int D\beta\, {\cal A}(\alpha,\beta) \frac{1}{N^4
}\sum_{p_c}\sum_{q_c} \cos(2\theta(p_c,q_c))
\sum_{\tilde{n}\in \mathbb{Z}^d}\sum_{\tilde{m}\in \mathbb{Z}^d}
e^{i L \tilde{n}(\alpha-p_c)} e^{i L \tilde{m}(\beta-q_c)}
\ee
This can be factorised as follows:
$$\mathcal{I}_{NP}(\mathbf{L})= \sum_{\tilde{n}}\sum_{\tilde{m}} G(\tilde{n},\tilde{m})
H(\tilde{n},\tilde{m})
$$ 
where 
$$G(\tilde{n},\tilde{m})= \int D\alpha\, \int D\beta\, {\cal
A}(\alpha,\beta) e^{i L \tilde{n}\alpha} e^{i L
\tilde{m}\beta}
$$ 
and 
$$H(\tilde{n},\tilde{m})=- \frac{1}{N^4}\sum_{p_c}\sum_{q_c} \cos(2\theta(p_c,q_c))
e^{-i L \tilde{n} p_c} e^{-i L \tilde{m}q_c}
$$
Let us now evaluate this function. This can be done by realizing that 
$$ 
e^{2i \theta(p_c,q_c)}
e^{-i L \tilde{n} p_c} e^{-i L \tilde{m}q_c}=
4N \Tr(\Gamma^\dagger(\tilde{n})\HG(p_c)\Gamma(\tilde{n}) \Gamma^\dagger(\tilde{m})\HG(q_c)\Gamma(\tilde{m}) \HG^\dagger(p^c)
\HG^\dagger(q^c))
$$
with 
$$
\Gamma(\tilde{n})=\Gamma_0^{\tilde{n}_0} \cdots \Gamma_{d-1}^{\tilde{n}_{d-1}}
$$
Now when summing over $p_c$ and $q_c$ and making use of the completeness relation Eq.~\eqref{completeness} we find 
$$
\sum_{p_c}\sum_{q_c} e^{2i \theta(p_c,q_c)}
e^{-i L \tilde{n} p_c} e^{-i L \tilde{m}q_c}= N
\Tr(\Gamma(\tilde{m})\Gamma(\tilde{n})\Gamma^\dagger(\tilde{m})\Gamma^\dagger(\tilde{n})=N^2\exp\{-2 \pi i n_{\mu \nu} \tilde{m}_\mu \tilde{n}_\nu/N\}
$$
Repeating the calculation with the complex conjugate phase and summing both results we end up with 
\be
H(\tilde{n},\tilde{m})=-\frac{1}{N^2} \cos(2 \pi  n_{\mu \nu} \tilde{m}_\mu \tilde{n}_\nu/N)
\ee
which only depends on $\tilde{m}_\mu$ or $\tilde{n}_\nu$ modulo N. 
Thus, our final expression becomes 
\be 
\mathcal{I}_{NP}(\mathbf{L})= -\frac{1}{N^2} \sum_{\tilde{n}}\sum_{\tilde{m}} G(\tilde{n},\tilde{m})
\cos(2 \pi  n_{\mu \nu} \tilde{m}_\mu \tilde{n}_\nu/N)
\ee
The previously obtained infinite volume result corresponds to taking $ \tilde{m}_\mu=\tilde{n}_\mu=0$ for all $\mu$. Excluding this value from the sum we get the finite volume correction.

Let us process our result a bit more by realizing that the cosine only depends on the arguments modulo N. Hence, one can split the integers as follows $ \tilde{m}_\mu= l_\mu+ N \hat{n}_\mu$. Indeed for the symmetric  twist cases the argument applies with $\hL$ replacing $N$. Thus,   we can rewrite 
\be
\mathcal{I}_{NP}(\mathbf{L})= \int D\alpha\, \int D\beta\, {\cal A}(\alpha,\beta) \mathcal{H}(L\alpha,L\beta,\hL,n_{\mu \nu}) 
\sum_{\hat{n}}\sum_{\hat{m}}
e^{i \LL \hL \hat{n}\alpha} e^{i \LL \hL \hat{m}\beta}
\ee
where 
\be
 \mathcal{H}(L\alpha,L\beta,\hL,n_{\mu \nu}) =-\frac{1}{N^2}\sum_{l} \sum_{l'} e^{ilL\alpha+i l'L\beta} \cos(2 \pi  n'_{\mu \nu} l_\mu l'_
 \nu/\hL)
\ee
The formula is valid for any twist if we identify $\hL$ with $N$ and $n'_{\mu \nu}$ with the twist tensor. However, for the four dimensional symmetric twist it is more convenient to take $\hL=\sqrt{N}$ and $n'_{\mu \nu}=n_{\mu \nu}/\hL$.

The function  $\mathcal{H}$ is an oscillatory function with periods proportional to $1/L$. Let us now restrict ourselves to the symmetric twist case in a symmetric box of size $L$ in both 2 and 4 dimensions. In that case the tensor $n'_{\mu \nu}=k\epsilon_{\mu \nu}$ where $k$ is an integer coprime with $\hL$ and $\epsilon_{\mu \nu}$
is an invertible antisymmetric matrix. We might redefine $\tilde{l}_\mu=\epsilon_{\mu \nu} l'_\nu$. Due to the invertibility, the range over which  $\tilde{l}_\mu$ runs coincides with that of $l'_\mu$. We can also change variables from $\beta$ to $\tilde{\beta}$ given by 
\be 
\tilde{\beta}_\mu=-\tilde{\epsilon}_{\mu \nu} \beta_\nu
\ee
where $\tilde{\epsilon}$ is the inverse of $\epsilon$. After these changes the function $\mathcal{H}$ takes a factorizable form
\be
 \mathcal{H}(L\alpha,L\beta,\hL,n_{\mu \nu}) =\prod_\mu \chi(L\alpha_\mu,L\tilde{\beta}_\mu,\hL,k)
\ee
where 
\be
\chi(x,y,\hL,k)=-\frac{1}{\hL}\sum_{l=0}^{\hL-1}
\sum_{l'=0}^{\hL-1} e^{ i (lx+l'y)} \cos(2\pi l l' k/\hL) 
\ee

\bibliography{wilson.bib}

\providecommand{\href}[2]{#2}\begingroup\raggedright\begin{thebibliography}{10}

\bibitem{GonzalezArroyo:1981vw}
A.~Gonz\'alez-Arroyo, J.~Jurkiewicz, and C.~Korthals-Altes, {\it {Ground state
  metamorphosis for Yang-Mills fields on a finite periodic lattice}},  {\em
  Freiburg ASI 1981:0339} (1981).

\bibitem{Luscher:1982ma}
M.~Luscher, {\it {Some Analytic Results Concerning the Mass Spectrum of
  Yang-Mills Gauge Theories on a Torus}},  {\em Nucl.Phys.} {\bf B219} (1983)
  233--261.

\bibitem{Luscher:1983gm}
M.~Luscher and G.~Munster, {\it {Weak Coupling Expansion of the Low Lying
  Energy Values in the SU(2) Gauge Theory on a Torus}},  {\em Nucl.Phys.} {\bf
  B232} (1984) 445.

\bibitem{Coste:1985mn}
A.~Coste, A.~Gonz\'alez-Arroyo, J.~Jurkiewicz, and C.~Korthals~Altes, {\it
  {Zero momentum contribution to Wilson loops in periodic boxes}},  {\em
  Nucl.Phys.} {\bf B262} (1985) 67.

\bibitem{vanBaal:1986ag}
P.~van Baal and J.~Koller, {\it {{QCD} on a Torus and Electric Flux Energies
  From Tunneling}},  {\em Annals Phys.} {\bf 174} (1987) 299.

\bibitem{Koller:1987fq}
J.~Koller and P.~van Baal, {\it {A Nonperturbative Analysis in Finite Volume
  Gauge Theory}},  {\em Nucl.Phys.} {\bf B302} (1988) 1.

\bibitem{Heller:1984hx}
U.~M. Heller and F.~Karsch, {\it {One Loop Perturbative Calculation of Wilson
  Loops on Finite Lattices}},  {\em Nucl.Phys.} {\bf B251} (1985) 254.

\bibitem{DiGiacomo:1981wt}
A.~Di~Giacomo and G.~Rossi, {\it {Extracting the Vacuum Expectation Value of
  the Quantity $\langle (\alpha / \pi) G G \rangle$ from Gauge Theories on a
  Lattice}},  {\em Phys.Lett.} {\bf B100} (1981) 481.

\bibitem{Weisz:1982zw}
P.~Weisz, {\it {Continuum Limit Improved Lattice Action for Pure Yang-Mills
  Theory. 1.}},  {\em Nucl.Phys.} {\bf B212} (1983) 1.

\bibitem{Weisz:1983bn}
P.~Weisz and R.~Wohlert, {\it {Continuum Limit Improved Lattice Action for Pure
  Yang-Mills Theory. 2.}},  {\em Nucl.Phys.} {\bf B236} (1984) 397.

\bibitem{Wohlert:1984hk}
R.~Wohlert, P.~Weisz, and W.~Wetzel, {\it {Weak Coupling Perturbative
  Calculations of the Wilson Loop for the Standard Action}},  {\em Nucl.Phys.}
  {\bf B259} (1985) 85.

\bibitem{Alles:1998is}
B.~Alles, A.~Feo, and H.~Panagopoulos, {\it {Asymptotic scaling corrections in
  QCD with Wilson fermions from the three loop average plaquette}},  {\em Phys.
  Lett.} {\bf B426} (1998) 361--366,
  [\href{http://xxx.lanl.gov/abs/hep-lat/9801003}{{\tt hep-lat/9801003}}].
  [Erratum: Phys. Lett.B553,337(2003)].

\bibitem{'tHooft:1979uj}
G.~'t~Hooft, {\it {A Property of Electric and Magnetic Flux in Nonabelian Gauge
  Theories}},  {\em Nucl.Phys.} {\bf B153} (1979) 141.

\bibitem{'tHooft:1980dx}
G.~'t~Hooft, {\it {Confinement and Topology in Nonabelian Gauge Theories}},
  {\em Acta Phys.Austriaca Suppl.} {\bf 22} (1980) 531--586.

\bibitem{Groeneveld:1980tt}
J.~Groeneveld, J.~Jurkiewicz, and C.~Korthals~Altes, {\it {Twist as a Probe for
  Phase Structure}},  {\em Phys.Scripta} {\bf 23} (1981) 1022.

\bibitem{Eguchi:1982nm}
T.~Eguchi and H.~Kawai, {\it {Reduction of Dynamical Degrees of Freedom in the
  Large N Gauge Theory}},  {\em Phys.Rev.Lett.} {\bf 48} (1982) 1063.

\bibitem{Bhanot:1982sh}
G.~Bhanot, U.~M. Heller, and H.~Neuberger, {\it {The Quenched Eguchi-Kawai
  Model}},  {\em Phys.Lett.} {\bf B113} (1982) 47.

\bibitem{Gross:1982at}
D.~J. Gross and Y.~Kitazawa, {\it {A Quenched Momentum Prescription for Large N
  Theories}},  {\em Nucl.Phys.} {\bf B206} (1982) 440.

\bibitem{Parisi:1982gp}
G.~Parisi, {\it {A Simple Expression for Planar Field Theories}},  {\em Phys.
  Lett.} {\bf 112B} (1982) 463--464.

\bibitem{GonzalezArroyo:1982ub}
A.~Gonz\'alez-Arroyo and M.~Okawa, {\it {A Twisted Model for Large $N$ Lattice
  Gauge Theory}},  {\em Phys.Lett.} {\bf B120} (1983) 174.

\bibitem{GonzalezArroyo:1982hz}
A.~Gonz\'alez-Arroyo and M.~Okawa, {\it {The Twisted Eguchi-Kawai Model: A
  Reduced Model for Large N Lattice Gauge Theory}},  {\em Phys.Rev.} {\bf D27}
  (1983) 2397.

\bibitem{GonzalezArroyo:1983ac}
A.~Gonz\'alez-Arroyo and C.~Korthals~Altes, {\it {Reduced Model for Large $N$
  Continuum Field Theories}},  {\em Phys.Lett.} {\bf B131} (1983) 396.

\bibitem{Eguchi:1982ta}
T.~Eguchi and R.~Nakayama, {\it {Simplification of Quenching Procedure for
  Large $N$ Spin Models}},  {\em Phys.Lett.} {\bf B122} (1983) 59.

\bibitem{Connes:1997cr}
A.~Connes, M.~R. Douglas, and A.~S. Schwarz, {\it {Noncommutative geometry and
  matrix theory: Compactification on tori}},  {\em JHEP} {\bf 02} (1998) 003,
  [\href{http://xxx.lanl.gov/abs/hep-th/9711162}{{\tt hep-th/9711162}}].

\bibitem{Douglas:2001ba}
M.~R. Douglas and N.~A. Nekrasov, {\it {Noncommutative field theory}},  {\em
  Rev. Mod. Phys.} {\bf 73} (2001) 977--1029,
  [\href{http://xxx.lanl.gov/abs/hep-th/0106048}{{\tt hep-th/0106048}}].

\bibitem{Ambjorn:1999ts}
J.~Ambjorn, Y.~M. Makeenko, J.~Nishimura, and R.~J. Szabo, {\it {Finite N
  matrix models of noncommutative gauge theory}},  {\em JHEP} {\bf 11} (1999)
  029, [\href{http://xxx.lanl.gov/abs/hep-th/9911041}{{\tt hep-th/9911041}}].

\bibitem{Ambjorn:2000nb}
J.~Ambjorn, Y.~M. Makeenko, J.~Nishimura, and R.~J. Szabo, {\it
  {Nonperturbative dynamics of noncommutative gauge theory}},  {\em Phys.
  Lett.} {\bf B480} (2000) 399--408,
  [\href{http://xxx.lanl.gov/abs/hep-th/0002158}{{\tt hep-th/0002158}}].

\bibitem{Ambjorn:2000cs}
J.~Ambjorn, Y.~M. Makeenko, J.~Nishimura, and R.~J. Szabo, {\it {Lattice gauge
  fields and discrete noncommutative Yang-Mills theory}},  {\em JHEP} {\bf 05}
  (2000) 023, [\href{http://xxx.lanl.gov/abs/hep-th/0004147}{{\tt
  hep-th/0004147}}].

\bibitem{Fabricius:1985jw}
K.~Fabricius and C.~P. Korthals~Altes, {\it {Reduction of fermion-gluon systems
  on extended lattices}},  {\em Nucl. Phys.} {\bf B269} (1986) 97--108.

\bibitem{Luscher:1985zq}
M.~Luscher and P.~Weisz, {\it {Computation of the Action for On-Shell Improved
  Lattice Gauge Theories at Weak Coupling}},  {\em Phys. Lett.} {\bf 158B}
  (1985) 250--254.

\bibitem{Luscher:1985wf}
M.~Luscher and P.~Weisz, {\it {Efficient Numerical Techniques for Perturbative
  Lattice Gauge Theory Computations}},  {\em Nucl.Phys.} {\bf B266} (1986) 309.

\bibitem{Coste:1986cb}
A.~Coste, A.~Gonz\'alez-Arroyo, C.~Korthals~Altes, B.~Soderberg, and
  A.~Tarancon, {\it {Finite Size Effects and Twisted Boundary Conditions}},
  {\em Nucl.Phys.} {\bf B287} (1987) 569.

\bibitem{Hansson:1986ia}
T.~H. Hansson, P.~van Baal, and I.~Zahed, {\it {Chromomagnetic Energy of SU(2)
  Gauge Fields on a Torus}},  {\em Nucl. Phys.} {\bf B289} (1987) 628--644.

\bibitem{GonzalezArroyo:1988dz}
A.~Gonz\'alez~Arroyo and C.~Korthals~Altes, {\it {The Spectrum of Yang-Mills
  Theory in a Small Twisted Box}},  {\em Nucl.Phys.} {\bf B311} (1988) 433.

\bibitem{Daniel:1989kj}
D.~Daniel, A.~Gonzalez-Arroyo, C.~P. Korthals~Altes, and B.~Soderberg, {\it
  {Energy Spectrum of SU(2) {Yang-Mills} Fields With Space - Like Symmetric
  Twist}},  {\em Phys. Lett.} {\bf B221} (1989) 136--142.

\bibitem{Daniel:1990iz}
D.~Daniel, A.~Gonzalez-Arroyo, and C.~P. Korthals~Altes, {\it {The Energy
  levels of lattice gauge theory in a small twisted box}},  {\em Phys. Lett.}
  {\bf B251} (1990) 559--566.

\bibitem{Snippe:1996bk}
J.~R. Snippe, {\it {Square Symanzik action to one loop order}},  {\em
  Phys.Lett.} {\bf B389} (1996) 119--120,
  [\href{http://xxx.lanl.gov/abs/hep-lat/9608146}{{\tt hep-lat/9608146}}].

\bibitem{Snippe:1997ru}
J.~R. Snippe, {\it {Computation of the one loop Symanzik coefficients for the
  square action}},  {\em Nucl.Phys.} {\bf B498} (1997) 347--396,
  [\href{http://xxx.lanl.gov/abs/hep-lat/9701002}{{\tt hep-lat/9701002}}].

\bibitem{Perez:2013dra}
M.~Garc\'{\i}a~P\'erez, A.~Gonz\'alez-Arroyo, and M.~Okawa, {\it {Spatial
  volume dependence for 2+1 dimensional SU(N) Yang-Mills theory}},  {\em JHEP}
  {\bf 1309} (2013) 003, [\href{http://xxx.lanl.gov/abs/1307.5254}{{\tt
  arXiv:1307.5254}}].

\bibitem{GarciaPerez:2016guo}
M.~Garcia~Perez, A.~Gonzalez-Arroyo, and M.~Okawa, {\it {Volume reduction
  through perturbative Wilson loops}},  {\em PoS} {\bf LATTICE2016} (2016) 329,
  [\href{http://xxx.lanl.gov/abs/1611.0720}{{\tt arXiv:1611.0720}}].

\bibitem{Kiskis:2003rd}
J.~Kiskis, R.~Narayanan, and H.~Neuberger, {\it {Does the crossover from
  perturbative to nonperturbative physics in QCD become a phase transition at
  infinite N}},  {\em Phys.Lett.} {\bf B574} (2003) 65--74,
  [\href{http://xxx.lanl.gov/abs/hep-lat/0308033}{{\tt hep-lat/0308033}}].

\bibitem{Narayanan:2003fc}
R.~Narayanan and H.~Neuberger, {\it {Large N reduction in continuum}},  {\em
  Phys.Rev.Lett.} {\bf 91} (2003) 081601,
  [\href{http://xxx.lanl.gov/abs/hep-lat/0303023}{{\tt hep-lat/0303023}}].

\bibitem{ishikawa:2003}
T.~Ishikawa and M.~Okawa, {\it {$Z_N$ symmetry breaking on the numerical
  simulation of twisted Eguchi-Kawai model}},  {\em talk given at the Annual
  Meeting of the Physical Society of Japan, March 28–31, Sendai, Japan}
  (2003).

\bibitem{Teper:2006sp}
M.~Teper and H.~Vairinhos, {\it {Symmetry breaking in twisted Eguchi-Kawai
  models}},  {\em Phys. Lett.} {\bf B652} (2007) 359--369,
  [\href{http://xxx.lanl.gov/abs/hep-th/0612097}{{\tt hep-th/0612097}}].

\bibitem{Vairinhos:2007qz}
H.~Vairinhos and M.~Teper, {\it {Structure and properties of the vacuum of the
  Twisted Eguchi-Kawai model}},  {\em PoS} {\bf LAT2007} (2007) 282,
  [\href{http://xxx.lanl.gov/abs/0710.3337}{{\tt arXiv:0710.3337}}].

\bibitem{Azeyanagi:2007su}
T.~Azeyanagi, M.~Hanada, T.~Hirata, and T.~Ishikawa, {\it {Phase structure of
  twisted Eguchi-Kawai model}},  {\em JHEP} {\bf 01} (2008) 025,
  [\href{http://xxx.lanl.gov/abs/0711.1925}{{\tt arXiv:0711.1925}}].

\bibitem{GonzalezArroyo:2010ss}
A.~Gonz\'alez-Arroyo and M.~Okawa, {\it {Large $N$ reduction with the Twisted
  Eguchi-Kawai model}},  {\em JHEP} {\bf 1007} (2010) 043,
  [\href{http://xxx.lanl.gov/abs/1005.1981}{{\tt arXiv:1005.1981}}].

\bibitem{Gonzalez-Arroyo:2014dua}
A.~Gonz\'alez-Arroyo and M.~Okawa, {\it {Testing volume independence of SU(N)
  pure gauge theories at large N}},  {\em JHEP} {\bf 1412} (2014) 106,
  [\href{http://xxx.lanl.gov/abs/1410.6405}{{\tt arXiv:1410.6405}}].

\bibitem{Perez:2014sqa}
M.~Garc\'{\i}a~P\'erez, A.~Gonz\'alez-Arroyo, and M.~Okawa, {\it {Volume
  independence for Yang-Mills fields on the twisted torus}},  {\em
  Int.J.Mod.Phys.} {\bf A29} (2014), no.~25 1445001,
  [\href{http://xxx.lanl.gov/abs/1406.5655}{{\tt arXiv:1406.5655}}].

\bibitem{Chamizo:2016msz}
F.~Chamizo and A.~Gonzalez-Arroyo, {\it {Tachyonic instabilities in 2+1
  dimensional Yang–Mills theory and its connection to number theory}},  {\em
  J. Phys.} {\bf A50} (2017), no.~26 265401,
  [\href{http://xxx.lanl.gov/abs/1610.0797}{{\tt arXiv:1610.0797}}].

\bibitem{Martin:1999aq}
C.~P. Martin and D.~Sanchez-Ruiz, {\it {The one loop UV divergent structure of
  U(1) Yang-Mills theory on noncommutative R**4}},  {\em Phys. Rev. Lett.} {\bf
  83} (1999) 476--479, [\href{http://xxx.lanl.gov/abs/hep-th/9903077}{{\tt
  hep-th/9903077}}].

\bibitem{Krajewski:1999ja}
T.~Krajewski and R.~Wulkenhaar, {\it {Perturbative quantum gauge fields on the
  noncommutative torus}},  {\em Int. J. Mod. Phys.} {\bf A15} (2000)
  1011--1030, [\href{http://xxx.lanl.gov/abs/hep-th/9903187}{{\tt
  hep-th/9903187}}].

\bibitem{SheikhJabbari:1999iw}
M.~M. Sheikh-Jabbari, {\it {Renormalizability of the supersymmetric Yang-Mills
  theories on the noncommutative torus}},  {\em JHEP} {\bf 06} (1999) 015,
  [\href{http://xxx.lanl.gov/abs/hep-th/9903107}{{\tt hep-th/9903107}}].

\bibitem{Minwalla:1999px}
S.~Minwalla, M.~Van~Raamsdonk, and N.~Seiberg, {\it {Noncommutative
  perturbative dynamics}},  {\em JHEP} {\bf 02} (2000) 020,
  [\href{http://xxx.lanl.gov/abs/hep-th/9912072}{{\tt hep-th/9912072}}].

\bibitem{Matusis:2000jf}
A.~Matusis, L.~Susskind, and N.~Toumbas, {\it {The IR / UV connection in the
  noncommutative gauge theories}},  {\em JHEP} {\bf 12} (2000) 002,
  [\href{http://xxx.lanl.gov/abs/hep-th/0002075}{{\tt hep-th/0002075}}].

\bibitem{Guralnik:2001pv}
Z.~Guralnik and J.~Troost, {\it {Aspects of gauge theory on commutative and
  noncommutative tori}},  {\em JHEP} {\bf 05} (2001) 022,
  [\href{http://xxx.lanl.gov/abs/hep-th/0103168}{{\tt hep-th/0103168}}].

\bibitem{Guralnik:2002ru}
Z.~Guralnik, R.~C. Helling, K.~Landsteiner, and E.~Lopez, {\it {Perturbative
  instabilities on the noncommutative torus, Morita duality and twisted
  boundary conditions}},  {\em JHEP} {\bf 05} (2002) 025,
  [\href{http://xxx.lanl.gov/abs/hep-th/0204037}{{\tt hep-th/0204037}}].

\bibitem{Bietenholz:2006cz}
W.~Bietenholz, J.~Nishimura, Y.~Susaki, and J.~Volkholz, {\it {A
  Non-perturbative study of 4-D U(1) non-commutative gauge theory: The Fate of
  one-loop instability}},  {\em JHEP} {\bf 10} (2006) 042,
  [\href{http://xxx.lanl.gov/abs/hep-th/0608072}{{\tt hep-th/0608072}}].

\bibitem{review}
A.~Gonz{\'a}lez-Arroyo, {\it {Yang-Mills fields on the 4-dimensional Torus.
  Part I: Classical Theory}},  {\em {World Scientific. Proceedings of the
  Pe\~niscola 1997 advanced school on non-perturbative quantum field physics}}
  (1998) {Singapore}, [\href{http://xxx.lanl.gov/abs/hep-th/9807108}{{\tt
  hep-th/9807108}}].

\bibitem{Groeneveld:1980zx}
J.~Groeneveld, J.~Jurkiewicz, and C.~Korthals~Altes, {\it {Local order
  parameter in twisted gauge fields}},  {\em Phys.Lett.} {\bf B92} (1980)
  312--314.

\bibitem{Dashen:1980vm}
R.~F. Dashen and D.~J. Gross, {\it {The Relationship Between Lattice and
  Continuum Definitions of the Gauge Theory Coupling}},  {\em Phys. Rev.} {\bf
  D23} (1981) 2340. [,246(1980)].

\bibitem{GonzalezArroyo:1981ce}
A.~Gonz\'alez-Arroyo and C.~Korthals-Altes, {\it {Asymptotic Freedom Scales for
  Any Lattice Action}},  {\em Nucl.Phys.} {\bf B205} (1982) 46--76.

\bibitem{Bali:2014fea}
G.~S. Bali, C.~Bauer, and A.~Pineda, {\it {Perturbative expansion of the
  plaquette to ${\cal O}(\alpha^{35})$ in four-dimensional SU(3) gauge
  theory}},  {\em Phys.Rev.} {\bf D89} (2014) 054505,
  [\href{http://xxx.lanl.gov/abs/1401.7999}{{\tt arXiv:1401.7999}}].

\bibitem{Bali:2002wf}
G.~S. Bali and P.~Boyle, {\it {Perturbative Wilson loops with massive sea
  quarks on the lattice}},  \href{http://xxx.lanl.gov/abs/hep-lat/0210033}{{\tt
  hep-lat/0210033}}.

\bibitem{Perez:2014jra}
M.~Garc\'{\i}a~P\'erez, A.~Gonz\'alez-Arroyo, M.~Koren, and M.~Okawa, {\it
  {Glueball masses in 2+1 dimensional SU(N) gauge theories with twisted
  boundary conditions}},  {\em PoS} {\bf LATTICE2014} (2014) 059,
  [\href{http://xxx.lanl.gov/abs/1411.5186}{{\tt arXiv:1411.5186}}].

\bibitem{Perez:2014isa}
M.~Garc\'{\i}a~P\'erez, A.~Gonz\'alez-Arroyo, L.~Keegan, and M.~Okawa, {\it
  {The $SU(\infty)$ twisted gradient flow running coupling}},  {\em JHEP} {\bf
  1501} (2015) 038, [\href{http://xxx.lanl.gov/abs/1412.0941}{{\tt
  arXiv:1412.0941}}].

\bibitem{Fabricius:1984wp}
K.~Fabricius and O.~Haan, {\it {Heat Bath Method for the Twisted {Eguchi-Kawai}
  Model}},  {\em Phys. Lett.} {\bf 143B} (1984) 459--462.

\bibitem{Perez:2015ssa}
M.~García~Pérez, A.~González-Arroyo, L.~Keegan, M.~Okawa, and A.~Ramos, {\it
  {A comparison of updating algorithms for large N reduced models}},  {\em
  JHEP} {\bf 06} (2015) 193, [\href{http://xxx.lanl.gov/abs/1505.0578}{{\tt
  arXiv:1505.0578}}].

\bibitem{Alles:1993dn}
B.~Alles, M.~Campostrini, A.~Feo, and H.~Panagopoulos, {\it {The Three loop
  lattice free energy}},  {\em Phys.Lett.} {\bf B324} (1994) 433--436,
  [\href{http://xxx.lanl.gov/abs/hep-lat/9306001}{{\tt hep-lat/9306001}}].

\bibitem{Kovtun:2007py}
P.~Kovtun, M.~Unsal, and L.~G. Yaffe, {\it {Volume independence in large N(c)
  QCD-like gauge theories}},  {\em JHEP} {\bf 06} (2007) 019,
  [\href{http://xxx.lanl.gov/abs/hep-th/0702021}{{\tt hep-th/0702021}}].

\bibitem{Kawai:1980ja}
H.~Kawai, R.~Nakayama, and K.~Seo, {\it {Comparison of the Lattice Lambda
  Parameter with the Continuum Lambda Parameter in Massless QCD}},  {\em Nucl.
  Phys.} {\bf B189} (1981) 40--62.

\bibitem{Catterall:2010gx}
S.~Catterall, R.~Galvez, and M.~Unsal, {\it {Realization of Center Symmetry in
  Two Adjoint Flavor Large-N Yang-Mills}},  {\em JHEP} {\bf 08} (2010) 010,
  [\href{http://xxx.lanl.gov/abs/1006.2469}{{\tt arXiv:1006.2469}}].

\bibitem{Azeyanagi:2010ne}
T.~Azeyanagi, M.~Hanada, M.~Unsal, and R.~Yacoby, {\it {Large-N reduction in
  QCD-like theories with massive adjoint fermions}},  {\em Phys. Rev.} {\bf
  D82} (2010) 125013, [\href{http://xxx.lanl.gov/abs/1006.0717}{{\tt
  arXiv:1006.0717}}].

\bibitem{Hietanen:2010fx}
A.~Hietanen and R.~Narayanan, {\it {Large-$N$ reduction of SU($N$) Yang–Mills
  theory with massive adjoint overlap fermions}},  {\em Phys. Lett.} {\bf B698}
  (2011) 171--174, [\href{http://xxx.lanl.gov/abs/1011.2150}{{\tt
  arXiv:1011.2150}}].

\bibitem{Hietanen:2009ex}
A.~Hietanen and R.~Narayanan, {\it {The large N limit of four dimensional
  Yang-Mills field coupled to adjoint fermions on a single site lattice}},
  {\em JHEP} {\bf 01} (2010) 079,
  [\href{http://xxx.lanl.gov/abs/0911.2449}{{\tt arXiv:0911.2449}}].

\bibitem{Lohmayer:2013spa}
R.~Lohmayer and R.~Narayanan, {\it {Weak-coupling analysis of the single-site
  large-N gauge theory coupled to adjoint fermions}},  {\em Phys. Rev.} {\bf
  D87} (2013), no.~12 125024, [\href{http://xxx.lanl.gov/abs/1305.1279}{{\tt
  arXiv:1305.1279}}].

\bibitem{Basar:2013sza}
G.~Basar, A.~Cherman, D.~Dorigoni, and M.~Ünsal, {\it {Volume Independence in
  the Large $N$ Limit and an Emergent Fermionic Symmetry}},  {\em Phys. Rev.
  Lett.} {\bf 111} (2013), no.~12 121601,
  [\href{http://xxx.lanl.gov/abs/1306.2960}{{\tt arXiv:1306.2960}}].

\bibitem{Bringoltz:2009kb}
B.~Bringoltz and S.~R. Sharpe, {\it {Non-perturbative volume-reduction of
  large-N QCD with adjoint fermions}},  {\em Phys. Rev.} {\bf D80} (2009)
  065031, [\href{http://xxx.lanl.gov/abs/0906.3538}{{\tt arXiv:0906.3538}}].

\bibitem{Koren:2013aya}
M.~Koren, {\em {Volume reduction in large-N lattice gauge theories [with
  adjoint fermions]}}.
\newblock PhD thesis, Jagiellonian U., 2013.
\newblock \href{http://xxx.lanl.gov/abs/1312.5351}{{\tt arXiv:1312.5351}}.

\bibitem{Bringoltz:2011by}
B.~Bringoltz, M.~Koren, and S.~R. Sharpe, {\it {Large-N reduction in QCD with
  two adjoint Dirac fermions}},  {\em Phys. Rev.} {\bf D85} (2012) 094504,
  [\href{http://xxx.lanl.gov/abs/1106.5538}{{\tt arXiv:1106.5538}}].

\bibitem{Gonzalez-Arroyo:2013bta}
A.~Gonz\'alez-Arroyo and M.~Okawa, {\it {Twisted space-time reduced model of
  large N QCD with two adjoint Wilson fermions}},  {\em Phys.Rev.} {\bf D88}
  (2013) 014514, [\href{http://xxx.lanl.gov/abs/1305.6253}{{\tt
  arXiv:1305.6253}}].

\bibitem{Sannino:2004qp}
F.~Sannino and K.~Tuominen, {\it {Orientifold theory dynamics and symmetry
  breaking}},  {\em Phys. Rev.} {\bf D71} (2005) 051901,
  [\href{http://xxx.lanl.gov/abs/hep-ph/0405209}{{\tt hep-ph/0405209}}].

\bibitem{Luty:2004ye}
M.~A. Luty and T.~Okui, {\it {Conformal technicolor}},  {\em JHEP} {\bf 09}
  (2006) 070, [\href{http://xxx.lanl.gov/abs/hep-ph/0409274}{{\tt
  hep-ph/0409274}}].

\bibitem{Bursa:2009we}
F.~Bursa, L.~Del~Debbio, L.~Keegan, C.~Pica, and T.~Pickup, {\it {Mass
  anomalous dimension in SU(2) with two adjoint fermions}},  {\em Phys. Rev.}
  {\bf D81} (2010) 014505, [\href{http://xxx.lanl.gov/abs/0910.4535}{{\tt
  arXiv:0910.4535}}].

\bibitem{DelDebbio:2010hx}
L.~Del~Debbio, B.~Lucini, A.~Patella, C.~Pica, and A.~Rago, {\it {The infrared
  dynamics of Minimal Walking Technicolor}},  {\em Phys. Rev.} {\bf D82} (2010)
  014510, [\href{http://xxx.lanl.gov/abs/1004.3206}{{\tt arXiv:1004.3206}}].

\bibitem{Perez:2015yna}
M.~Garc\'{\i}a~P\'erez, A.~Gonz\'alez-Arroyo, L.~Keegan, and M.~Okawa, {\it
  {Mass anomalous dimension of Adjoint QCD at large N from twisted volume
  reduction}},  {\em JHEP} {\bf 08} (2015) 034,
  [\href{http://xxx.lanl.gov/abs/1506.0653}{{\tt arXiv:1506.0653}}].

\end{thebibliography}\endgroup

\end{document}